\documentclass[pdflatex,iicol,sn-mathphys-num]{sn-jnl}

\usepackage{graphicx}%
\usepackage{multirow}%
\usepackage{amsmath,amssymb,amsfonts}%
\usepackage{amsthm}%
\usepackage{mathrsfs}%
\usepackage[title]{appendix}%
\usepackage{xcolor}%
\usepackage{textcomp}%
\usepackage{manyfoot}%
\usepackage{booktabs}%
\usepackage{algorithm}%
\usepackage{algorithmicx}%
\usepackage{algpseudocode}%
\usepackage{listings}%
\usepackage{orcidlink}

\newcommand{\rev}[1]{{#1}}

\begin{document}

\title[Reaction Rates for Proton-Rich Nuclides]{Astrophysical Reaction Rates for Charged-Particle Induced Reactions on Proton-Rich Nuclides}


\author[1,2]{\fnm{Thomas} \sur{Rauscher}\orcidlink{0000-0002-1266-0642}}\email{Thomas.Rauscher@unibas.ch}

\affil[1]{\orgdiv{Department of Physics}, \orgname{University of Basel}, \orgaddress{\street{Klingelbergstrasse 82}, \city{Basel}, \postcode{4056}, \country{Switzerland}}}

\affil[2]{\orgdiv{Centre for Astrophysics Research}, \orgname{University of Hertfordshire}, \orgaddress{\street{College Lane}, \city{Hatfield}, \postcode{AL10 9AB}, \country{United Kingdom}}}


\abstract{Astrophysical reaction rates for reactions with proton-rich \rev{isotopes of Ne to Bi} from stability to the proton dripline were calculated with an updated version of the SMARAGD statistical model (Hauser-Feshbach) code. Here, the focus was on reactions with protons or $\alpha$ particles as required for nucleosynthesis in proton-rich matter. For completeness, also neutron-induced reactions are provided for the same set of targets.
Some comments on dependencies of rates on various nuclear properties and on the appropriate way to compare to experiments are given.
The new rate set for charged-particle induced reactions provides a better description of experimental data than previously widely used rates, especially for reactions involving $\alpha$ particles.}




\maketitle

\section{Introduction}

Nuclear reaction rates provide the basis for all astrophysical investigations concerning energy generation and nucleosynthesis in the Early Universe as well as in hydrostatic and explosive nuclear burning of stars and stellar objects. They are derived from experimentally measured and/or theoretically predicted nuclear reaction cross sections. A direct experimental determination of reaction cross sections at the interaction energies relevant to astrophysical environments is possible only if the involved nuclei are accessible for reaction studies, i.e., are stable or have sufficiently long half-lives, cross sections are sufficiently large, and thermally excited states of the target nuclei do not contribute considerably to the reaction rate in an astrophysical plasma. Therefore the majority of astrophysical reaction rates used in astrophysical simulations are based on theory and may rely only indirectly on experimentally obtained information required to inform the theoretical treatment of the nuclear properties used in the reaction models. Depending on the relative energy and the level density of the interacting nuclei, different reaction models are employed. For target nuclides of intermediate and high mass number ($A\gtrapprox 20$) the relevant excitation energy of the compound system formed from the combination of projectile and target in most cases is high enough \rev{(see Section \ref{sec:statmodel}) to apply} the statistical model of nuclear reactions, originally proposed by Bohr \cite{1936Natur.137..344B} and later refined by \cite{1940PhRv...57..472W,1952PhRv...87..366H}. At low level density at the compound formation energy or at high interaction energy, individual resonances and direct reaction mechanisms have to be considered, respectively (see, e.g., \cite{2011IJMPE..20.1071R,2020entn.book.....R} for an overview).

The statistical model approach following \cite{1952PhRv...87..366H} (also called Hauser-Feshbach model) has been applied
in the calculation of thermonuclear reaction rates for astrophysical
purposes by many researchers, with early investigations only making use of ground-state properties of nuclei \cite{1966CaJPh..44..151T}. Soon after, the importance of an inclusion of excited nuclear states was pointed out \cite{1972A&A....19...92A}. Extended compilations have been provided by \cite{1976ADNDT..18..305H,1978ADNDT..22..371W}. More recently, further compilations have been published, based on the codes SMOKER \cite{1986ana..work..525T}, NON-SMOKER \cite{2000ADNDT..75....1R,2001ADNDT..79...47R}, and TALYS \cite{2008A&A...487..767G}. Most recently, the code SMARAGD
\cite{2011IJMPE..20.1071R} has been developed as a successor to
NON-SMOKER, with improvements in the treatment of charged particle and photon transmissions, in
the nuclear level density, and with inclusion of recent experimental information on
low-lying excited states. Reaction rate compilations obtained from calculations using
NON-SMOKER and TALYS
are presently the ones utilized
in large scale applications in all subfields of nuclear astrophysics,
when experimental information is unavailable. SMARAGD has been used to analyse
experimental data and for improvements of constrained sets of reaction rates \cite{2010ApJS..189..240C} but
a large-scale set of reaction rates has not been published yet. In this work, astrophysical reaction rates from SMARAGD for charged-particle induced reactions are presented, superseding and extending the preliminary evaluation given in \cite{2010ApJS..189..240C}. Differences to previous calculations are summarised and general remarks on the dependence of rates on various nuclear properties, also important in the comparison of experimental and theoretical rates, are provided.

\section{Definition of Astrophysical Reaction Rates}
\label{sec:ratedef}

Astrophysical reaction rates count the number of reactions of a specific combination of projectile and target nucleus occurring per volume and time in a plasma with given temperature $T$. The energy distribution of atomic nuclei in astrophysical plasmas is given by the Maxwell-Boltzmann distribution. To obtain a rate, the energy- and temperature-dependent reaction \textit{stellar} cross sections $\sigma^*$ have to be weighted by this energy distribution. Multiplying this reactivity (i.e., rate per particle pair) with the respective number densities $n_\mathrm{a}$, $n_\mathrm{A}$ of the interacting nuclides, the two-body reaction rate
\begin{eqnarray}
    r_\mathrm{a+A}&=&\frac{n_\mathrm{a} n_\mathrm{A}}{1+\tilde{\delta}_\mathrm{aA}}  \langle \sigma^* v \rangle \nonumber \\
    &=&\frac{n_\mathrm{a} n_\mathrm{A}}{1+\tilde{\delta}_\mathrm{aA}}  \nonumber \\
    &\times& \sqrt{\frac{8}{\pi \mu}} \left( \frac{1}{k_\mathrm{B} T} \right)^{\frac{3}{2}} \int \limits_0^\infty \sigma^*(E,T) E e^{-\frac{E}{k_\mathrm{B}T}}\,\mathrm{d}E \nonumber \\
    \label{eq:rate}
\end{eqnarray}
is obtained \cite {2020entn.book.....R}, with $\tilde{\delta}$ being the Kronecker delta to avoid double counting, $\mu$ being the reduced mass of projectile and target nuclide, and $k_\mathrm{B}$ is the Boltzmann constant. The reactivity (rate per particle pair) is defined using angle brackets signifying the averaging over the energy (or velocity) distribution of the interacting nuclides. (For practical application, there is a limited energy range across which the integration has to be performed because no significant contributions to the integral values come from outside this range \cite{2010PhRvC..81d5807R}. This is called the Gamow window.)

The quantity $\sigma^*(E,T)$ is the \textit{stellar} reaction cross section \cite{1974QJRAS..15...82F,1976ADNDT..18..305H,2011IJMPE..20.1071R,2020entn.book.....R}, including reactions not only on the ground states of the interacting nuclei but also on their excited states which are populated by thermal excitation in the plasma,
\begin{equation}
    \sigma^*(E,T)=\frac{1}{G_0(T)}\sum_i \sum_j \frac{2J_i+1}{2J_0+1} \frac{E_i}{E} \sigma_{ij}(E_i)\quad,\label{eq:stellcs}
\end{equation}
summing over reactions leading from target levels $i$ to final levels $j$ with cross sections $\sigma_{ij}$ with $E_i=E-E^\mathrm{x}_i$ being the center-of-mass energy relative to excited level $i$ (the ground state is given by $i=0$). Following \cite{1974QJRAS..15...82F}, $\sigma^{ij}=0$ for $E_i\leq 0$. The spins of the target levels $i$ populated by thermal excitation are denoted by $J_i$ and $G_0(T)$ is the partition function of the target nuclide normalised to the ground state,
\begin{equation}
G_0(T)=\frac{1}{2J_0+1} \sum_i (2J_i+1) e^{-\frac{E^\mathrm{x}_i}{k_\mathrm{B}T}} \quad,
\label{eq:partf}
\end{equation}
with $E^\mathrm{x}_i$ being the excitation energy of level $i$.

\rev{In the stellar cross section, the contributions of excited states are weighted by $(2J_i+1)(E-E^\mathrm{x}_i)/E=(2J_i+1)(1-E^\mathrm{x}_i/E)$. This looks differently compared to the usual Boltzmann population factors exhibiting an exponential suppression depending on excitation energy $(2J_i+1)\exp(-E^\mathrm{x}_i/(k_\mathrm{B}T))$. The stellar cross section in this form is obtained when transforming the Boltzmann-weighted sum of reaction rate integrals for each excited state to the single integral shown in Equation (\ref{eq:rate}). This not only involves an exchange of summation and integration but also requires a substitution of the integration variables, leading to the transformed weights \cite{1974QJRAS..15...82F,1976ADNDT..18..305H,2011IJMPE..20.1071R,2020entn.book.....R}. The use of the stellar cross section not only simplifies the expression for the reactivity but also allows to judge the importance of excited state contributions better than by just examining Boltzmann population factors multiplied by integrals. The single integration of a stellar cross section further provides better numerical accuracy in the calculation of a rate and that is why it has been used in theoretical approaches since \cite{1974QJRAS..15...82F} (see, e.g., \cite{1976ADNDT..18..305H,1978ADNDT..22..371W,1986ana..work..525T,2000ADNDT..75....1R,2008A&A...487..767G,2011IJMPE..20.1071R}).}

Note that the temperature dependence of the stellar cross section solely stems from the temperature dependence of the partition function, reflecting the increasing population of levels at higher excitation energy with increasing temperature. Also note that Equation~(\ref{eq:stellcs}) assumes that the projectile is in its ground state (with its spin degeneracy included in $\sigma_{ij}$). Reactions considered here are two-body reactions with light projectiles, specifically protons and $\alpha$ particles, for which this assumption is justified.

\rev{Depending on the nucleus and the temperature, the sum in Equation (\ref{eq:partf}) has to be extended into a region of excitation energy with insufficient knowledge of discrete excited states. In this case, above the excitation energy $E^\mathrm{x}_\mathrm{last}$ of a last discrete level included, an integration over the density $\rho$ of nuclear levels with spin $J$ and parity $\Pi$ has to be invoked:
\begin{eqnarray}
    G_0(T)&=&\frac{1}{2J_0+1} \left[ \left\{ \sum_{i=0}^{i_\mathrm{last}} (2J_i+1) e^{-\frac{E^\mathrm{x}_i}{k_\mathrm{B}T}} \right\} \right. \nonumber \\
   &+& \left. \int\limits_{E^\mathrm{x}_\mathrm{last}}^\infty \sum_{J\Pi} (2J+1) e^{-\frac{E'}{k_\mathrm{B}}}\rho(E',J,\Pi)\, \mathrm{d}E' \right]. \nonumber \\
\end{eqnarray}
(In practice, there is a finite energy for the upper limit of the integration \cite{2003ApJS..147..403R}.)}

\section{The Statistical Model (Hauser-Feshbach)}
\label{sec:statmodel}

To assure the applicability of the statistical model of nuclear reactions, the nuclear level density at the compound-nucleus formation energy has to be high enough to \rev{assume overlapping resonances that} permit the use of averaged resonance widths. \rev{As can be seen from the sum appearing in Equation (\ref{eq:hf}) below, an additional assumption is that in principle states with all spins and parities are present at the compound formation energy. The statistical model cross section is smooth without exhibiting resonance features. Due to the fact that the calculation of the reaction \textit{rate} involves an integration over an energy range, i.e., an energy average of cross sections (see Equation~\ref{eq:rate}), the criterion for the applicability of the statistical model is more relaxed when applied to stellar rates than to reaction cross sections. A smaller number of contributing resonances can be sufficient when applying the energy average of a statistical model cross section because resonance features in the actual cross section are smoothed out, anyway \cite{1997PhRvC..56.1613R}.}
For the majority of reactions of astrophysical interest across the nuclear chart, it has been shown that the statistical model is applicable to the calculation of astrophysical reaction rates \cite{1997PhRvC..56.1613R}. Exceptions are target nuclides with low mass number, at shell closures (for rates at low $T$), and towards the driplines.

The statistical model cross section for a reaction a+A$^i$$\rightarrow$C$^c$$\rightarrow$B$^j$+b (including transitions from the $i$th level in target nuclide A to the $j$th level in the final nucleus B via compound states c in nuclide C) reads
\begin{eqnarray}
    \sigma^{ij}_\mathrm{HF}(E)&=&\frac{\pi}{k_i^2}
\frac{1+\delta_{Aa}}{(2J_i+1)(2J_\mathrm{a}+1)} \nonumber \\
&&\times \sum_{J\Pi \overline{J}\ell \overline{J}'\ell'}
(2J+1) \frac{\mathcal{T}^i_{J\Pi\ell\overline{J}}(E_i) \mathcal{T}^j_{J\Pi\ell'\overline{J}'}(E_i)}{\sum_{c \ell \overline{J}} \hat{\mathcal{T}}^c_{Jlj}(E_i)} \nonumber \\
&&\times W(E_i) \quad,
\label{eq:hf}
\end{eqnarray}
with the transmission coefficients $\mathcal{T}$ for particle or photon transitions at the compound formation energy $E_i=E+S_\mathrm{a}$ connecting a compound level with spin $J$ and parity $\Pi$ with initial state $i$ and final state $j$, respectively. The separation energy of the projectile in the compound nucleus is denoted by $S_\mathrm{a}$. All quantum mechanically allowed channel spins $\overline{J}$ and partial waves $\ell$ are included in the summation. The denominator contains a sum over total transmission coefficients $\hat{\mathcal{T}}^c$ including all energetically possible transitions from the compound states to any final states (this includes the ones in the nuclides A and B but also in all other nuclides reached by photon and particle emission). The (energy-dependent) factor $W$ is the width fluctuation correction and includes non-statistical correlations between initial and final states. It is of the order of $W=1$ for most energies except close to openings of additional reaction channels.

There is a simple relation between averaged resonance widths $\overline{\Gamma}$ and transmission coefficients,
\begin{equation}
    \mathcal{T}_{J\Pi\ell\overline{J}}(E_i)=2\pi\,\rho(E_i,J,\Pi)\,\overline{\Gamma}_{J\Pi\ell\overline{J}}(E_i)\quad,
    \label{eq:transwidths}
\end{equation}
with the nuclear level density $\rho$ taken at the compound formation energy $E_i$.

As for the partition function (Equation~\ref{eq:partf}), also in the transmission coefficients sums over discrete levels have to be replaced by integrations over a level density above a last known, discrete level. Therefore the level density enters in two ways: first, the level density in the compound nucleus at the compound formation energy determines whether the statistical model actually is applicable; second, the level densities in the various nuclides reachable from the compound levels determine (in part) the transmission coefficients. The impact in the latter case depends on the particle separation energies, as the difference between compound formation energy and particle separation energies determines whether the region of (low-lying) discrete states has to be supplemented with a region covered by using the level density. With a small energy difference, transitions to discrete states dominate the transmission coefficients and the impact of a level density will be small. For a study on the sensitivity of the astrophysical reaction rate to variation of the level density, see \cite{2012ApJS..201...26R}.

\section{Implementation in SMARAGD}
\label{sec:implementation}

Codes for calculating statistical model cross sections mainly differ in the treatment of the nuclear properties entering the calculation of transmission coefficients, as well as in the masses and excited state properties used. Particle transmission coefficients require the use of optical potentials describing the effective interaction between particles absorbed and emitted by a nucleus. Electromagnetic de-excitation of the compound nucleus through emission of $\gamma$ radiation is described by photon-transmission coefficients. These require knowledge of the photon strength-function, specifying the strength of an electromagnetic transition emitting a photon with given energy. For details on the calculation of the transmission coefficients for the two cases, see, e.g., \cite{1976ADNDT..18..305H,2000ADNDT..75....1R,2011IJMPE..20.1071R,2020entn.book.....R}.

Here, results obtained with version v0.42.0s of the code SMARAGD are presented.
The code SMARAGD (Statistical Model for Astrophysical Reactions And Global Direct reactions) is written completely in FORTRAN90/95 (with exception of the routines handling a web interface, which are written in C) and has a modular structure, making changes easy. The latest nuclear data can also swiftly be included through files or web downloads. Code versions below v1.0.0s do not yet use the global approach for including direct capture. This will be presented in future versions. Direct capture is important close to driplines where the particle separation energy $S_\mathrm{a}$ is low and where also few, widely-spaced resonances or no resonances are present. Due to the higher relevant energy range for astrophysical reaction rates with charged particles than for neutrons, translating to higher excitation energy of the compound nucleus, direct capture generally is more important for neutron captures than for charged particle captures in astrophysics \cite{2011IJMPE..20.1071R}.

In addition to an updated implementation of nuclear properties listed below, some numerical improvements have been implemented in SMARAGD. Of particular importance is the implementation of an improved method to calculate charged-particle transmission coefficients, improving the method previously used \cite{1972CoPhC...3...73B} in the NON-SMOKER code by implementing a Fox-Goodwin algorithm (see, e.g., \cite{1988SchmidSpitzLosch,2020entn.book.....R}) to solve the Schr\"odinger equation combined with a reliable routine to compute Coulomb wavefunctions that is stable also far below the Coulomb barrier. This affects transmission coefficients for charged particle reactions below the Coloumb barrier and therefore most of these reactions in astrophysical environments.

Masses were taken from the 2020 Atomic Mass Evaluation (AME2020) \cite{2021ChPhC..45c0002H,2021ChPhC..45c0003W}. Experimentally unknown masses (and microscopic corrections for the calculation of the nuclear level density) were taken from \cite{2016ADNDT.109....1M}.
Experimentally known spins and parities of ground states and low-lying states were taken from \rev{the 2021 ENSDF+Nubase snapshot as provided by RIPL-3 \cite{2009NDS...110.3107C}}. Up to 40 levels were included but only up to the last level for which an unambiguous assignment of spin and parity is present and up to which the level scheme can be considered as being complete. Above the last included discrete level with energy $E_\mathrm{last}^\mathrm{x}$, a theoretical nuclear level density description was employed. The description is based on the approach described in \cite{1997PhRvC..56.1613R,2000ADNDT..75....1R} combining a constant temperature formula for low excitation energies with a backshifted Fermi-gas formula having a level density parameter that depends on the excitation energy \cite{1997PhRvC..56.1613R}. In this previous description, the parities were assumed to be equi-distributed. Here, a parity distribution also depending on the excitation energy is implemented \cite{2007PhRvC..75d5805M}.

The calculation of particle transmission coefficients requires the use of optical model potentials to describe the effective interaction between projectile and target nucleus. For neutrons, the microscopic potential of Lejeune (1980) \cite{1980PhRvC..21.1107L} was used. This is based on the microscopic potential of \cite{1977PhRvC..16...80J} but specifically adapted for low-energy scattering as encountered in astrophysical applications. This approach was used for neutrons and protons also in the SMOKER and NON-SMOKER codes. In SMARAGD, a modified version of the Lejeune (1980) potential is used for protons. Comparisons to experimental data motivated an increase of the depth of the imaginary part \cite{2008PhRvL.101s1101K,2009PhRvC..80c5801R,2010JPhCS.202a2013R}. This may also be adequate for the neutron potential but since no comparison to neutron data has been made so far, it was decided to use the Lejeune (1980) potential for neutrons as previously.

Defining an optical potential for $\alpha$ particles that is at the same time applicable to a large range of nuclides and also at the low, subCoulomb energies encountered in astrophysics has been a long-standing problem. Many suggestions for such general, low-energy $\alpha$+nucleus potentials can be found in literature (see, e.g. \cite{2025EPJA...61...89M} for an overview). The SMOKER and NON-SMOKER codes had used the global potential by \cite{1966NucPh..84..177M}. Despite its overall success, over the past decades it became apparent that improvements are required for low energies. Most recently, the ATOMKI-V2 potential has shown promise in reproducing modern experimental data at low energies \cite{2020PhRvL.124y2701M,MOHR2021101453} and is also implemented in SMARAGD. The real part of the ATOMKI-V2 potential is a folding potential and requires the knowledge of the nuclear matter density distribution in nuclides. Therefore, the folding potentials of \cite{MOHR2021101453} were used where available and theoretical nuclear density distributions \cite{2007PhRvC..75f4312G} were used to compute the required folding potential otherwise.

In the calculation of transmission coefficients for electromagnetic transitions a $\gamma$-strength function is folded with the number of available levels (discrete levels or integrated over a nuclear level density) for given $\gamma$ energies. Included were E1, M1, and E2 transitions. The strength function for the dominant E1 transitions (Giant Dipole Resonance, GDR) was computed using a Lorentzian shape but with a modification of the width at low energy. For deformed nuclides (with deformation taken from \cite{2016ADNDT.109....1M}) the GDR is split into two contributions in the well-established manner (see also \cite{2000ADNDT..75....1R}). Experimentally known GDR energies $E_\mathrm{GDR}$ and widths $\Gamma_\mathrm{GDR}$ were taken from RIPL-3 \cite{2009NDS...110.3107C}. Beyond the experimental knowledge the description of \cite{thielemannarnouldGDR} was used. A straighforward application of the Lorentzian shape would overestimate the $\gamma$-strength at low energy $E_\gamma$. Therefore an energy-dependent GDR width $\Gamma_\mathrm{GDR}'$ was applied, following \cite{1981PhRvC..23.1394M},
\begin{equation}
    \Gamma_\mathrm{GDR}'(E_\gamma)=\Gamma_\mathrm{GDR}\sqrt{\frac{E_\gamma}{E_\mathrm{GDR}}}\quad.
\end{equation}

For the M1 component of the $\gamma$ transitions a simple single-particle approach was used, leading to a dependence on the cube of the $\gamma$ energy \cite{1976ADNDT..18..305H}. The Giant Quadrupole Resonance (GQR) for obtaining E2 transitions was treated as given in \cite{1984ZPhyA.315..103P} (width and peak cross-section) and \cite{1981RPPh...44..719S} (energy).

\section{Results for charged-particle induced reactions}
\label{sec:results}

The datasets obtained with the code SMARAGD are available online, see Section \ref{sec:conclusion}. They include machine-readable tables of
reaction cross sections, astrophysical reaction reactivities, and fit parameters of reactivities for target nuclides from Ne to Bi for isotopes from stability to the proton dripline. The full extension of the calculation can be seen in Figure~\ref{fig:qualifit_proton} (left), where the considered target nuclides are marked by open green squares for unstable targets and open black squares for stable ones.

Astrophysical reaction reactivities (including the effects of thermal excitation of the interacting nuclides) are provided for temperatures from 0.1 to 10 GK. Reaction cross sections for target nuclides in the ground state (g.s.) are provided for centre-of-mass energies up to 14 MeV, as these cover the range of astrophysically important energies \footnote{It is important to note that g.s. cross sections are not suitable for calculating astrophysical reaction rates.}. Fits of astrophysical reaction reactivities are given in the standard REACLIB parameterisation, using seven fit parameters $a_0 - a_6$, with the reactivity to be computed from
\begin{eqnarray}
    &&N_\mathrm{A} \langle \sigma^* v \rangle_\mathrm{a+A\rightarrow b+B}=\exp \left( a_0 + \frac{a_1}{T_9} + \frac{a_2}{T_9^{1/3}}\right. \nonumber \\
    &&\left. + a_3 T_9^{1/3} + a_4 T_9 + a_5 T_9^{5/3} + a_6  \log (T_9) \right)\quad,
    \label{eq:rateparam}
\end{eqnarray}
with $N_\mathrm{A}$ being the Avogadro constant.
This yields $N_\mathrm{A} \langle \sigma^* v \rangle$ in cm$^3$ mole$^{-1}$ s$^{-1}$ when the temperature $T_9$ is given in GK. Fit coefficients for the reactivity of the reverse reaction are also given. Note that for the reverse reactivity the value obtained by the REACLIB parameterisation has to be multiplied by the ratio of the partition functions $G_0^\mathrm{A}/G_0^\mathrm{B}$. Finally, to obtain the actual rate, both for the forward and the reverse reaction, the values for $\langle \sigma^* v \rangle$ have to be multiplied by $(n_\mathrm{a} n_\mathrm{A})/(1+\tilde{\delta}_\mathrm{aA})$ or $(n_\mathrm{b} n_\mathrm{B})/(1+\tilde{\delta}_\mathrm{bB})$, respectively.

Although the emphasis here is on charged-particle induced reactions, the online tables also include neutron-induced reactions on nuclides in the same element and isotope range as for the charged-particle induced reactions. This is to provide a complete dataset for astrophysical applications. Moreover, reaction rates should always be fitted in the direction of positive $Q$ value (except for captures \cite{2009PhRvC..80c5801R,2013JPhCS.420a2138R}) and therefore it is necessary to provide values for the cases where the (p,n) and ($\alpha$,n) reactions have negative reaction $Q$-value.

\section{Comparison to data}

\subsection{Some comments on comparing data to predictions}
\label{sec:sensis}

The use of experimental data is always preferable to using theoretical predictions and a reaction rate set for implementation in nuclear network calculations usually includes a mix of experimental and theoretical reaction rates. However, it is important to realise that only a small subset of the required rates, especially for nuclides in the intermediate and heavy mass range, may be fully constrained by data. On one hand, reaction cross-sections below the Coulomb barrier become tiny and are difficult or currently impossible to measure, even for stable targets. On the other hand, a measurement of g.s.~cross sections, as usually performed in laboratory experiments, may only constrain a small contribution to the astrophysical reaction rate because the contributions of reactions on excited states of the target nuclei often dominate the rate, especially at the elevated temperatures encountered in explosive nucleosynthesis \cite{2011IJMPE..20.1071R,2012ApJS..201...26R,2014AIPA....4d1012R,2015AIPC.1681e0003R}. In such cases, the results of the measurement have to be supplemented or replaced by theory, with the data \rev{providing the means for assessing the validity of the theoretical treatment \textit{only of the g.s.~contribution}}. The online tables also include the g.s.~contribution to the total astrophysical rate as guidance.

\begin{figure*}
    \centering
    \includegraphics[width=\columnwidth]{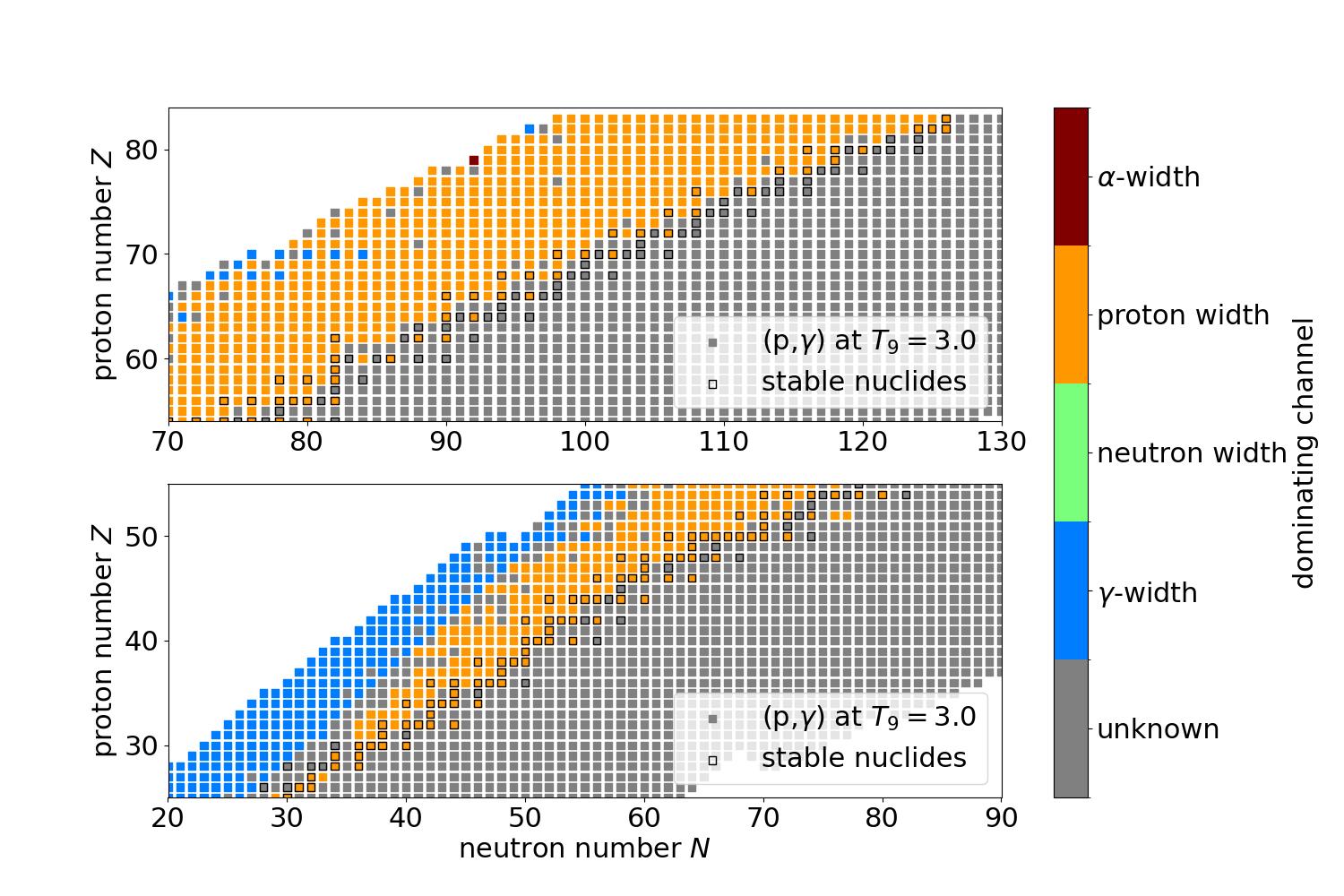}
    \includegraphics[width=\columnwidth]{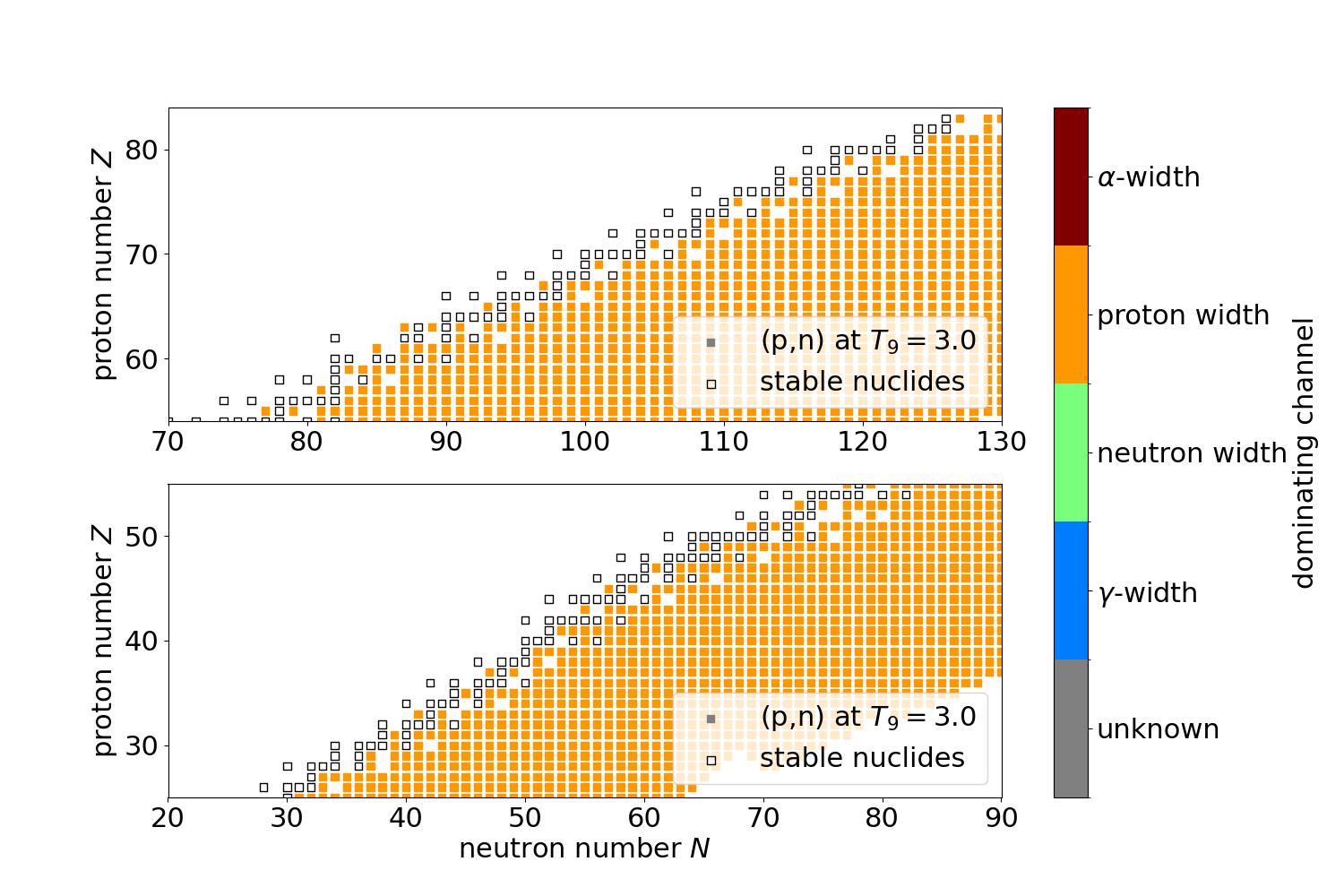}
    \includegraphics[width=\columnwidth]{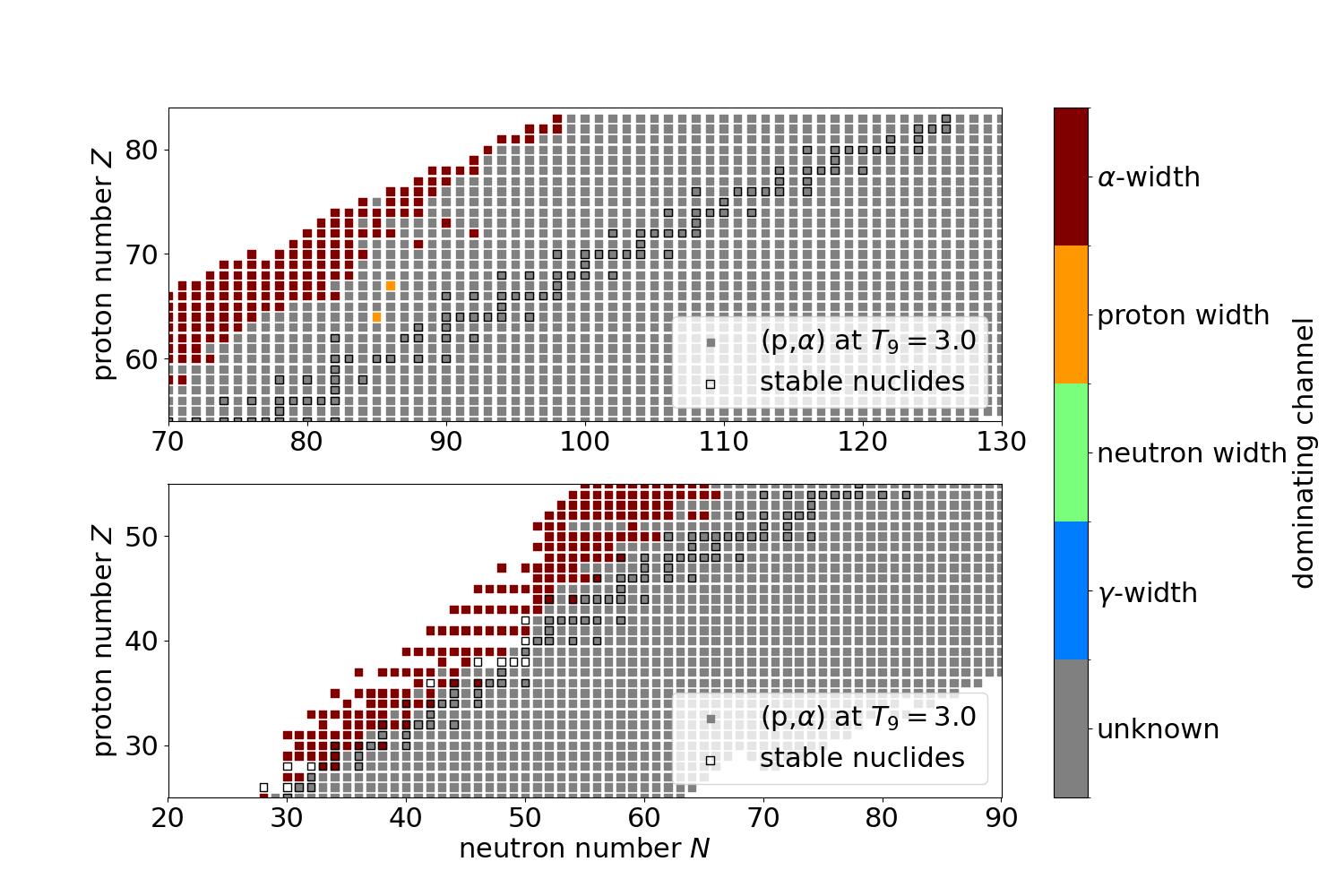}
    \caption{Type of reaction width dominating the size of proton-induced astrophysical reaction rates at 3 GK; the dominating width is identified by the colour of the square for the target nuclide. The assignment "unknown" means that the rate is sensitive to more than one width of comparable size. Except for capture reactions, only reactions with positive $Q$ value are shown. This figure was created using the rate sensitivities given in \cite{2012ApJS..201...26R}.}
    \label{fig:domi_proton}
\end{figure*}

\begin{figure*}
    \centering
    \includegraphics[width=\columnwidth]{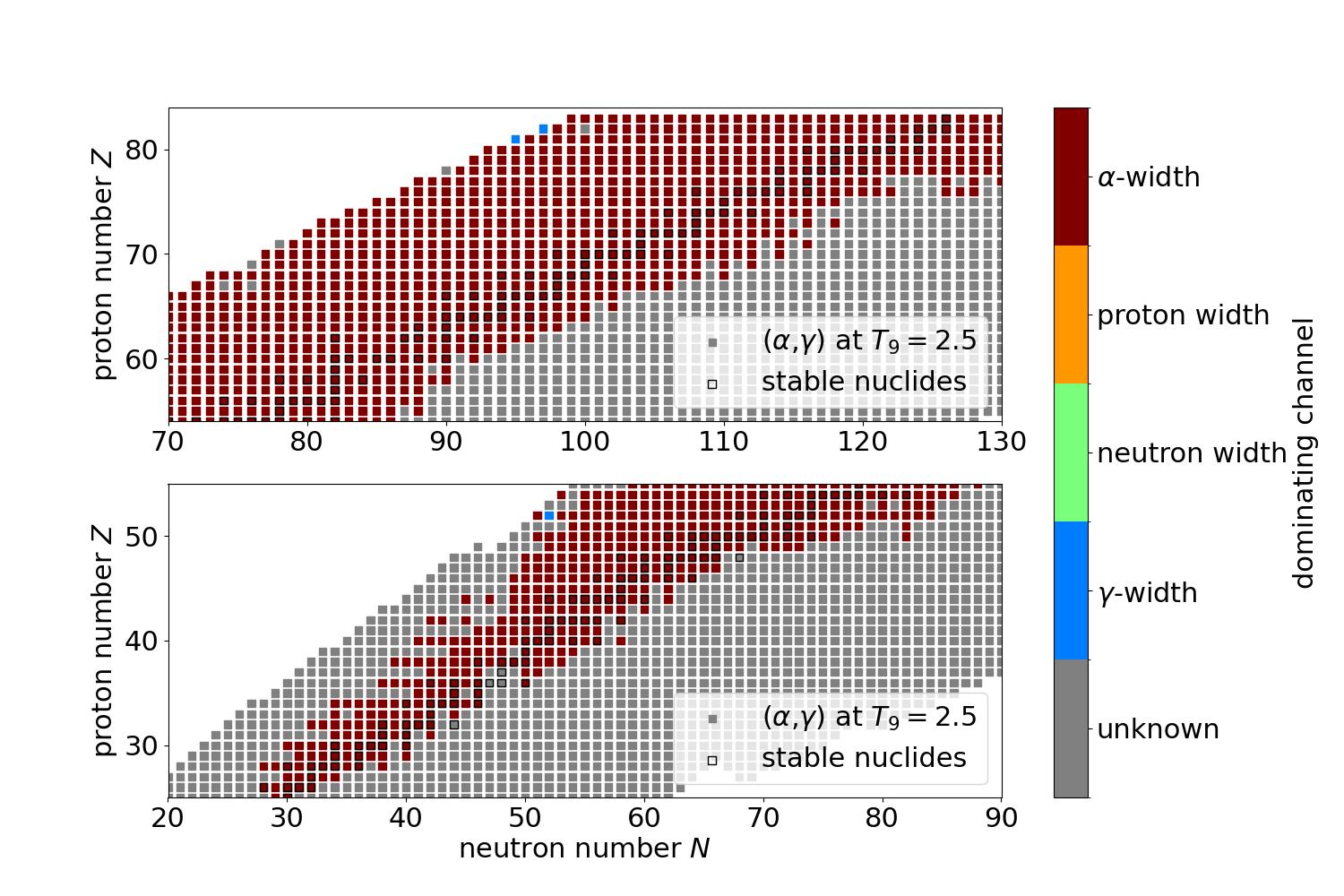}
    \includegraphics[width=\columnwidth]{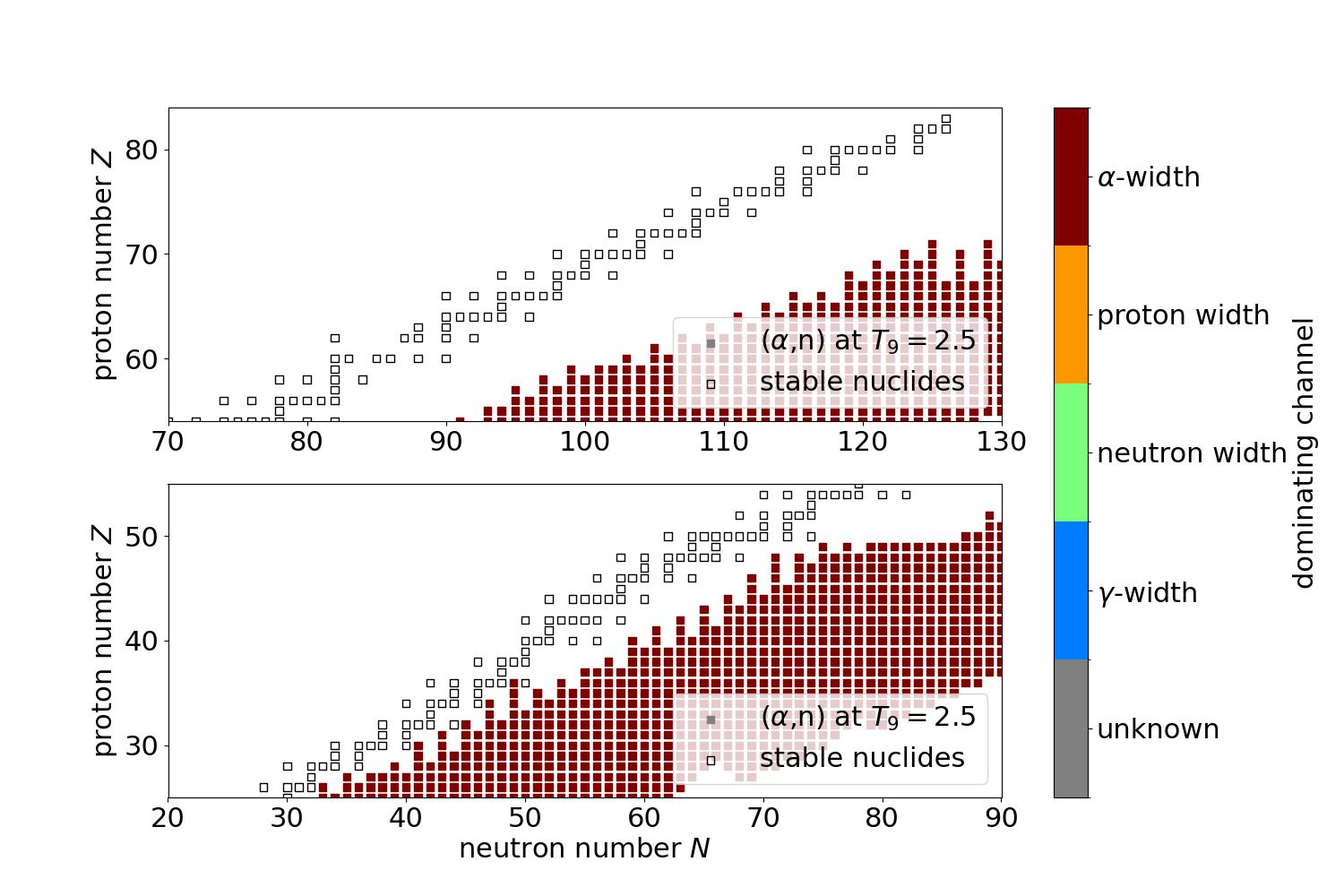}
    \includegraphics[width=\columnwidth]{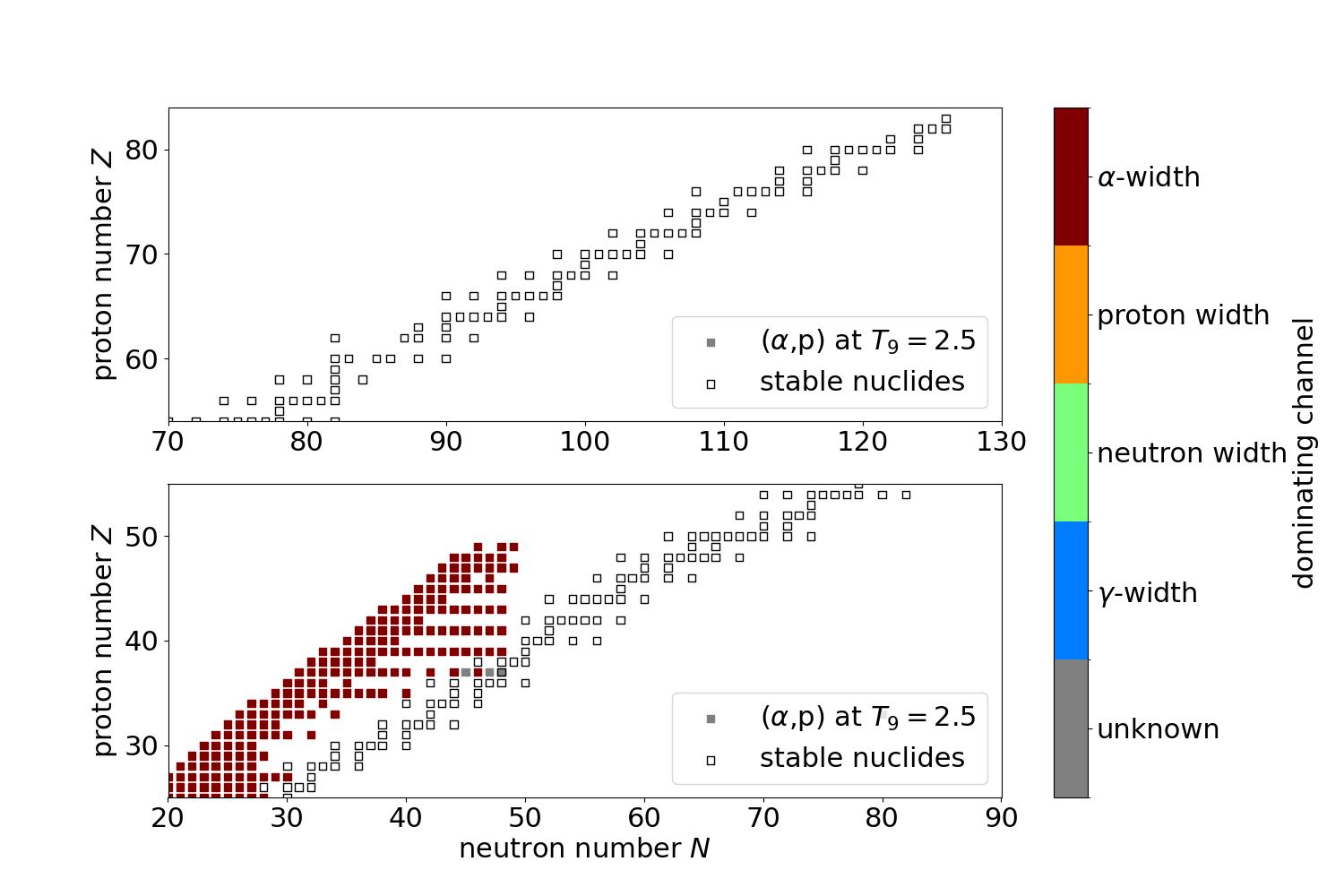}
    \caption{Same as Figure~\ref{fig:domi_proton} but for $\alpha$-induced astrophysical reaction rates at 2.5 GK.}
    \label{fig:domi_alpha}
\end{figure*}

\begin{figure*}
    \centering
    \includegraphics[width=\columnwidth]{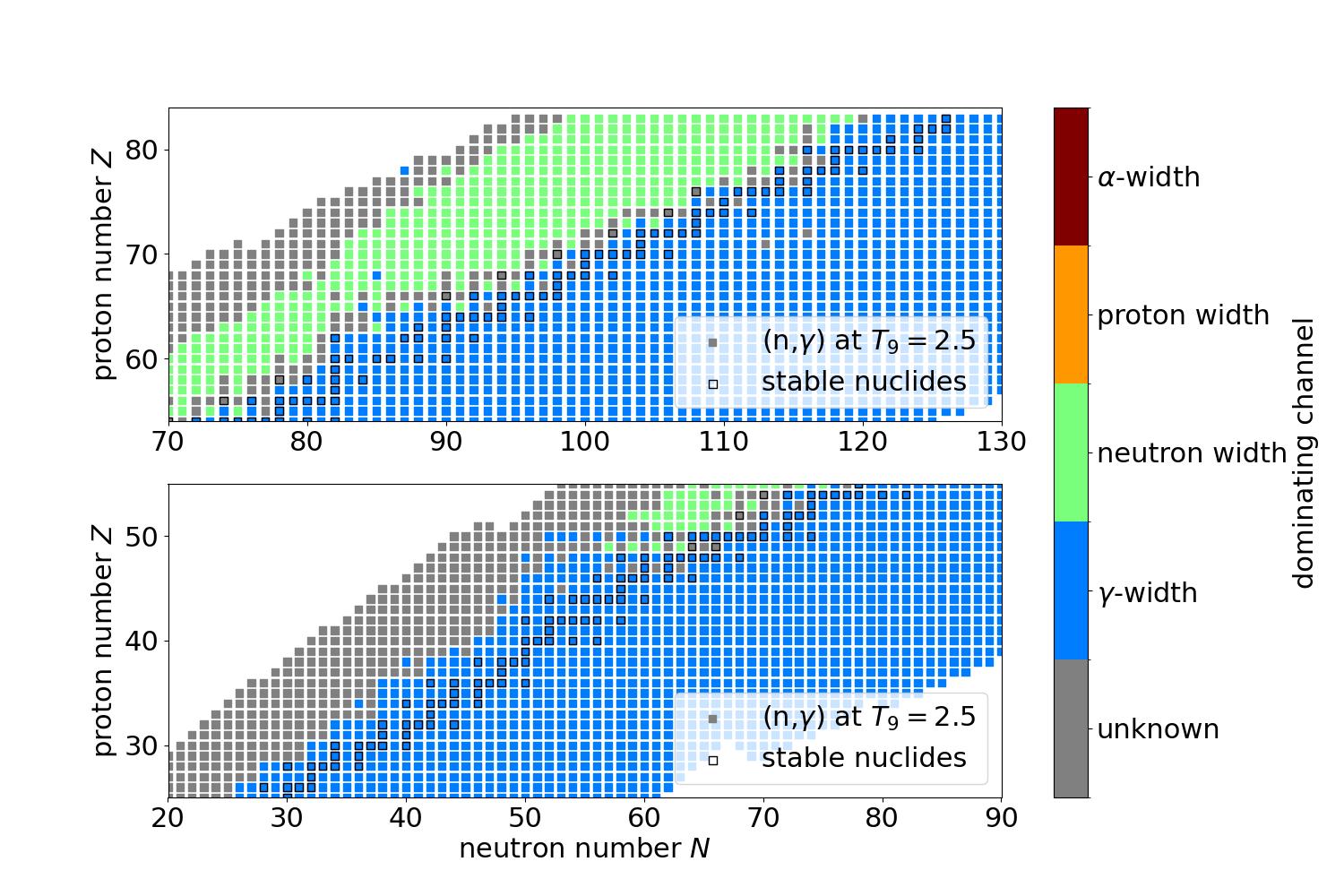}
    \includegraphics[width=\columnwidth]{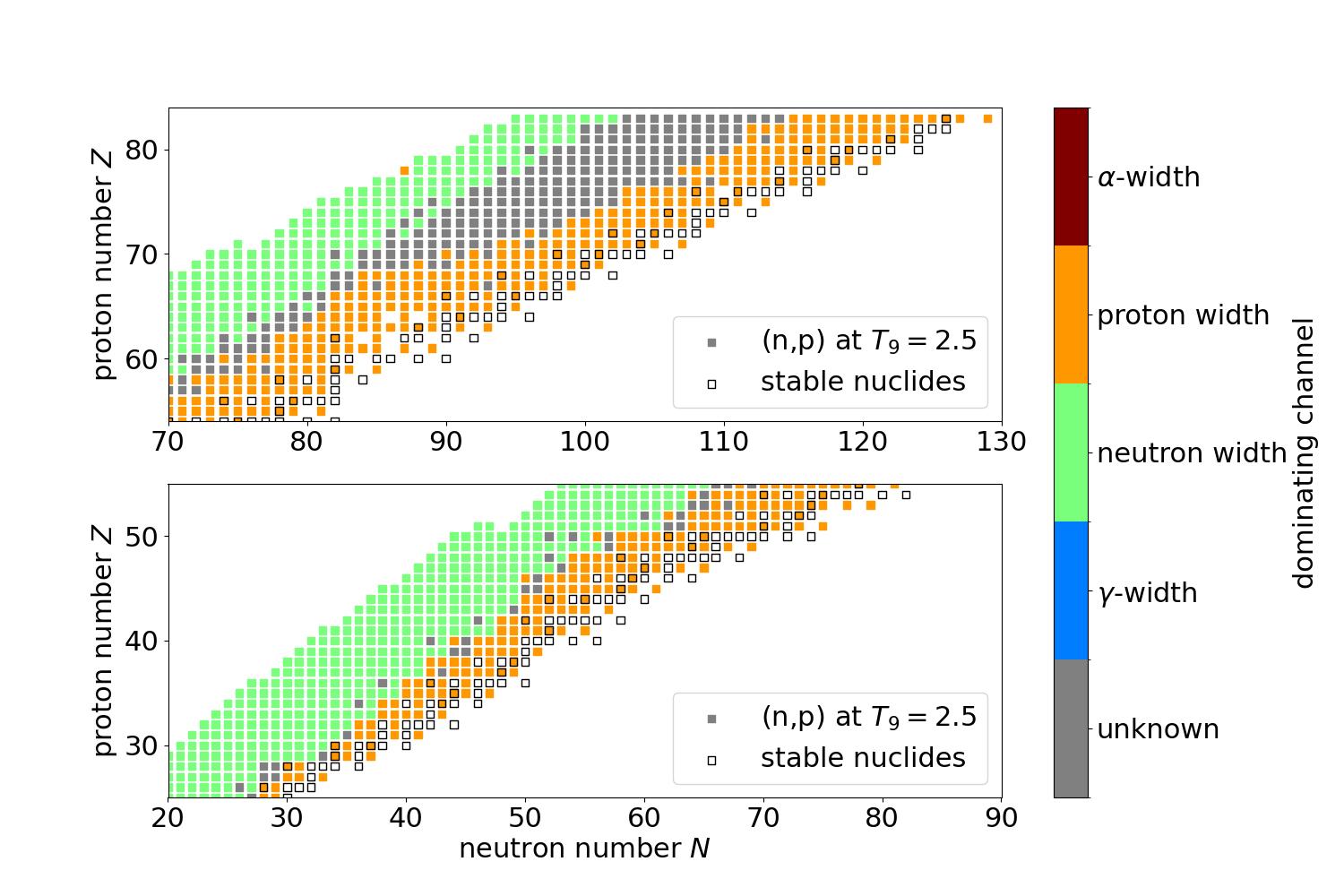}
    \includegraphics[width=\columnwidth]{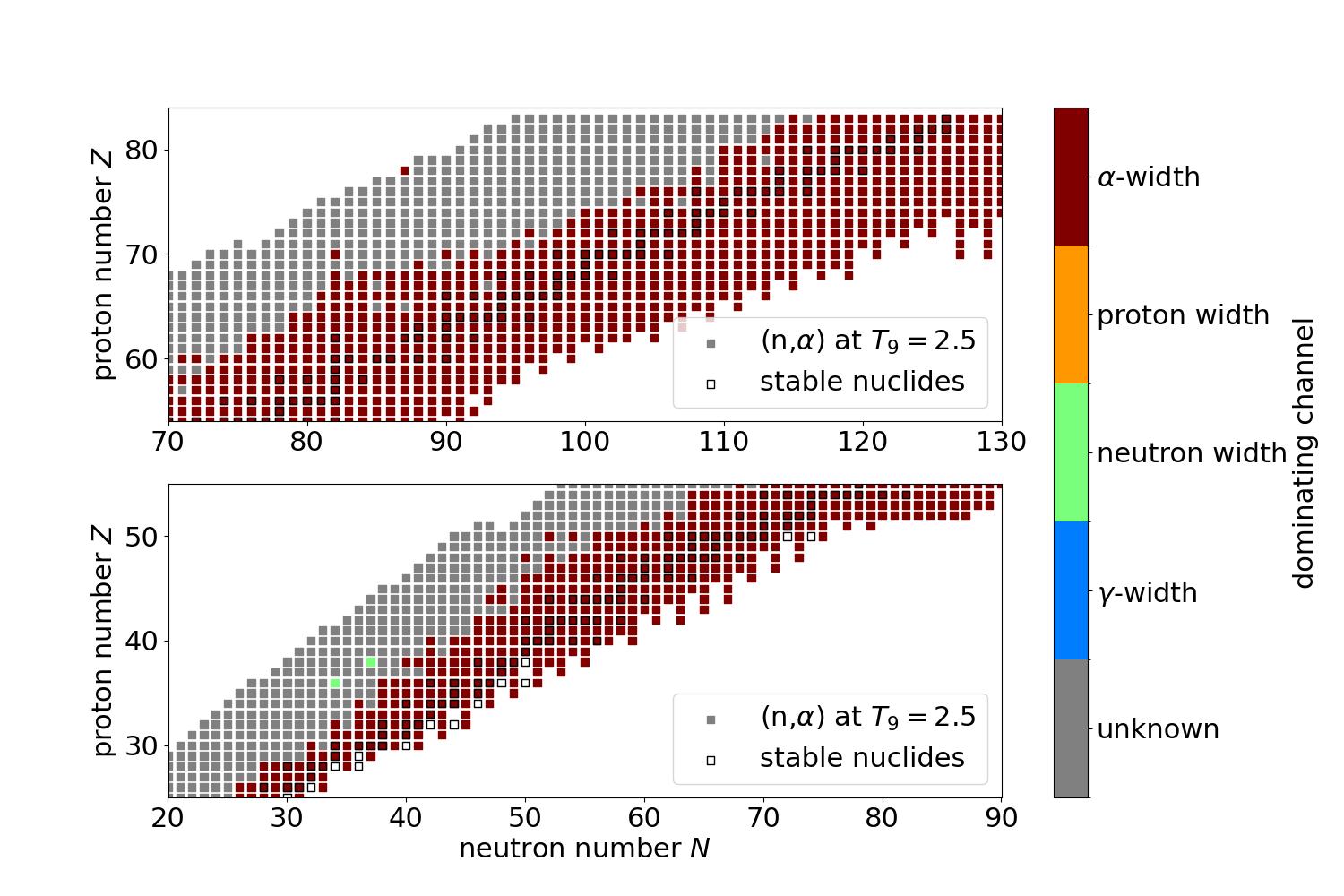}
    \caption{Same as Figure~\ref{fig:domi_proton} but for neutron-induced astrophysical reaction rates at 2.5 GK.}
    \label{fig:domi_neutronhi}
\end{figure*}

\begin{figure*}
    \centering
    \includegraphics[width=\columnwidth]{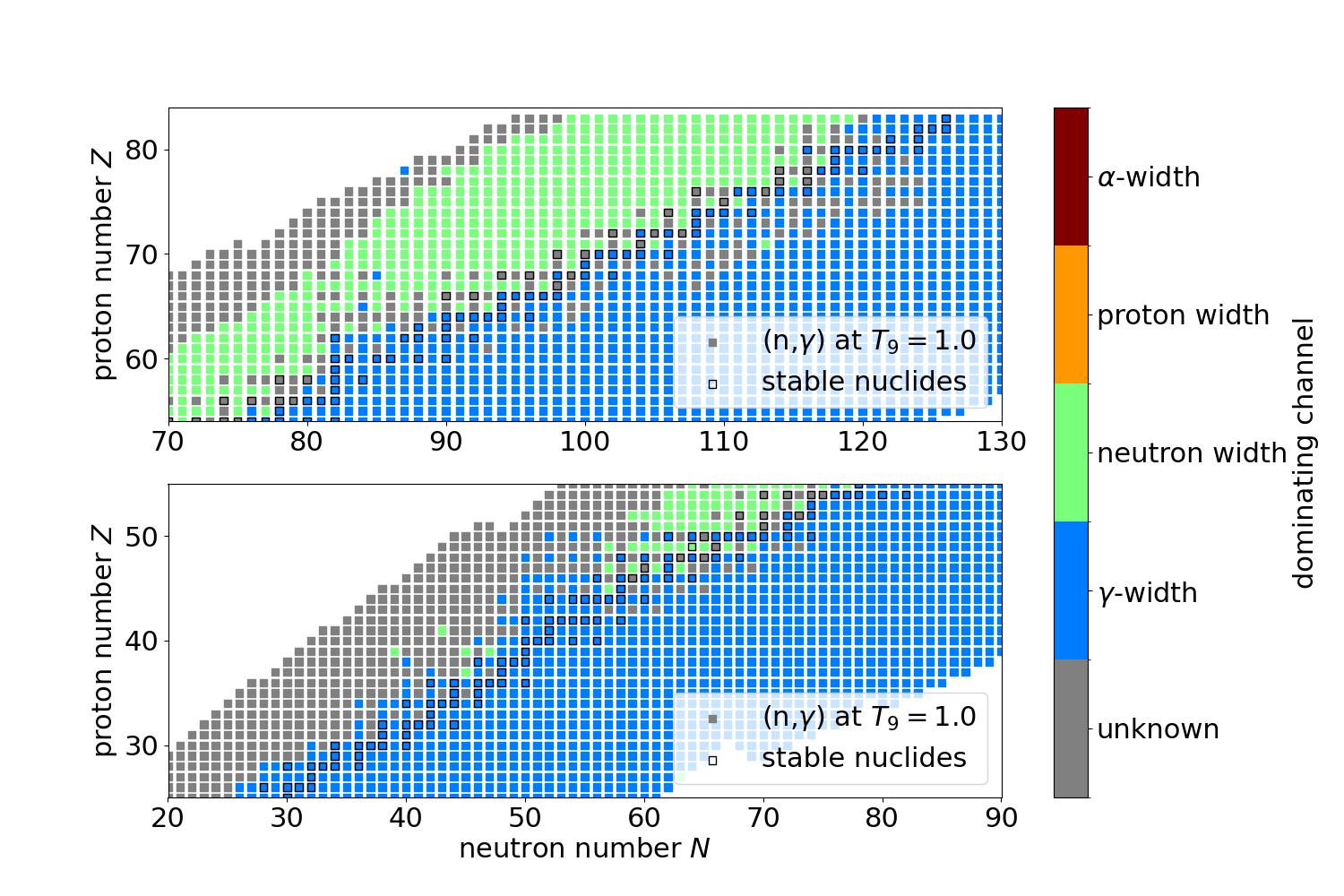}
    \includegraphics[width=\columnwidth]{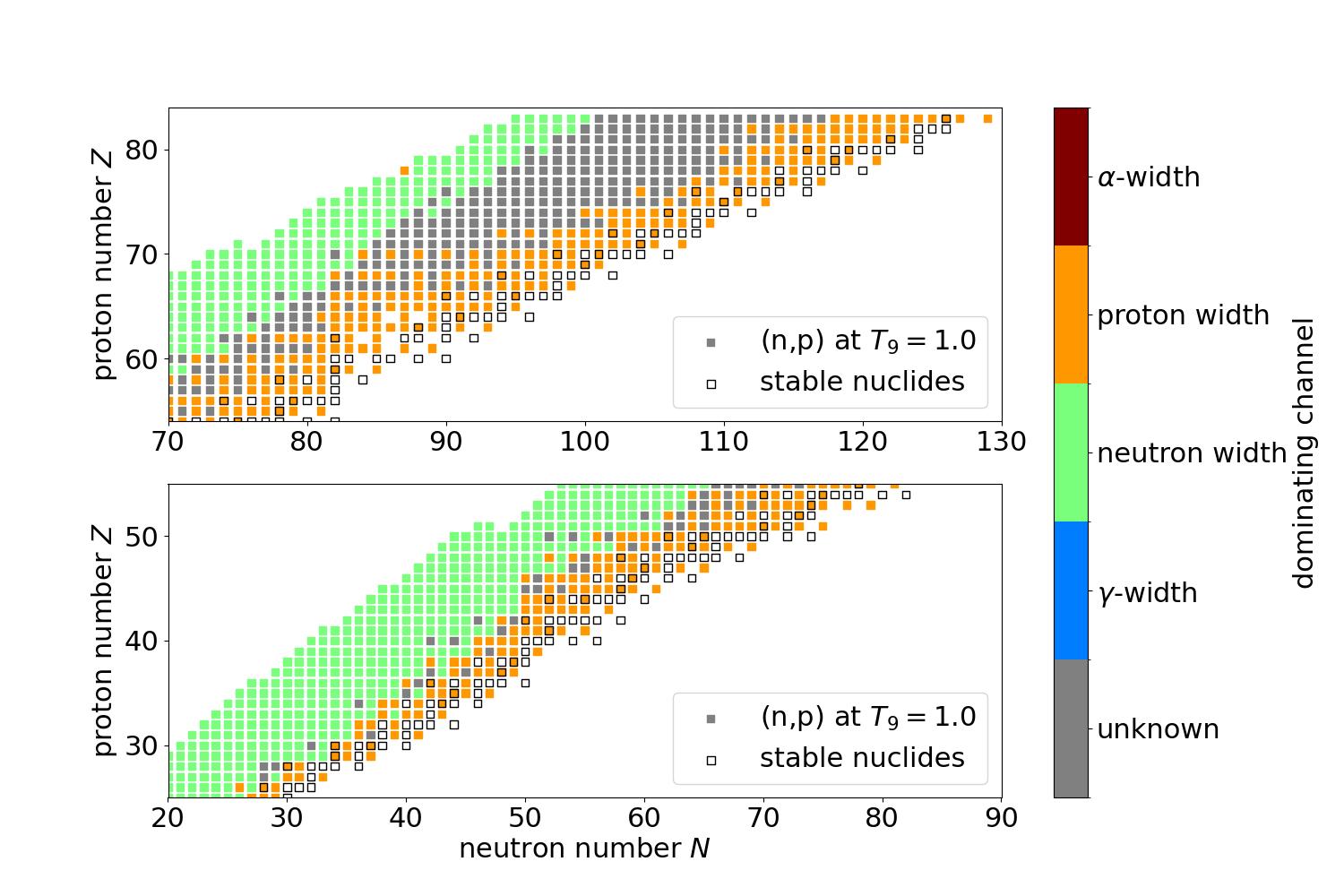}
    \includegraphics[width=\columnwidth]{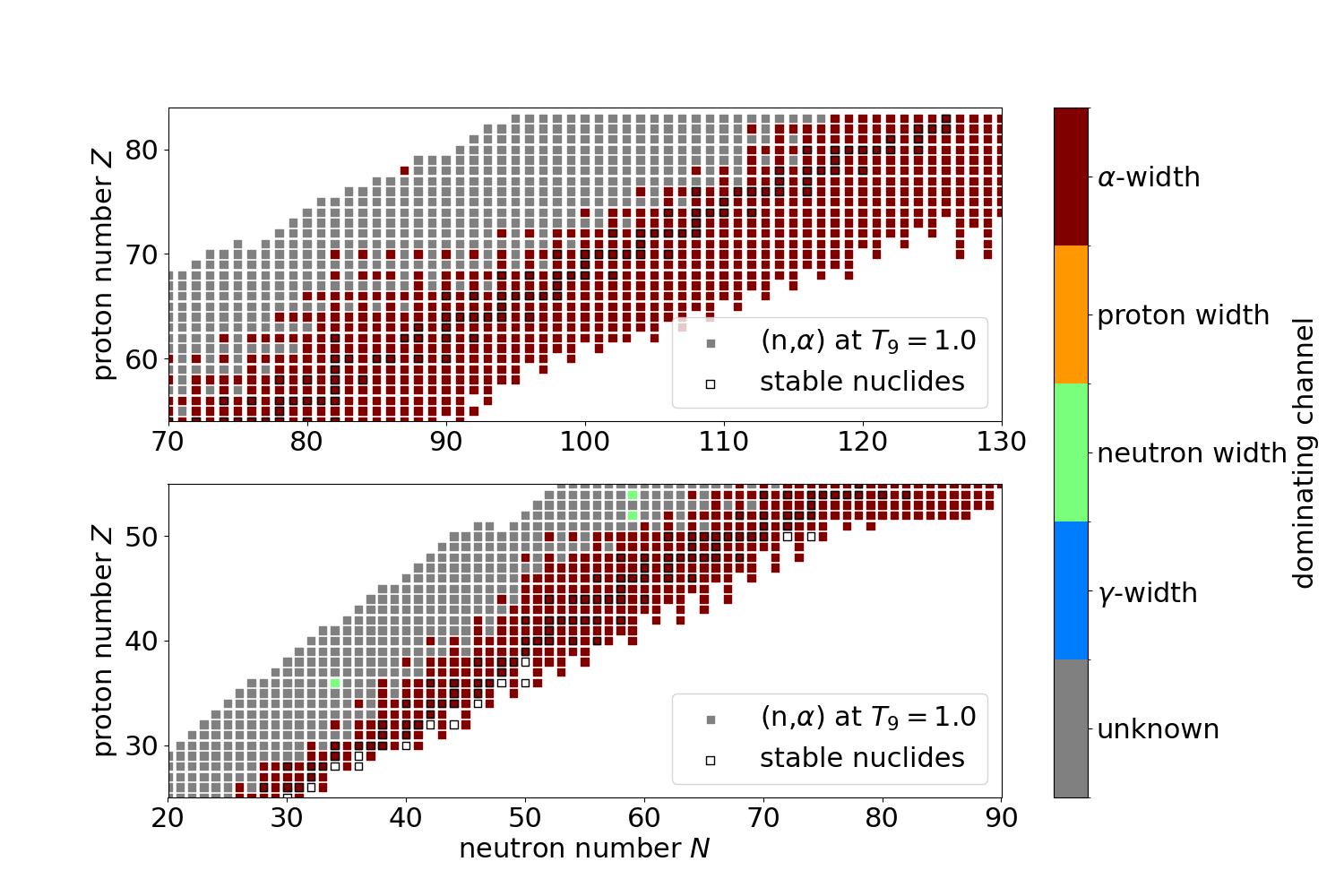}
    \caption{Same as Figure~\ref{fig:domi_proton} but for neutron-induced astrophysical reaction rates at 1 GK.}
    \label{fig:domi_neutronlo}
\end{figure*}

Since the calculation of the astrophysical reaction rate involves an integration over energy ranges (the relevant relative interaction energies for the reactions on the g.s. and the excited states) the sensitivity of a rate to change in a nuclear property may be different from the sensitivity of the (laboratory) cross section. To guide experimentalists, tables of sensitivities of cross sections and rates have been published in \cite{2012ApJS..201...26R}. For convenience, Figures~\ref{fig:domi_proton}$-$\ref{fig:domi_neutronlo} show an overview of the (averaged) reaction widths dominating the respective reaction rates. As shown in Equation (\ref{eq:transwidths}), the widths are directly related to the transmission coefficients entering the Hauser-Feshbach formula. From Equation (\ref{eq:hf}) it can be inferred that the cross section will be sensitive to the smaller transmission coefficient appearing in the numerator \cite{2011IJMPE..20.1071R,2012ApJS..201...26R,2013JPhCS.420a2138R}. The same holds for reaction rates with "astrophysical" transmission coefficients accounting for excited state contributions. Widths of comparable size in the numerator impact the cross section or rate equally. Widths of comparable size may also lead to branchings in a reaction path when several reaction channels compete (see, e.g., \cite{2006PhRvC..73a5804R,2013RPPh...76f6201R}). If the total width (in the denominator) is dominated by a large width not appearing in the numerator, the reaction rate will also be sensitive to a variation of that width. Therefore the rate can be sensitive to more than one width. These cases are marked as "unknown" in the Figures~\ref{fig:domi_proton}$-$\ref{fig:domi_neutronlo}. To further identify the contributing widths it will be necessary to consult the detailed tables of \cite{2012ApJS..201...26R}.

\rev{The fact that the widths depend on the relative energy (and also the spin of the involved levels) has the important consequence that the excited state contributions to the stellar rate may exhibit different sensitivities than the contribution from reactions on the g.s.\ of a nucleus. This, in turn, implies that a comparison between data and theory for g.s.\ reactions may not provide the required information to improve the prediction of the stellar rate. This is especially relevant at the elevated temperatures of explosive nucleosynthesis with their tiny g.s.~contributions to the stellar rate (see the online tables and Refs.~\cite{2011IJMPE..20.1071R,2012ApJS..201...26R,2014AIPA....4d1012R,2015AIPC.1681e0003R}).}

Also noteworthy is the fact that the particle widths not only scale with the relative interaction energy of absorbed or emitted particles but also are affected by the proton to neutron ratio. For example, Figures~\ref{fig:domi_neutronhi} and \ref{fig:domi_neutronlo} show that neutron captures on neutron-rich nuclides are mostly sensitive to the $\gamma$-width whereas for proton-rich nuclides either the dependence on the neutron width is dominating or neutron and $\gamma$ width are both affecting the rate (label "unknown"). This is because the $\gamma$ width is much smaller than the neutron width for neutron-rich nuclides whereas the neutron width becomes comparable to or smaller than the $\gamma$ width for proton-rich nuclides.

Obviously, predictions can and should be compared to data. It is crucial, however, to perform the comparison in an appropriate and informed manner, considering several important aspects. As mentioned above, when comparing experimental and theoretical g.s.~cross sections, a "translation" of the result into an impact on an astrophysical rate may not be straightforward. When comparing rates, it is important to identify whether the rate has been derived from a measured g.s.~cross section or whether it includes the (temperature dependent) contributions of excited states.

Perhaps less obvious, it is important to note whether the comparison involves actually calculated rates or fits of such rates (e.g., in the REACLIB parameterisation introduced above). Fits may deviate from the actually predicted rates and a disagreement between data and fit may hint at a problem with the fit and not necessarily with the actual prediction. On the other hand, if such fits are used in reaction networks in astrophysical simulations, it is important to identify (and rectify) such a problem as it may impact the results of the simulation. Evaluating the quality of fits is not straightforward, either. A larger deviation is permissible for temperature regions where forward and reverse rate would be in equilibrium and especially in environments exhibiting equilibria between a group of rates or all rates. For these cases, the individual rates are not important and abundances are calculated from equilibrium formulas (see, e.g., \cite{2020entn.book.....R}). This is the case for temperatures $T\geq 5$ GK, although some equilibration may already occur above 1 GK. Larger deviations between calculated rate and its fit are also permissible for situations in which the rate is extremely slow and therefore will not cause an appreciable change in abundance across a time window defined by the duration of a nucleosynthesis process. In explosive environments, the relevant timescale is on the order of a few seconds (plus the time to allow decays of short-lived products). This is relevant for reactions between charged particles for which the Coulomb barrier strongly suppresses the reaction at low temperature.

\begin{figure*}
    \centering
    \includegraphics[width=\columnwidth]{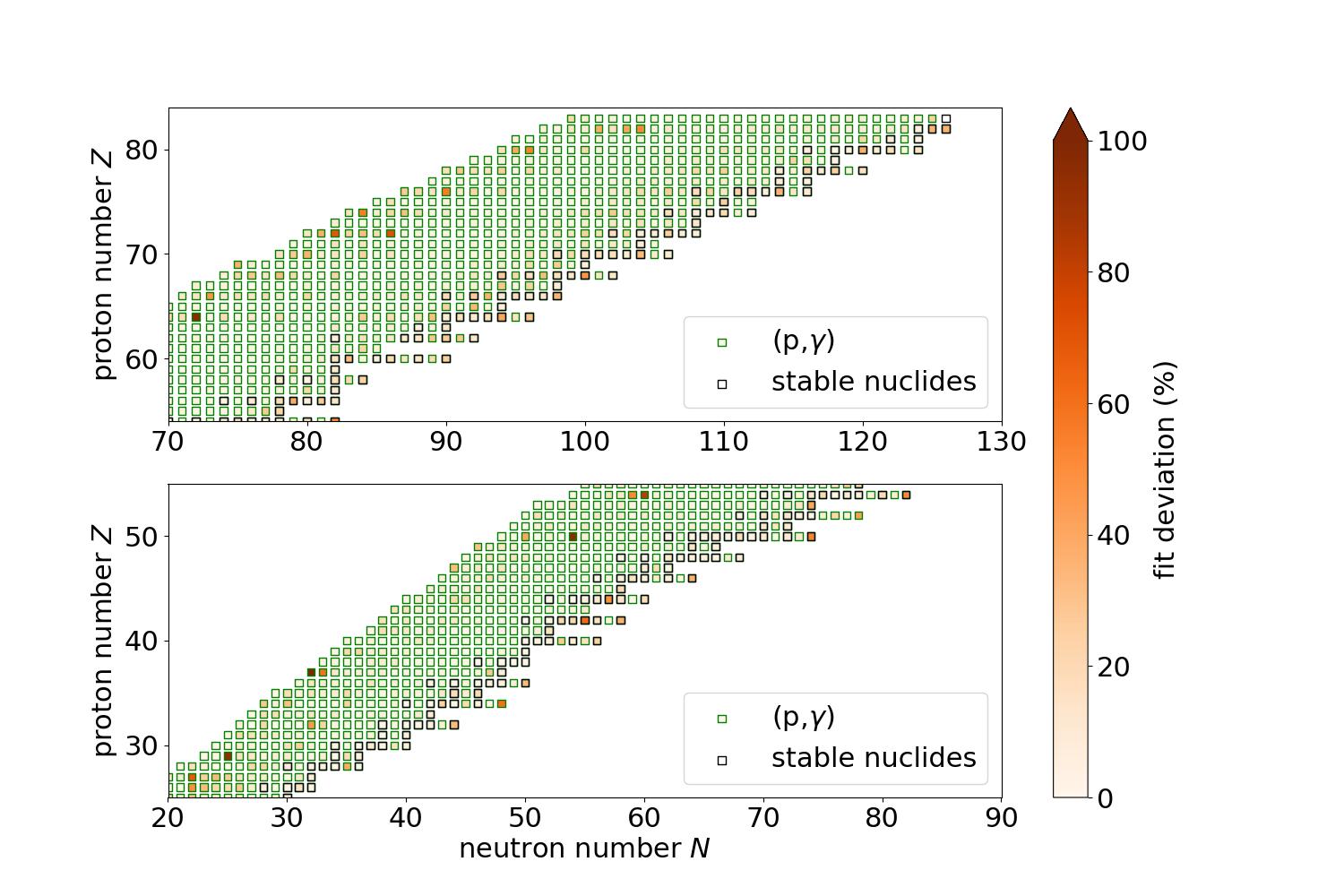}
    \includegraphics[width=\columnwidth]{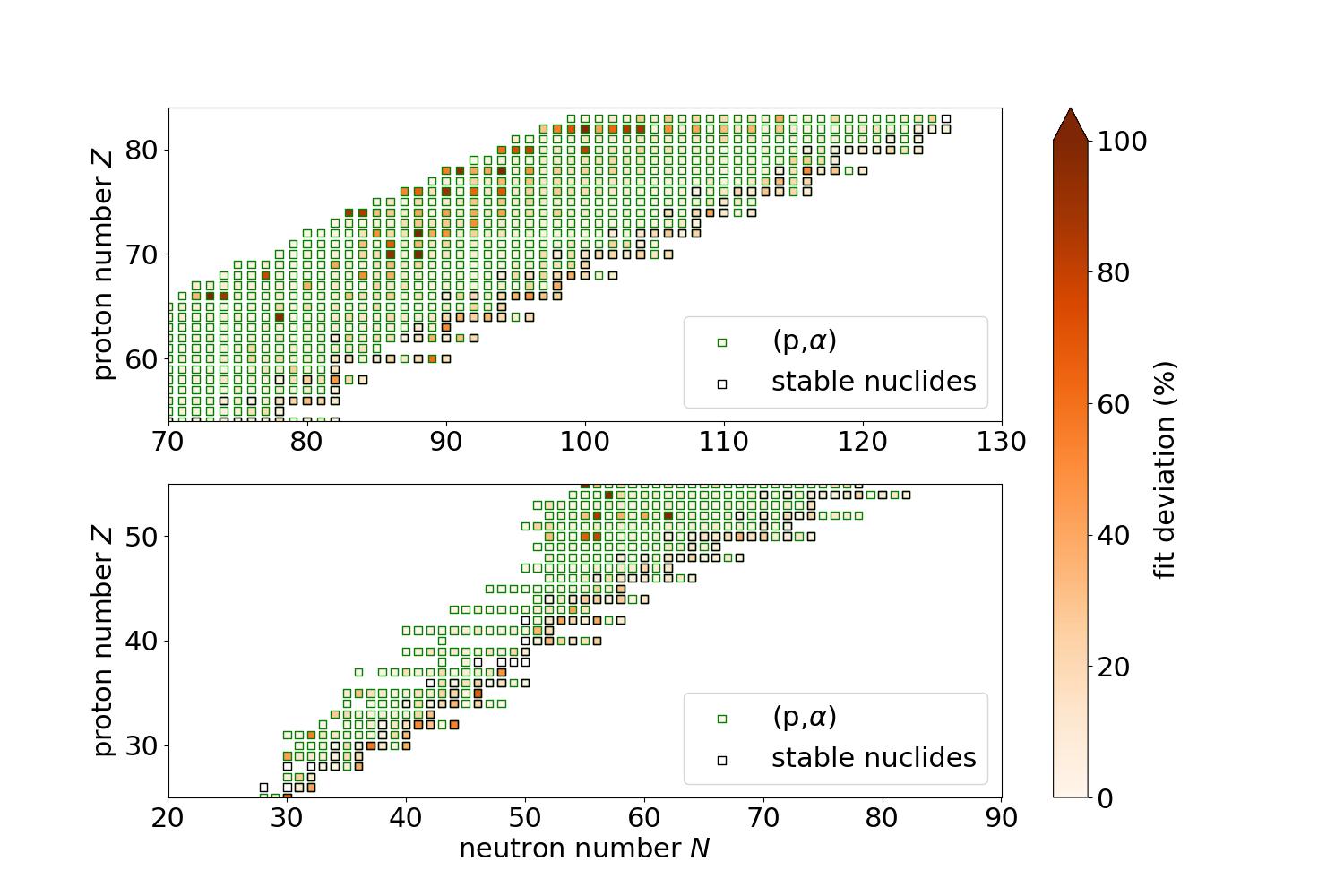}
    \caption{Fit quality for proton-induced reactions; shown is the overall deviation of the fit from the actual values within the relevant temperature region (see text for details) by the fill colour of the square identifying the target nuclide. For the (p,$\alpha$) reactions only reactions with positive $Q$ value are shown. Reactions off the neutron-rich side of stability are not included in the rate set presented here.}
    \label{fig:qualifit_proton}
\end{figure*}

\begin{figure*}
    \centering
    \includegraphics[width=\columnwidth]{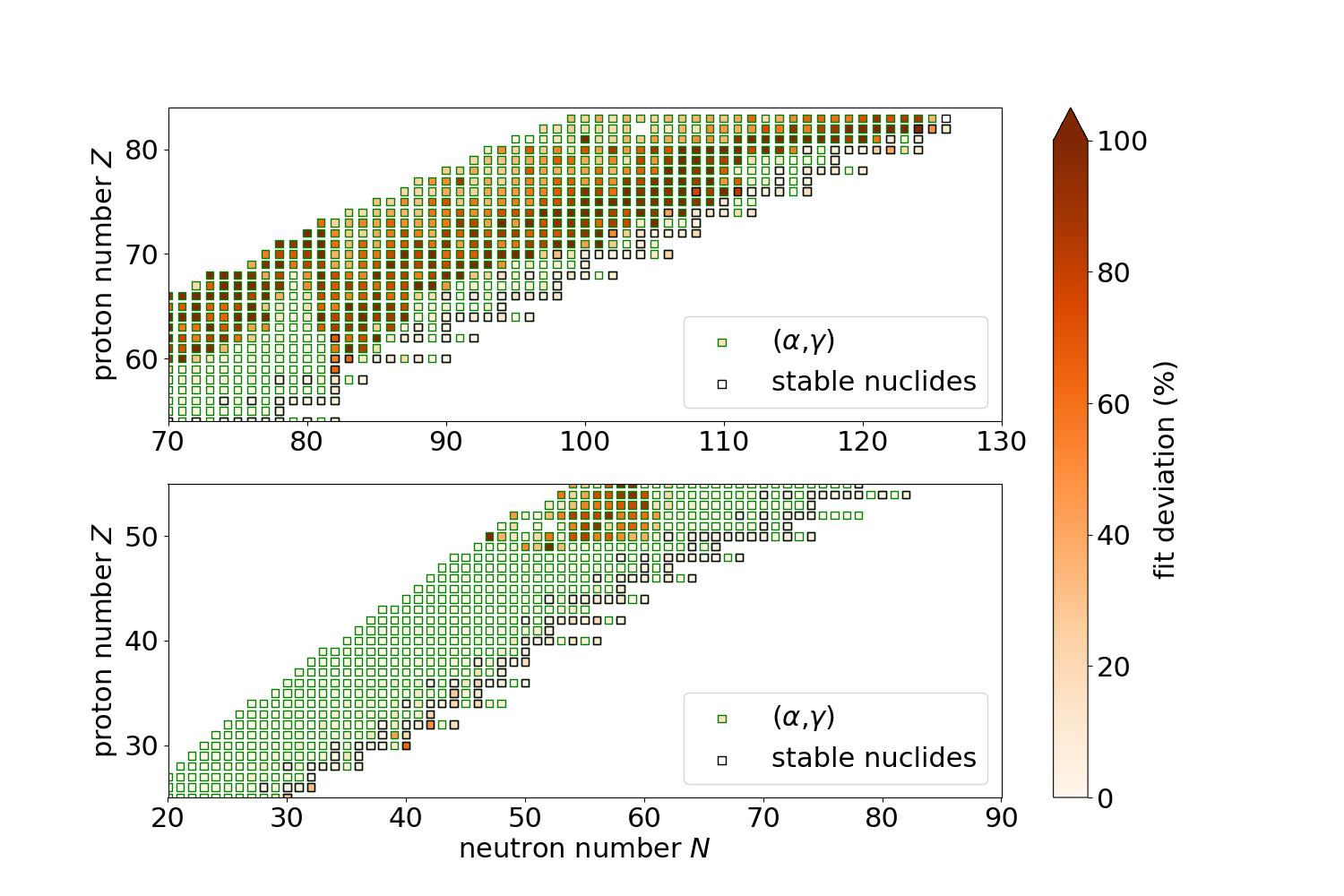}
    \includegraphics[width=\columnwidth]{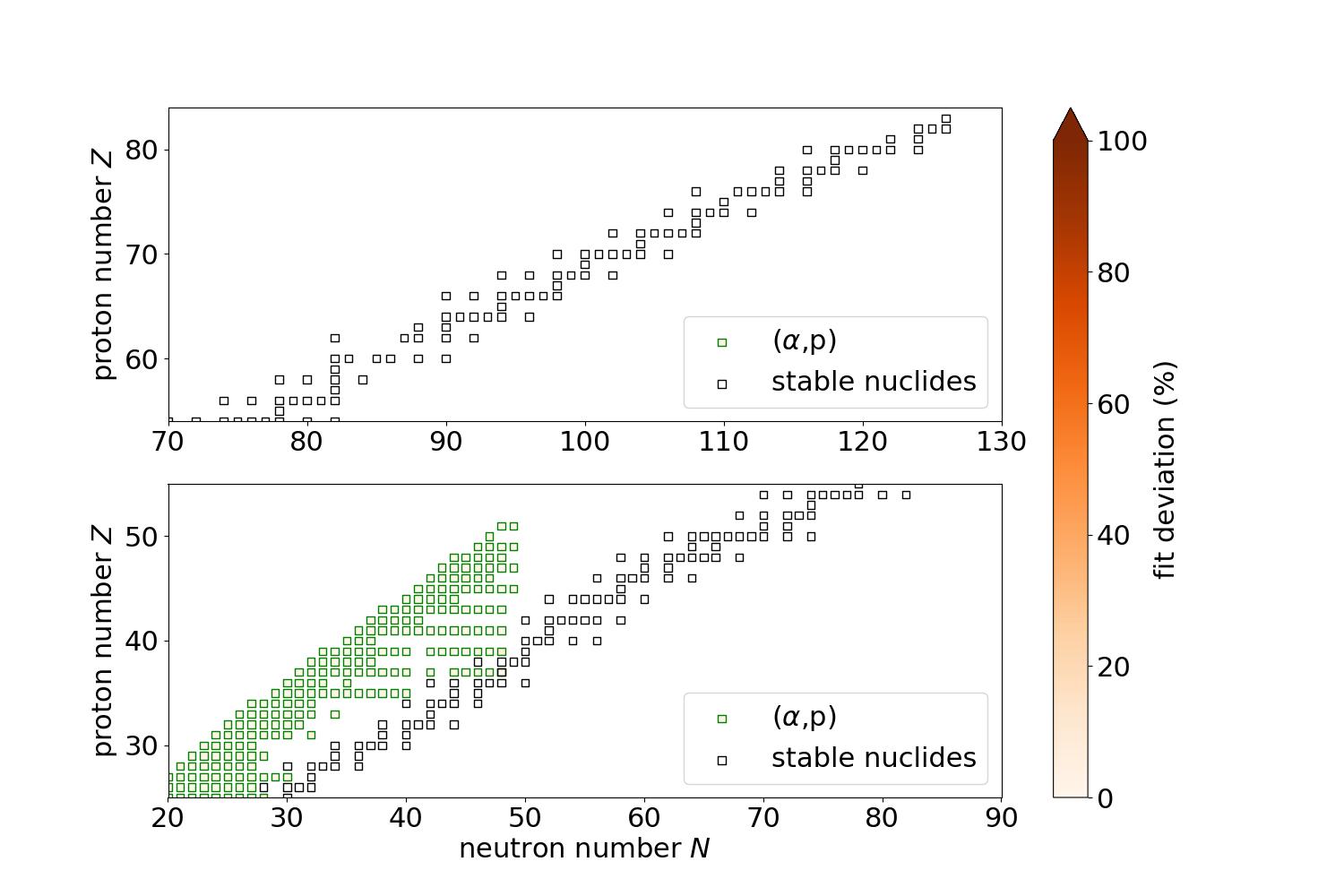}
    \caption{Same as Figure~\ref{fig:qualifit_proton} for $\alpha$-induced reactions; for the ($\alpha$,p) reactions only reactions with positive $Q$ value are shown.}
    \label{fig:qualifit_alpha}
\end{figure*}

\begin{figure*}
    \centering
    \includegraphics[width=\columnwidth]{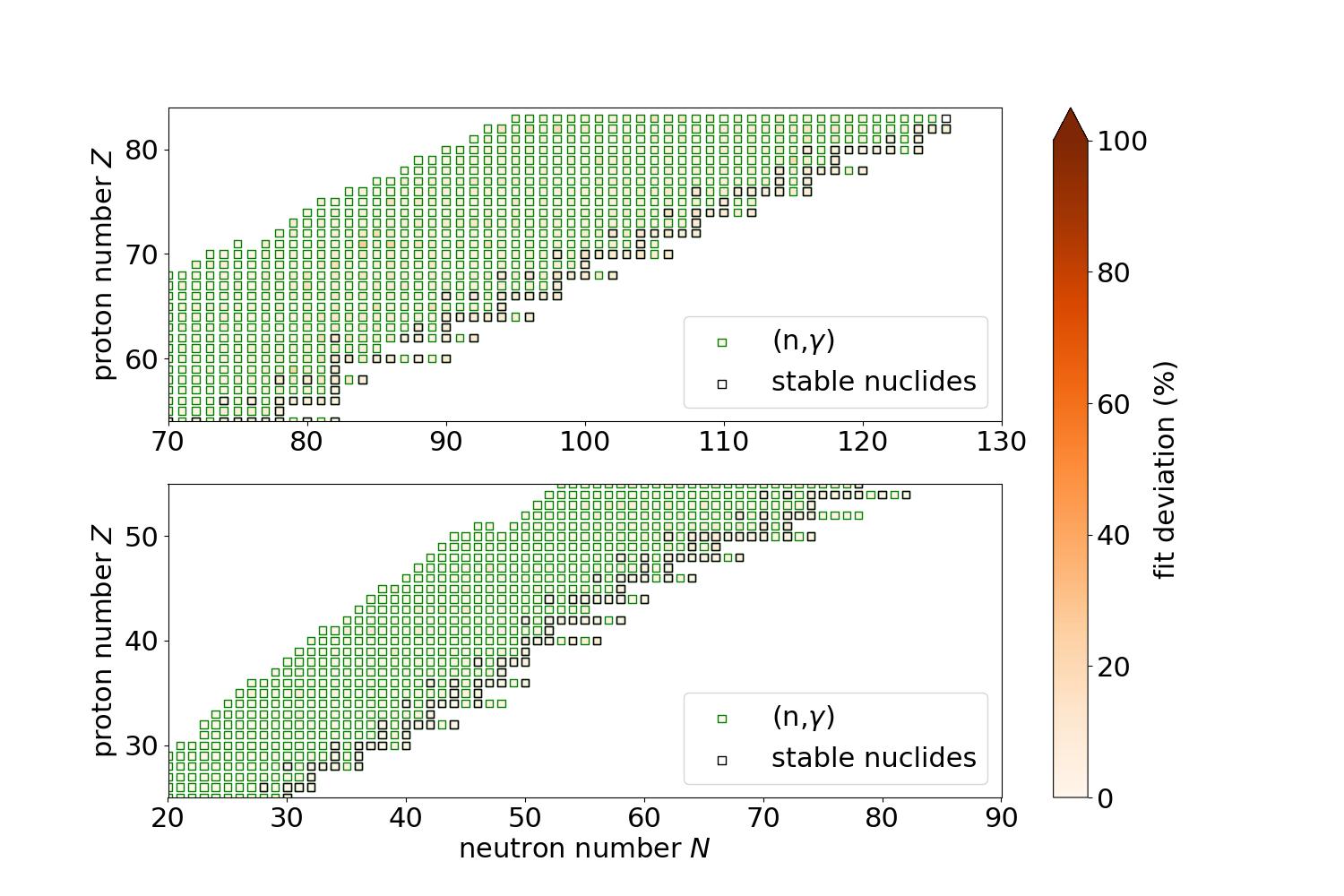}
    \includegraphics[width=\columnwidth]{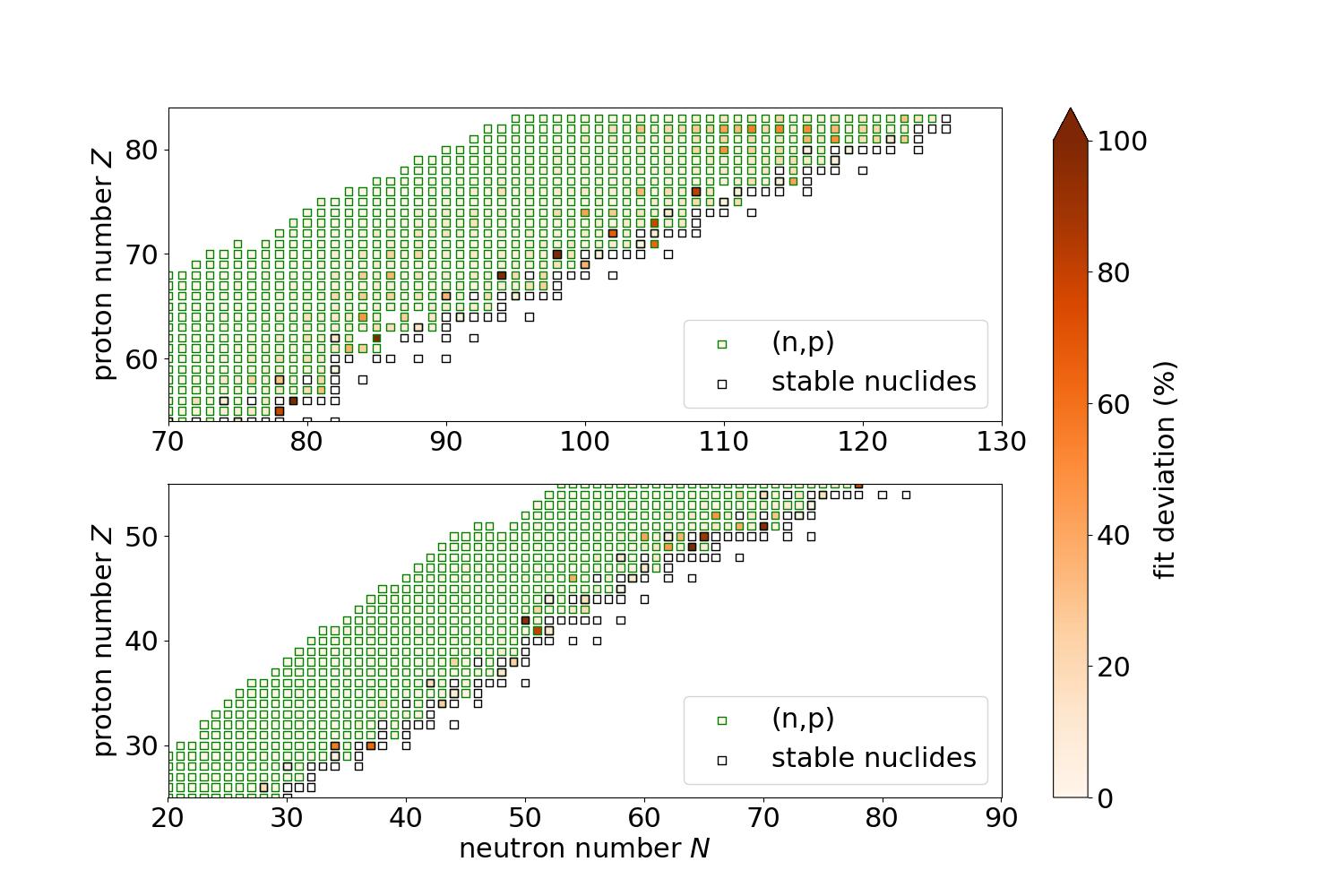}
    \includegraphics[width=\columnwidth]{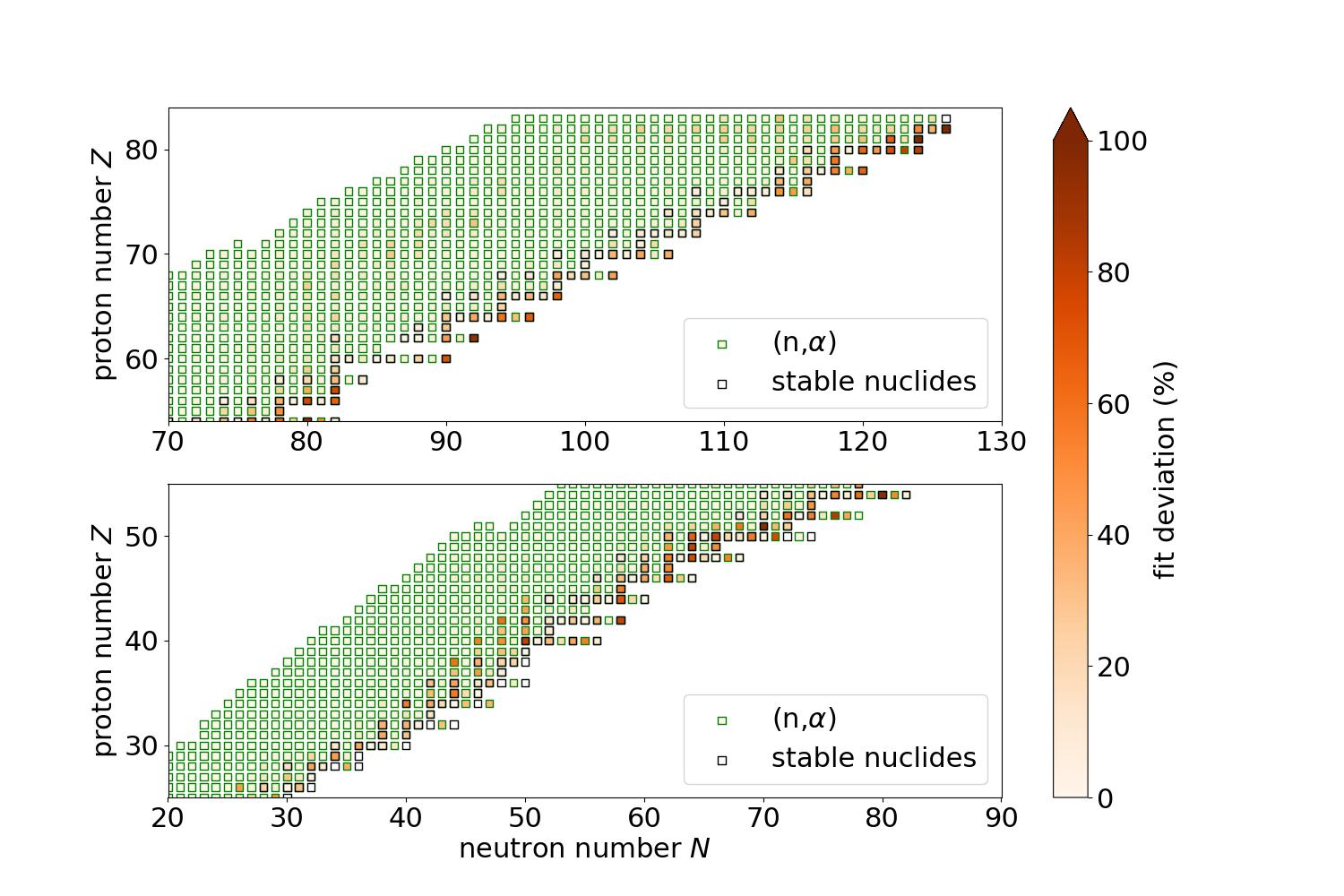}
    \caption{Same as Figure~\ref{fig:qualifit_proton} for neutron-induced reactions.}
    \label{fig:qualifit_neutron}
\end{figure*}

The present online tables of fit parameters also include a measure of the fit quality taking into account the criteria (relevant temperature window, reaction speed) described above. Due to the large number of rates, the fits were obtained using an automated fit procedure which worked well for the majority of cases but locally gave rise to larger deviations. An overview of the deviations is shown in Figures~\ref{fig:qualifit_proton}$-$\ref{fig:qualifit_neutron}. For the majority of the considered reactions, the deviation of the fit from the true value is below a few percent, displayed as open squares in the figures. More problematic are fits of $\alpha$-capture reactions in the region of spontaneous $\alpha$ emitters, where larger fit deviations of 40$-$80\% are found, with a few even exceeding these values. In consequence, tabular values would be preferred over fits in that region. For comparison of a few specific rates, a local refitting may be in order but this is beyond the scope of this work. The figures and the actual values for the fit quality are intended to guide the application of the fits.

Last but not less important is the consideration of the applicability of the model. The statistical model assumes that there is a "sufficiently large" number of overlapping resonances in the compound nucleus at the compound formation energy. When the nuclear level density at the compound formation is too low, the calculated model cross section will overestimate the actual cross section because inherent in the model is a summation over an extended number of spins assumed to be present at the relevant excitation energy of the compound nucleus. The inherent nuclear level density is low at shell closure and towards the dripline, the latter because of the lower proton and $\alpha$ separation energies. This means that for lower temperatures (meaning lower compound formation energies) the Hauser-Feshbach model may not be applicable to calculate reaction rates. This has been investigated previously \cite{1997PhRvC..56.1613R}. In that previous publication, temperature limits for the application have been derived. These limits have been shown in the tables of \cite{2000ADNDT..75....1R} and are also included in the present tables of reaction rates and their fits. Using values outside their range of applicability \rev{it is to be expected that deviations to, e.g., measured cross sections and rates, increase}.

There is also another regime in which the statistical model is not applicable: at high projectile energy direct processes -- occuring much faster than the compound formation and equilibration -- dominate the cross sections. For nucleosynthesis processes, however, this is less important because the projectile energies remain comparatively low even in explosive environments (much higher energies would actually lead to disintegration of nuclei through photodissociation). Direct processes, such as direct radiative capture, play a role, however, in systems with low level density because it scales differently than the statistical contribution and can provide a sizeable contribution to the total cross section \cite{1996PhRvC..53..469K,2011IJMPE..20.1071R,2015AIPC.1681e0003R}. Future versions of SMARAGD will account for such direct processes to provide more accurate rates close to the driplines and at magic nucleon numbers.

\subsection{Comparison to selected experimental data}
\label{sec:exp}

\begin{figure*}
    \centering
    \includegraphics[width=\columnwidth]{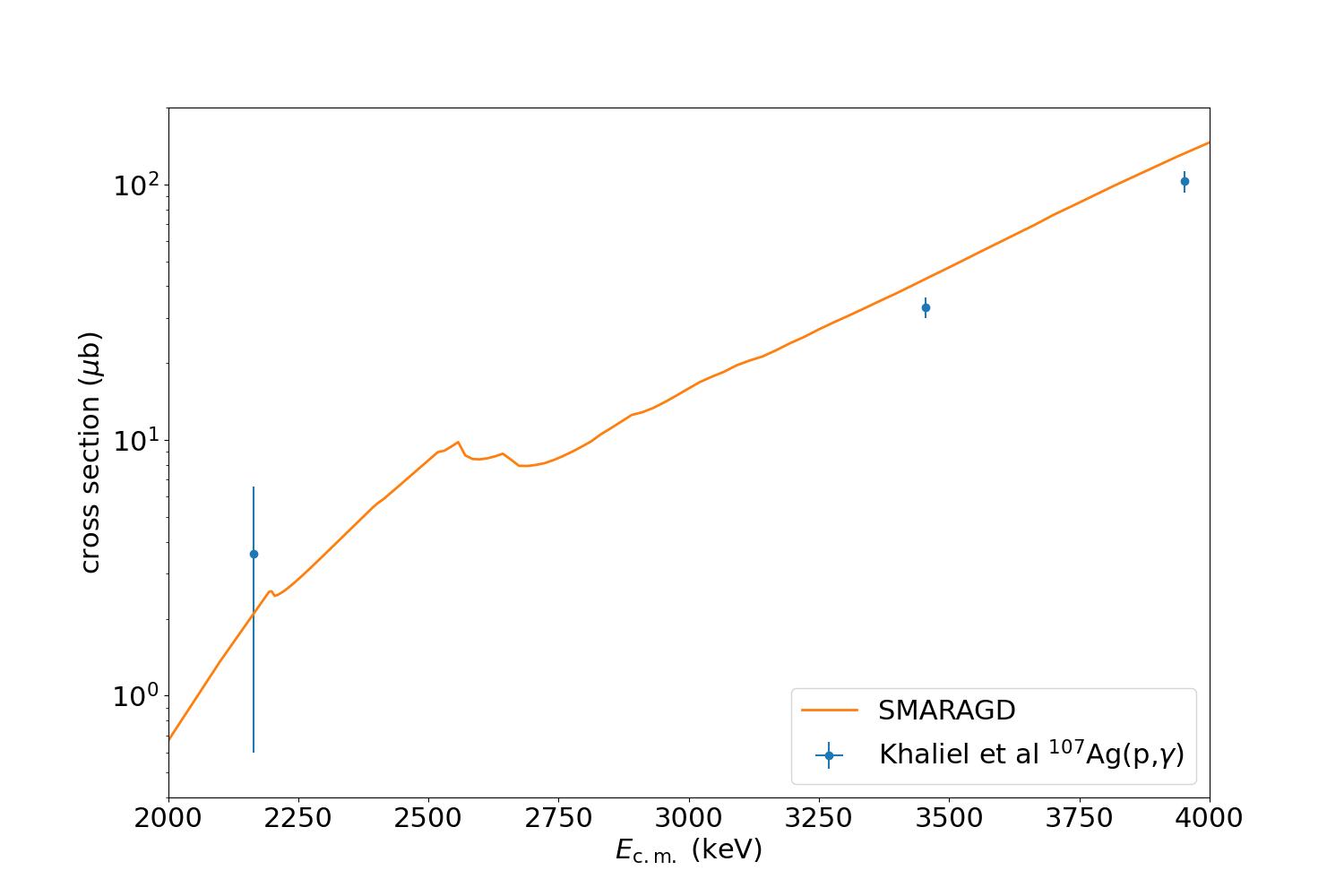}
    \includegraphics[width=\columnwidth]{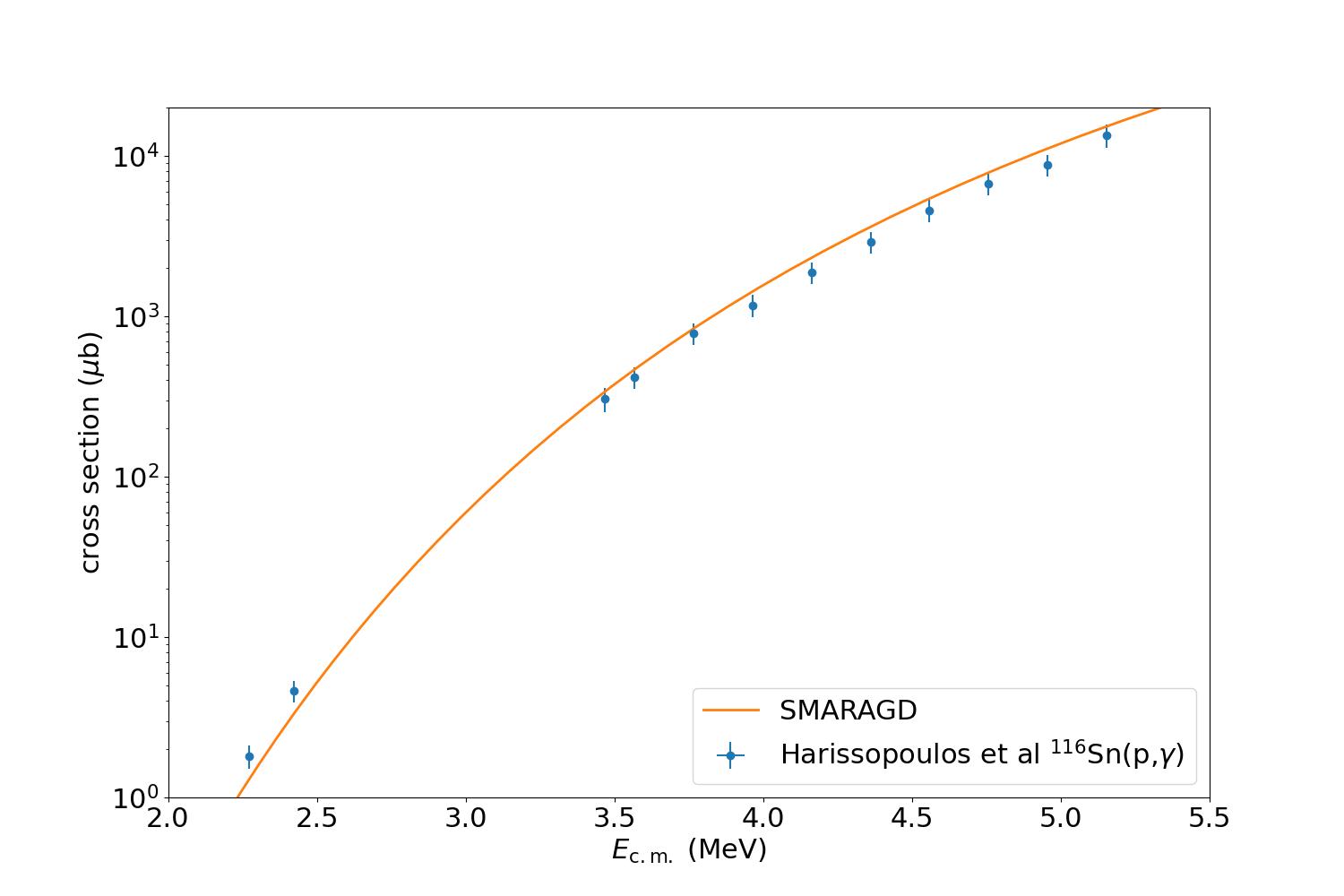}
    \includegraphics[width=\columnwidth]{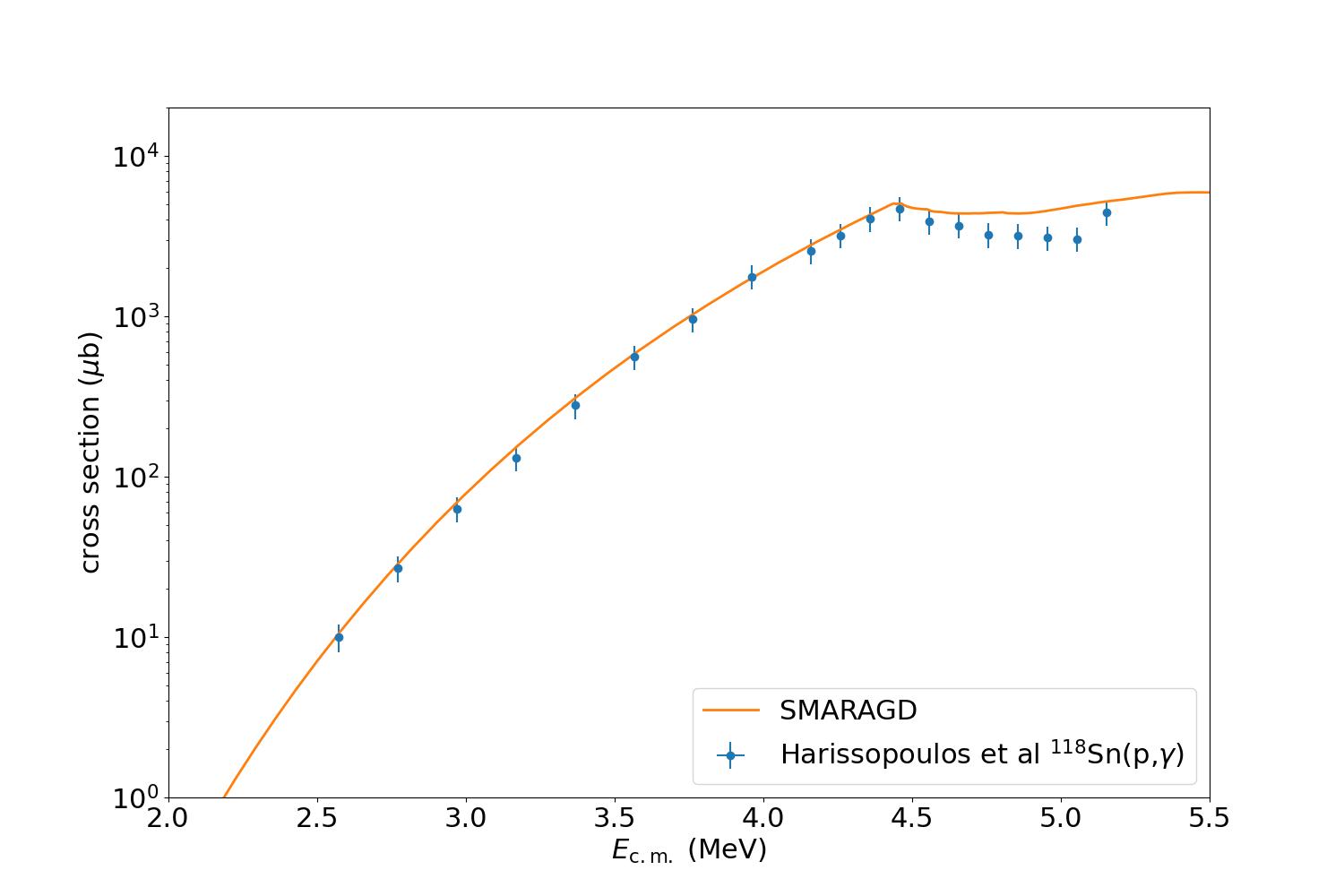}
    \includegraphics[width=\columnwidth]{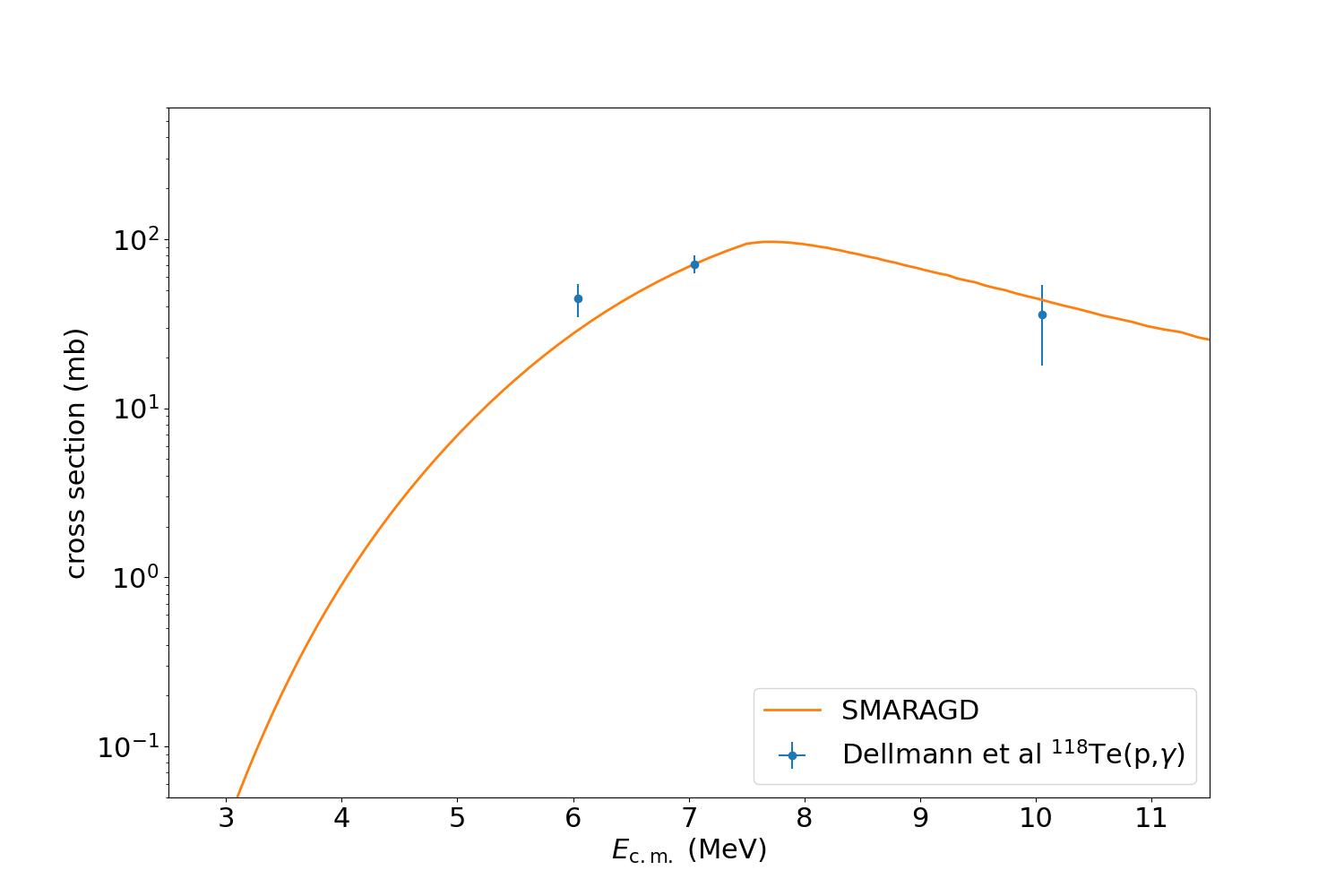}
    \caption{Comparison of proton-capture cross sections predicted by SMARAGD with experimental data by Khaliel et al \cite{2017PhRvC..96c5806K}, Harissopoulos et al \cite{2024PhRvC.110a5803H}, and Dellmann et al \cite{2025PhRvL.134n2701D}. Note the different scales on the vertical axes.}
    \label{fig:exp_proton}
\end{figure*}

\begin{figure*}
    \centering
    \includegraphics[width=\columnwidth]{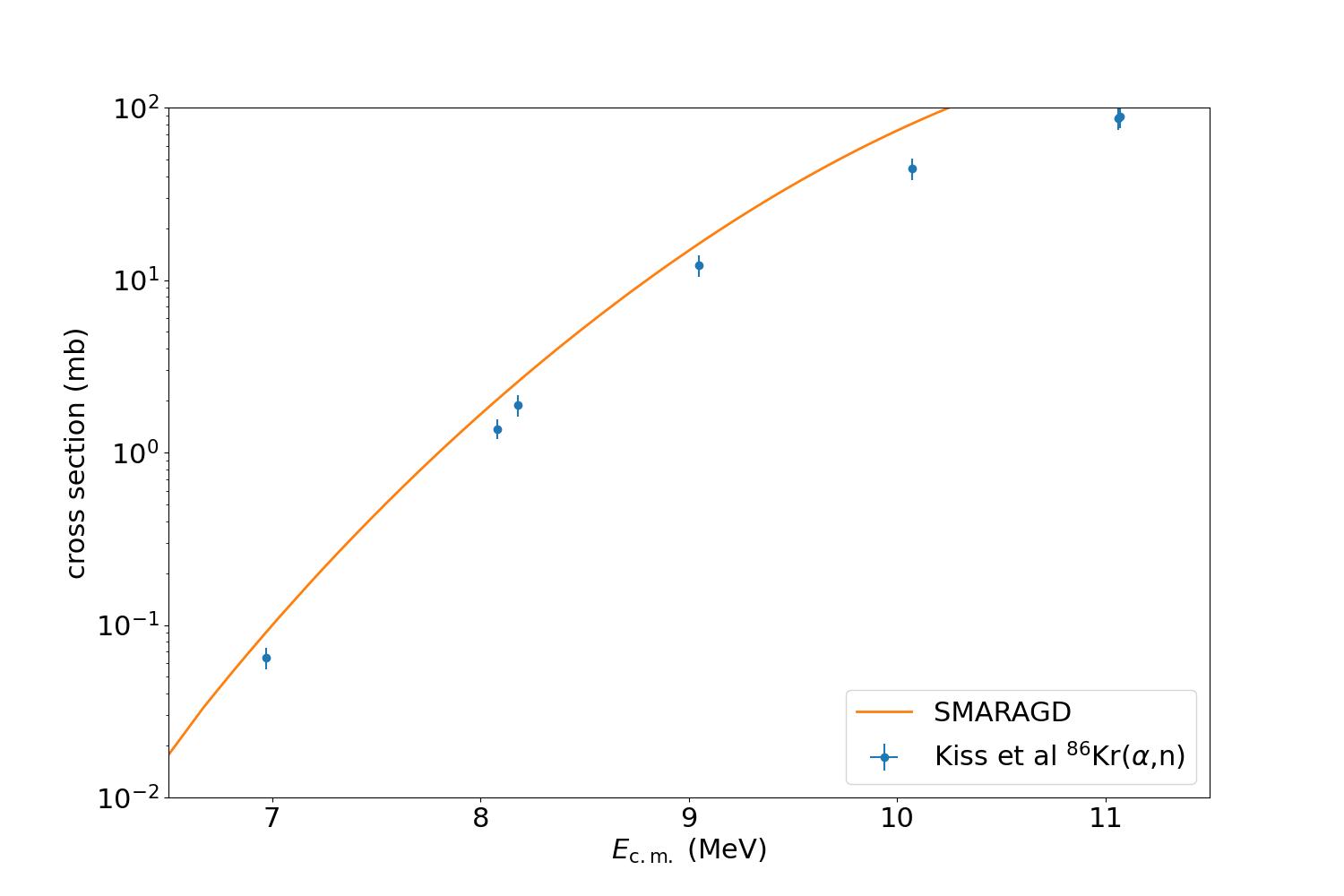}
    \includegraphics[width=\columnwidth]{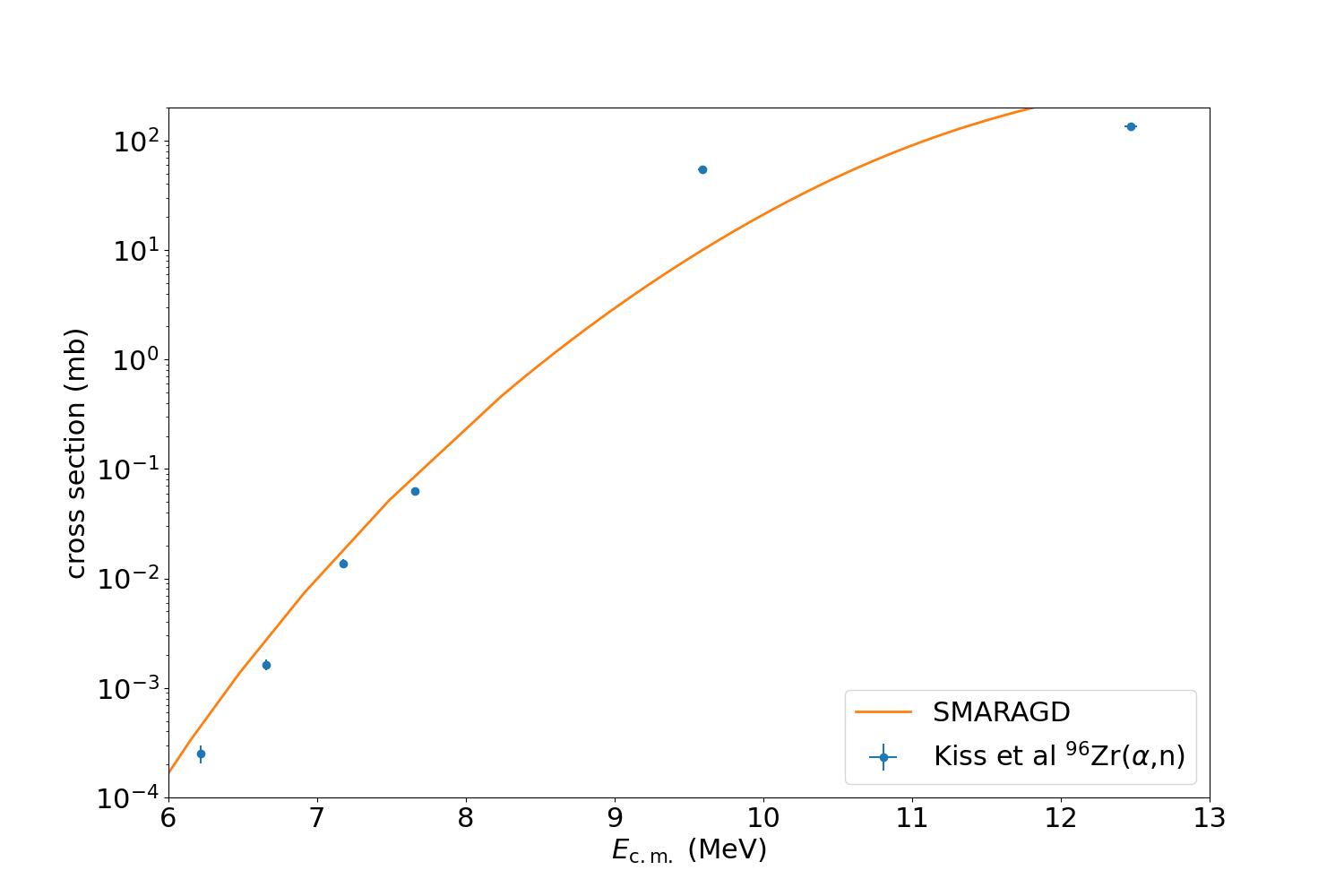}
    \includegraphics[width=\columnwidth]{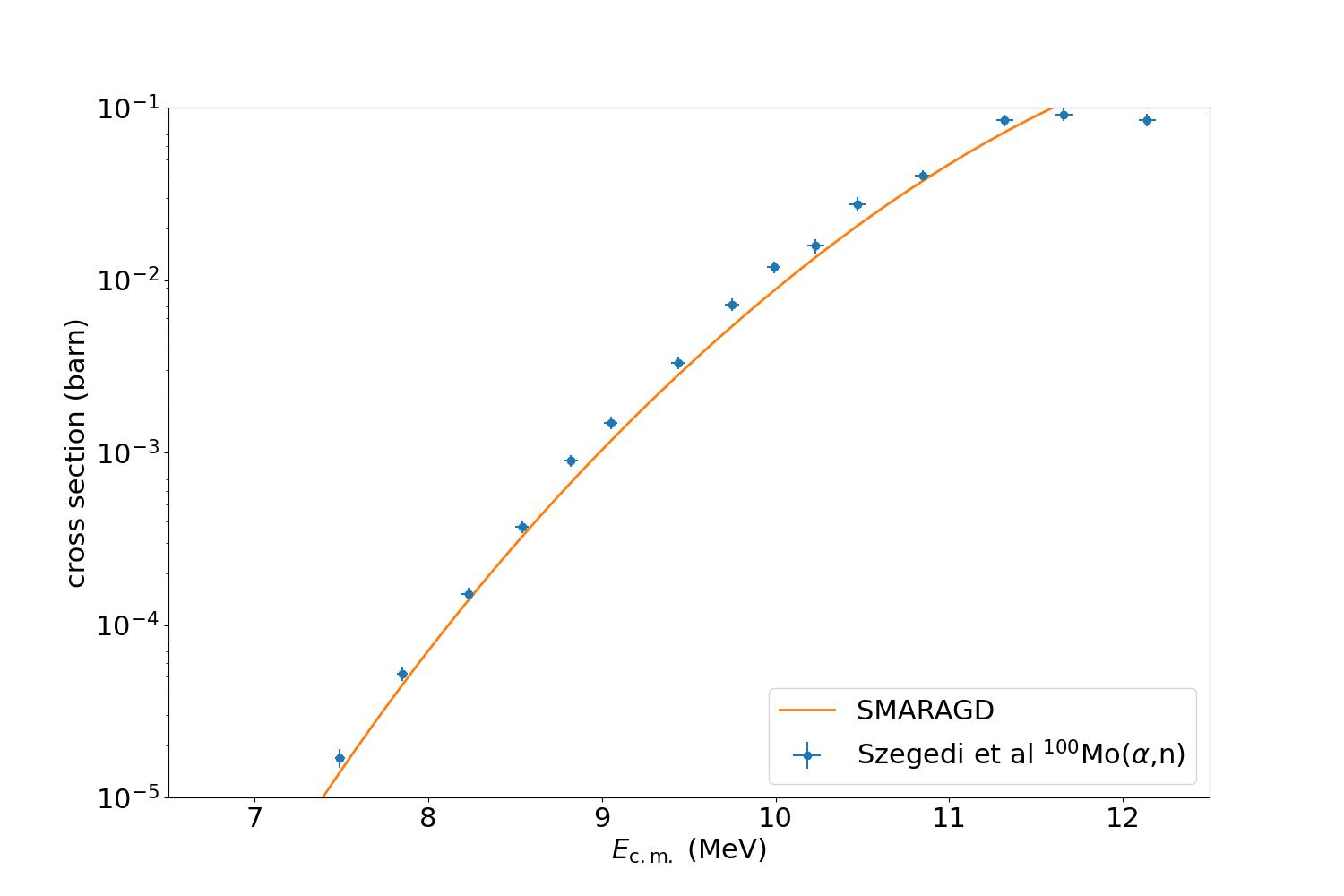}
    \caption{Comparison of ($\alpha$,n) cross sections predicted by SMARAGD with experimental data by Kiss et al (2025) ($^{86}$Kr($\alpha$,n) \cite{2025ApJ...988..170K}), Kiss et al (2021) ($^{96}$Zr($\alpha$,n) \cite{2021ApJ...908..202K}), and Szegedi et al \cite{2021PhRvC.104c5804S}. Note the different scales on the vertical axes.}
    \label{fig:exp_alpha}
\end{figure*}

Experimental data for charged-particle reactions within or close to the astrophysical energy range are scarce \cite{2013RPPh...76f6201R,MOHR2021101453}. This is especially true for $\alpha$-induced reactions due to the high Coulomb barriers strongly suppressing reaction cross sections at low interaction energies. For a few examples, comparisons between predictions with the current version of the SMARAGD code and data at low energy are shown in Figures \ref{fig:exp_proton} and \ref{fig:exp_alpha}. These relevant literature data are excellently reproduced by the predictions.

For the proton capture reactions shown in Figure~\ref{fig:exp_proton}, the use of the modified proton potential improved the reproduction of the experimental data, both regarding the energy dependence and the scale.

The ATOMKI-V2 potential proved essential for the reproduction of the ($\alpha$,n) data shown in Figure \ref{fig:exp_alpha}. There is no ($\alpha$,$\gamma$) data available close to the astrophysically interesting energy range but due to the fact that the ($\alpha$,n) cross sections are exclusively dependent on the $\alpha$ width (see Section \ref{sec:sensis}), the shown comparisons provide a good measure of the quality of the predictions for $\alpha$ widths.

As explained above \rev{in Section~\ref{sec:sensis}}, even a good reproduction of experimental \rev{g.s.} cross sections can only increase the confidence in the predictions but does not necessarily imply also accurate reaction rates. This is especially the case when the contribution of thermally excited target states is not negligible \rev{because, as explained above, reactions on excited target states may be sensitive to other widths than reactions on the ground state. Furthermore,} for measurements above the relevant energy window (Gamow window) it has to be assessed carefully whether the measured cross sections and the astrophysical reaction rate have a similar sensitivity to the contributing nuclear properties. At higher interaction energies, the sensitivity may be quite different from the one at astrophysical, lower energies (see \cite{2013JPhCS.420a2138R,2011IJMPE..20.1071R} for examples). 

For the shown cases, the sensitivities are similar to the sensitivities at astrophysical energies. The good reproduction of the energy dependence also provides confidence that the predictions hold at low energy. The contribution of excited states, though, is not negligible. Accounting for these requires the prediction to hold at even lower energies than given by the Gamow window.

\section{Comparison to calculations with different treatment of nuclear properties}

In addition to the different numerics for calculating transmission coefficients, the new rates also exhibit differences to previous calculations with NON-SMOKER due to changed nuclear input (experimentally determined spins and parities of discrete levels) and changed treatment of certain nuclear properties. To demonstrate the impact of those changes, the figures addressed below not only show comparisons between NON-SMOKER and SMARAGD but also of rates obtained using previous treatments with the new SMARAGD rates at typical temperatures for nucleosynthesis processes with charged particles (e.g., the $\gamma$ process \cite{2013RPPh...76f6201R}).

\begin{figure*}
    \centering
    \includegraphics[width=\columnwidth]{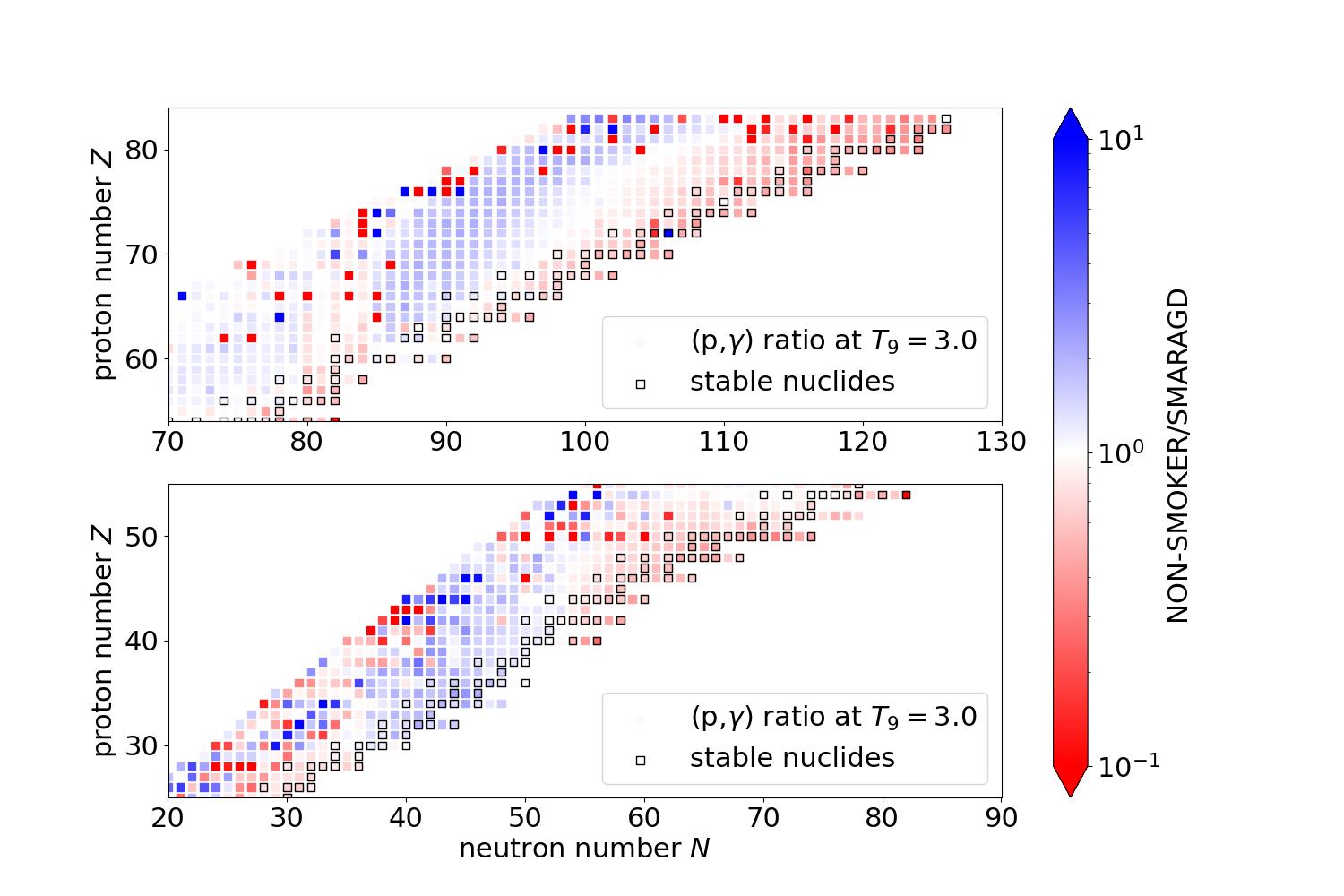}
    \includegraphics[width=\columnwidth]{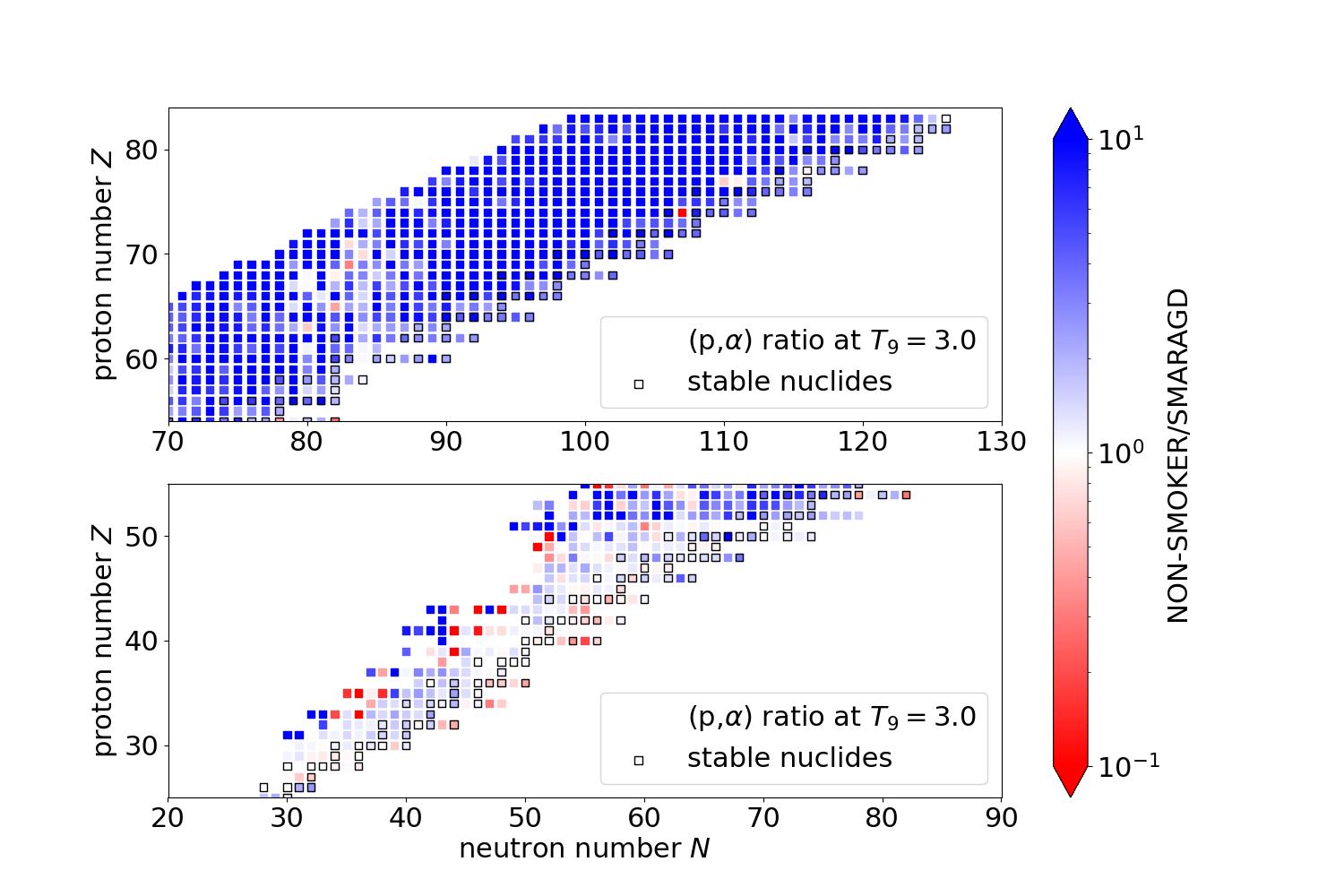}
    \caption{Comparison of NON-SMOKER \cite{2000ADNDT..75....1R} and SMARAGD rates for proton-induced reactions. Only (p,$\alpha$) reactions with positive $Q$ value are shown.}
    \label{fig:comp_proton}
\end{figure*}

\begin{figure*}
    \centering
    \includegraphics[width=\columnwidth]{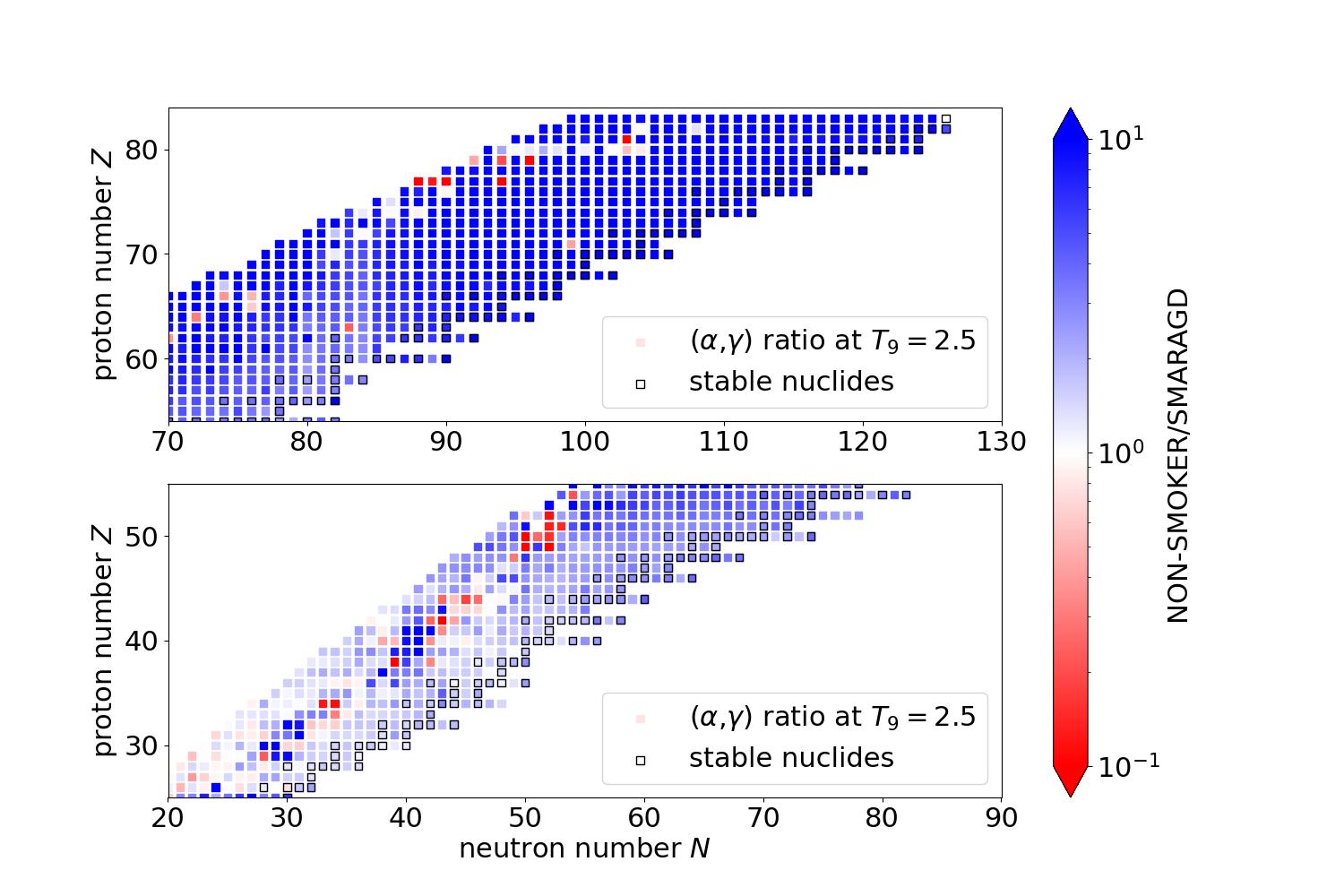}
    \includegraphics[width=\columnwidth]{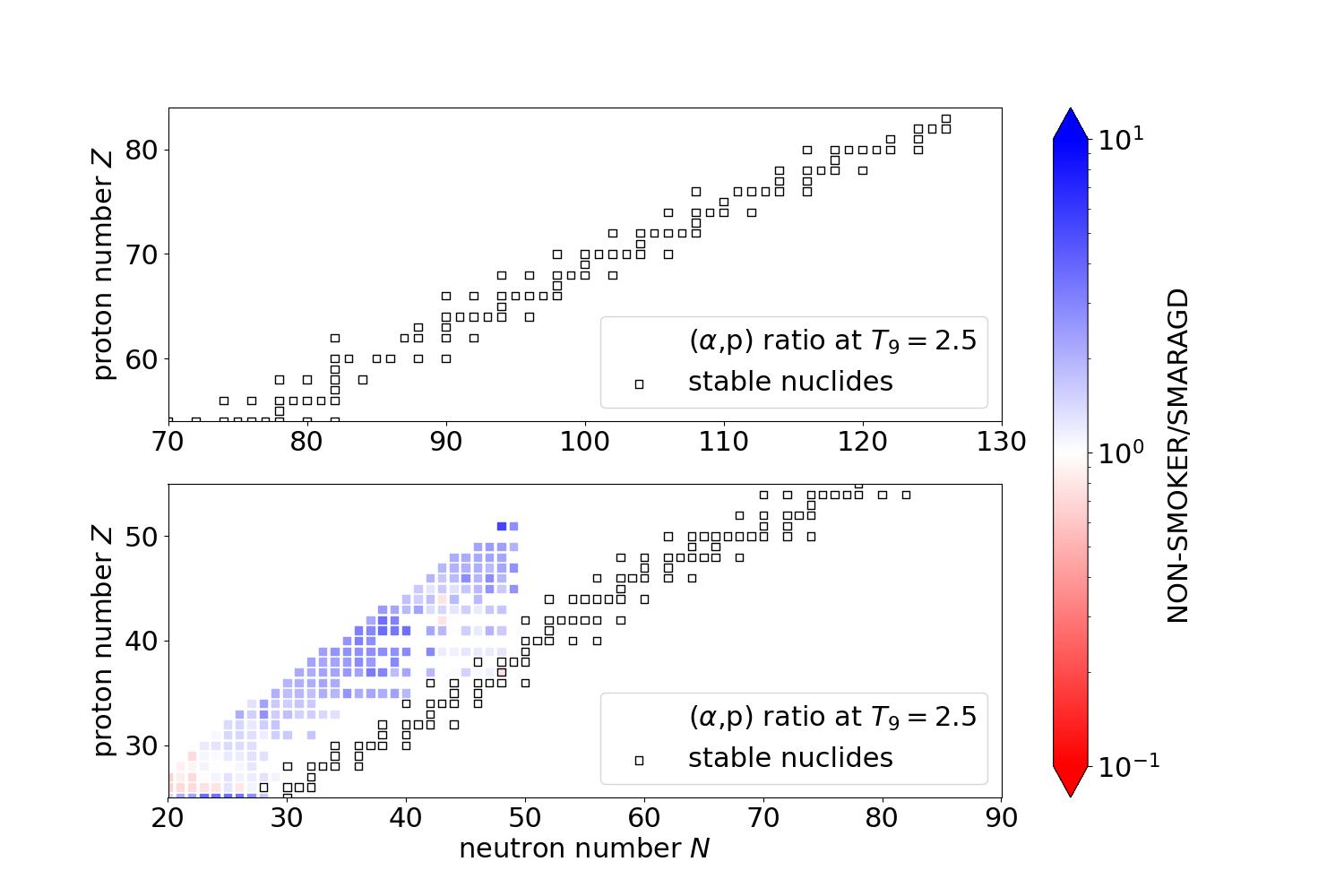}
    \caption{Comparison of NON-SMOKER \cite{2000ADNDT..75....1R} and SMARAGD rates for $\alpha$-induced reactions. Only ($\alpha$,p) reactions with positive $Q$ value are shown.}
    \label{fig:comp_alpha}
\end{figure*}

\begin{figure*}
    \centering
    \includegraphics[width=\columnwidth]{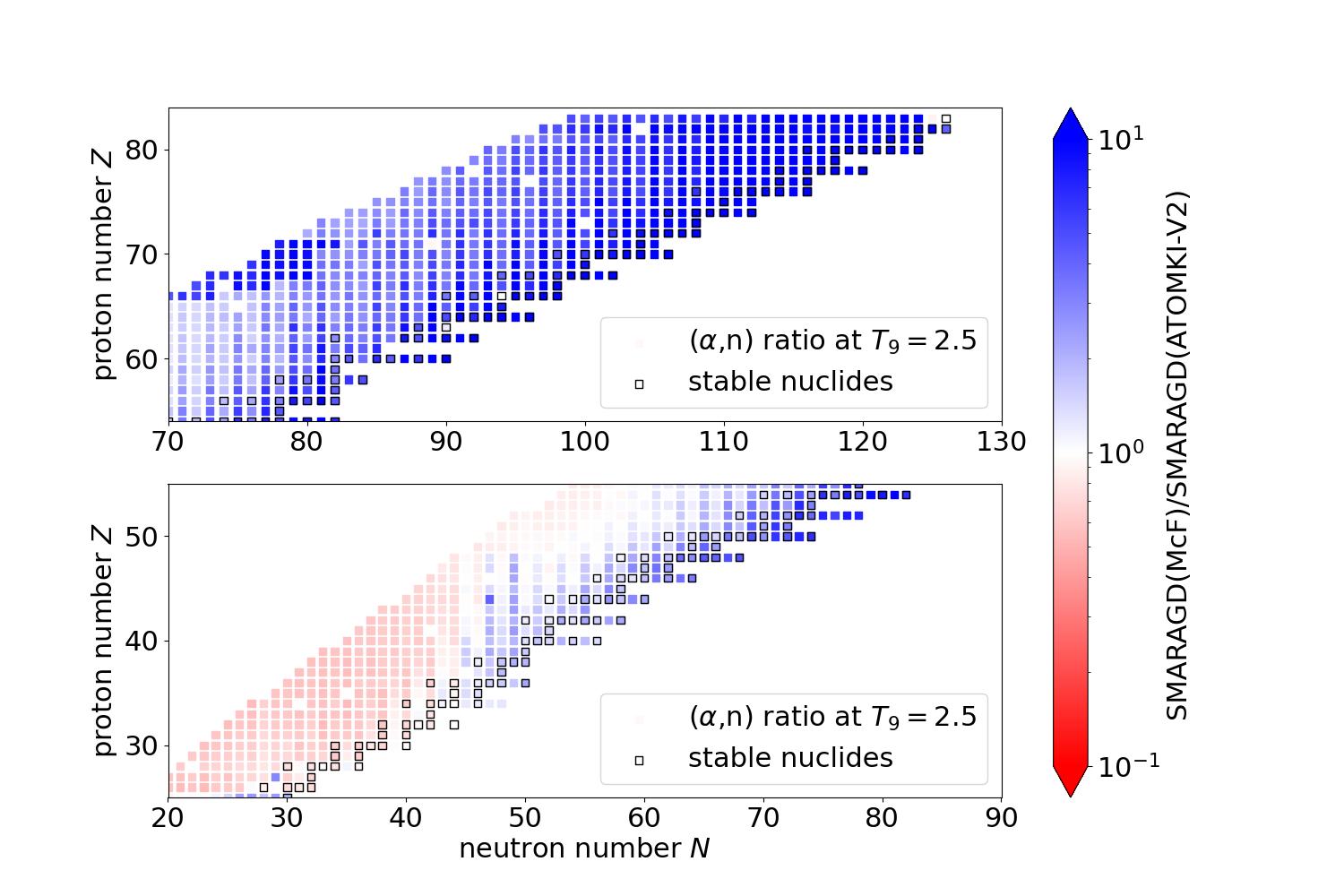}
    \includegraphics[width=\columnwidth]{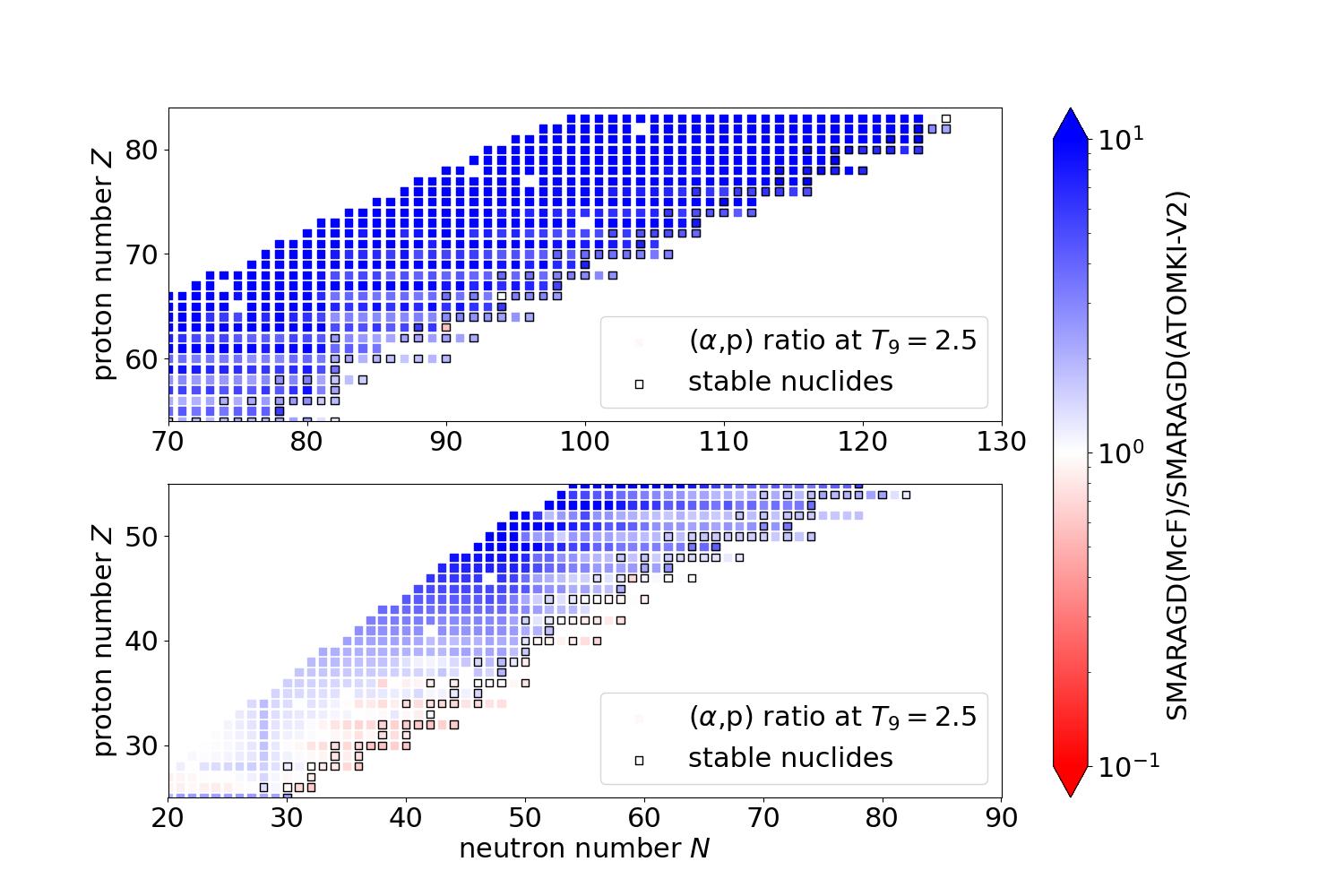}
    \caption{Comparison of rates obtained with the optical $\alpha$+nucleus potentials by \cite{1966NucPh..84..177M} (McF) and \cite{2020PhRvL.124y2701M,MOHR2021101453} (ATOMKI-V2).}
    \label{fig:comp_alphapot}
\end{figure*}

\begin{figure*}
    \centering
    \includegraphics[width=\columnwidth]{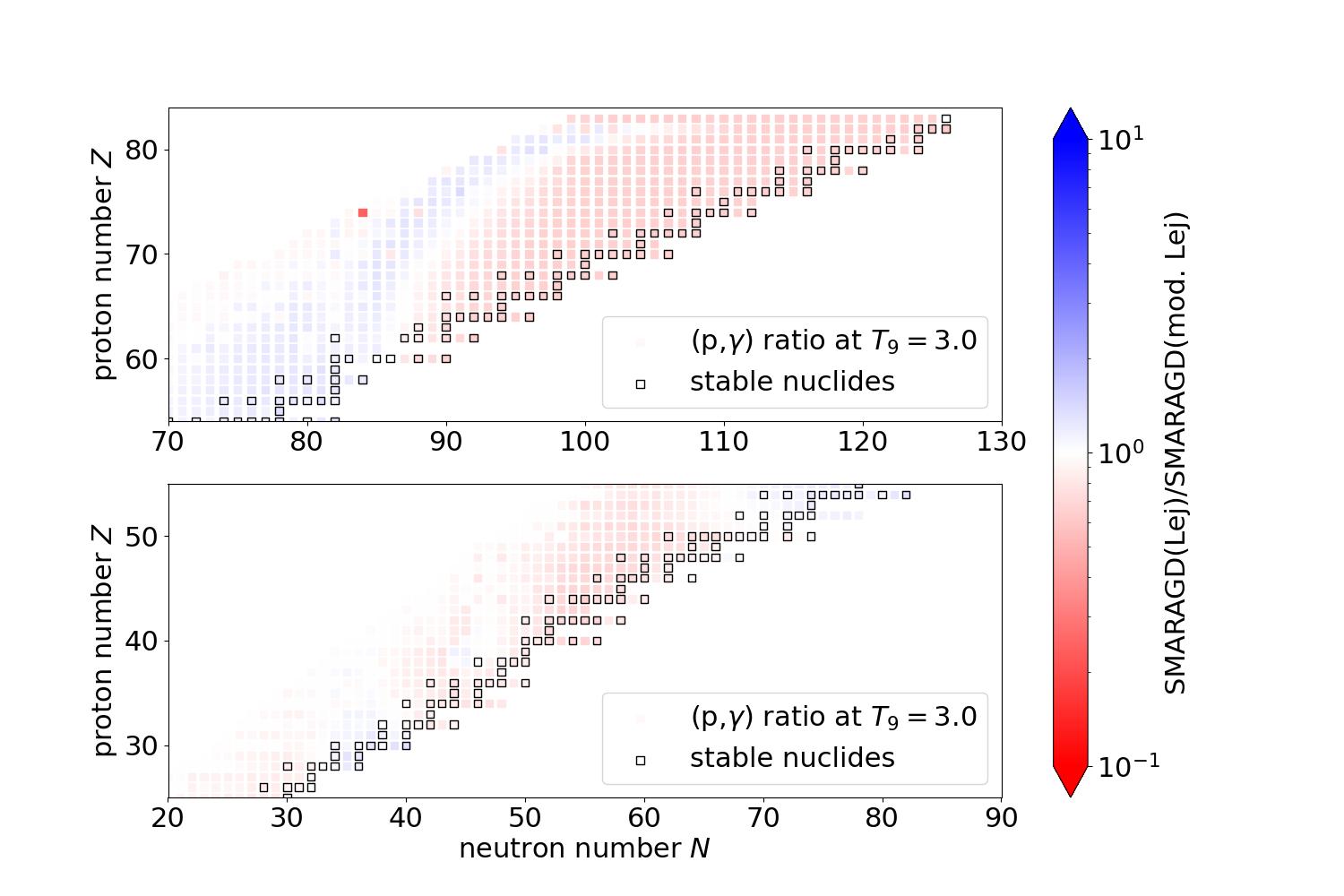}
    \includegraphics[width=\columnwidth]{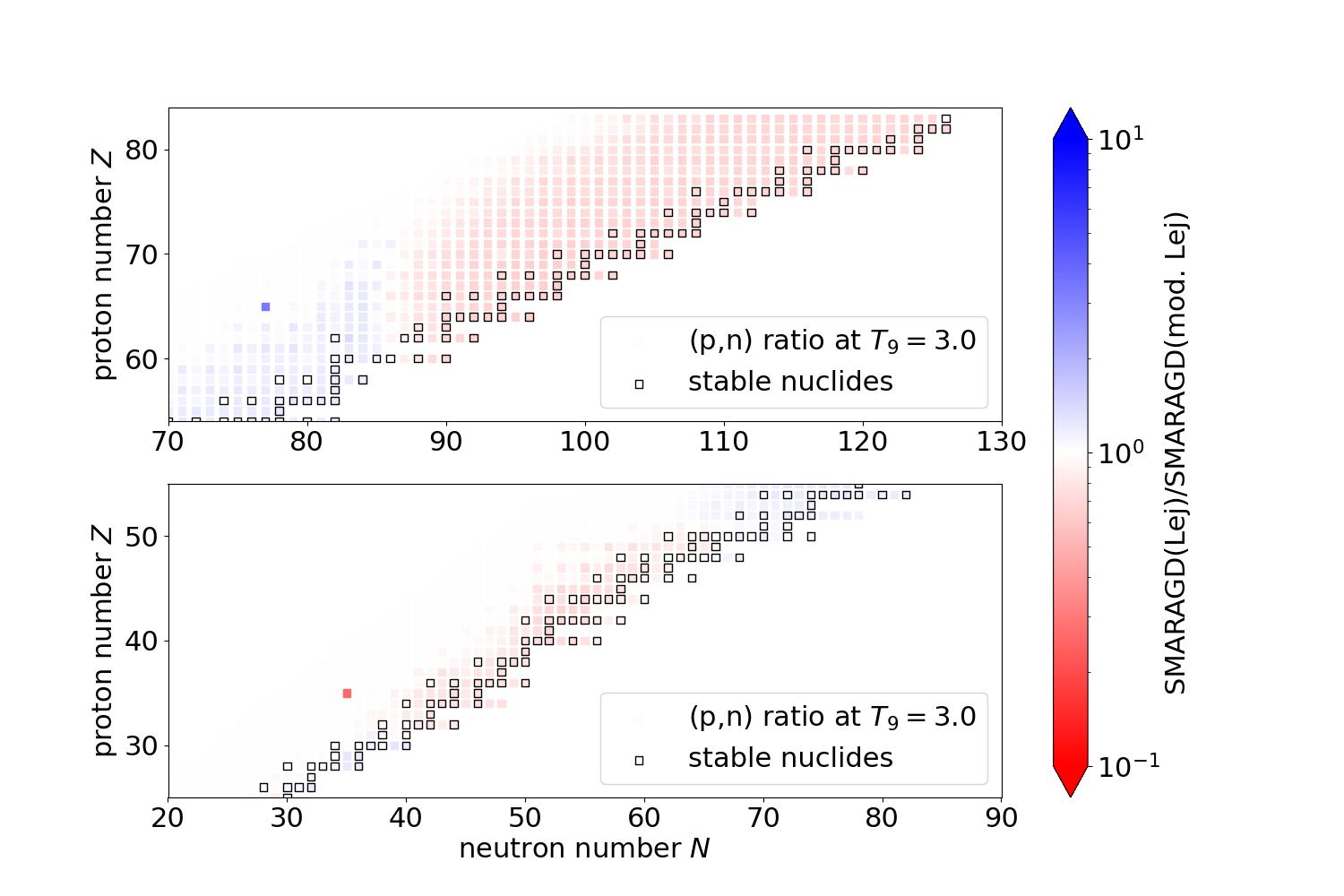}
    \caption{Comparison of rates obtained with the optical p+nucleus potentials by \cite{1980PhRvC..21.1107L} (Lej) and \cite{2008PhRvL.101s1101K,2009PhRvC..80c5801R,2010JPhCS.202a2013R} (mod Lej).}
    \label{fig:comp_protpot}
\end{figure*}

\begin{figure*}
    \centering
    \includegraphics[width=\columnwidth]{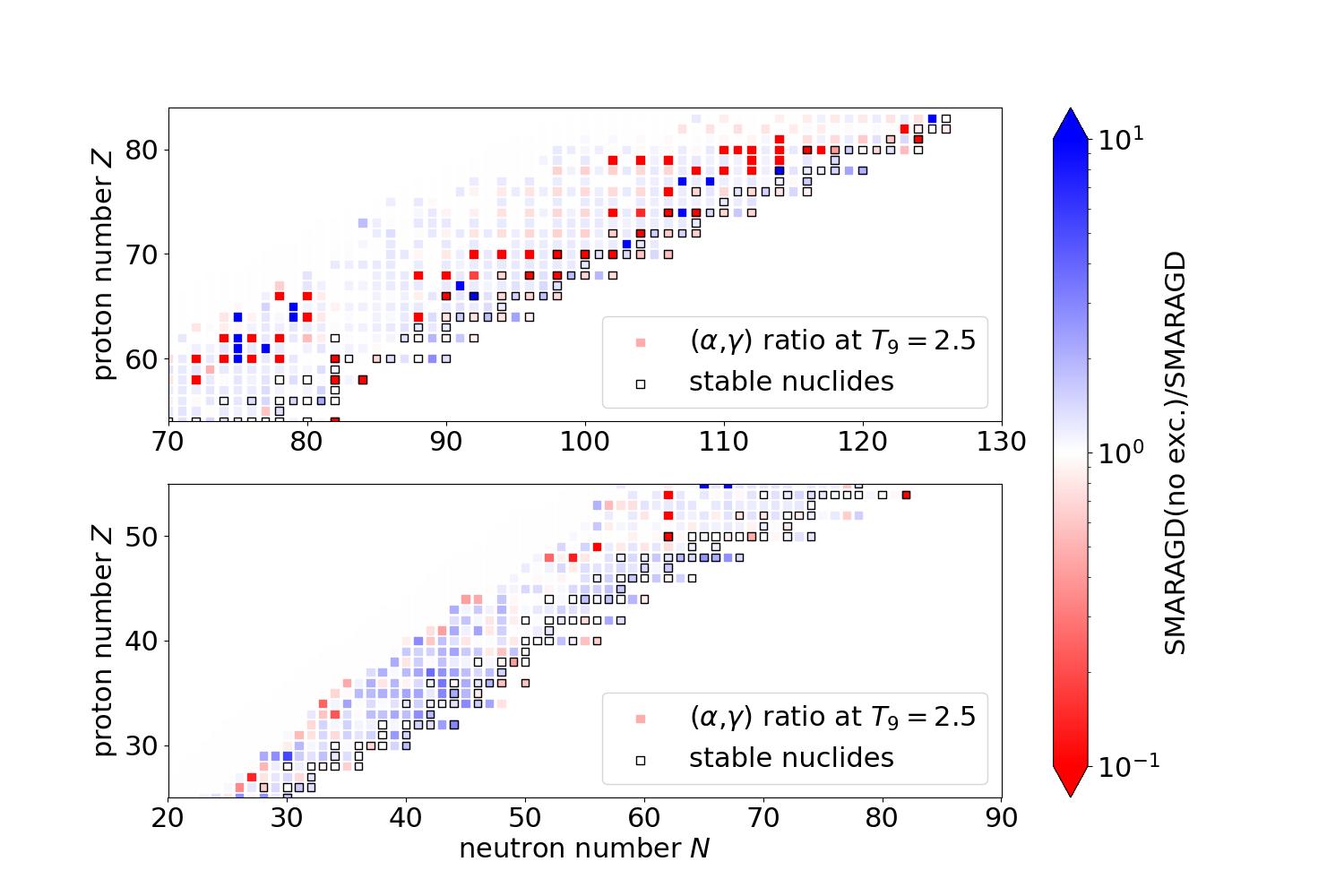}
    \includegraphics[width=\columnwidth]{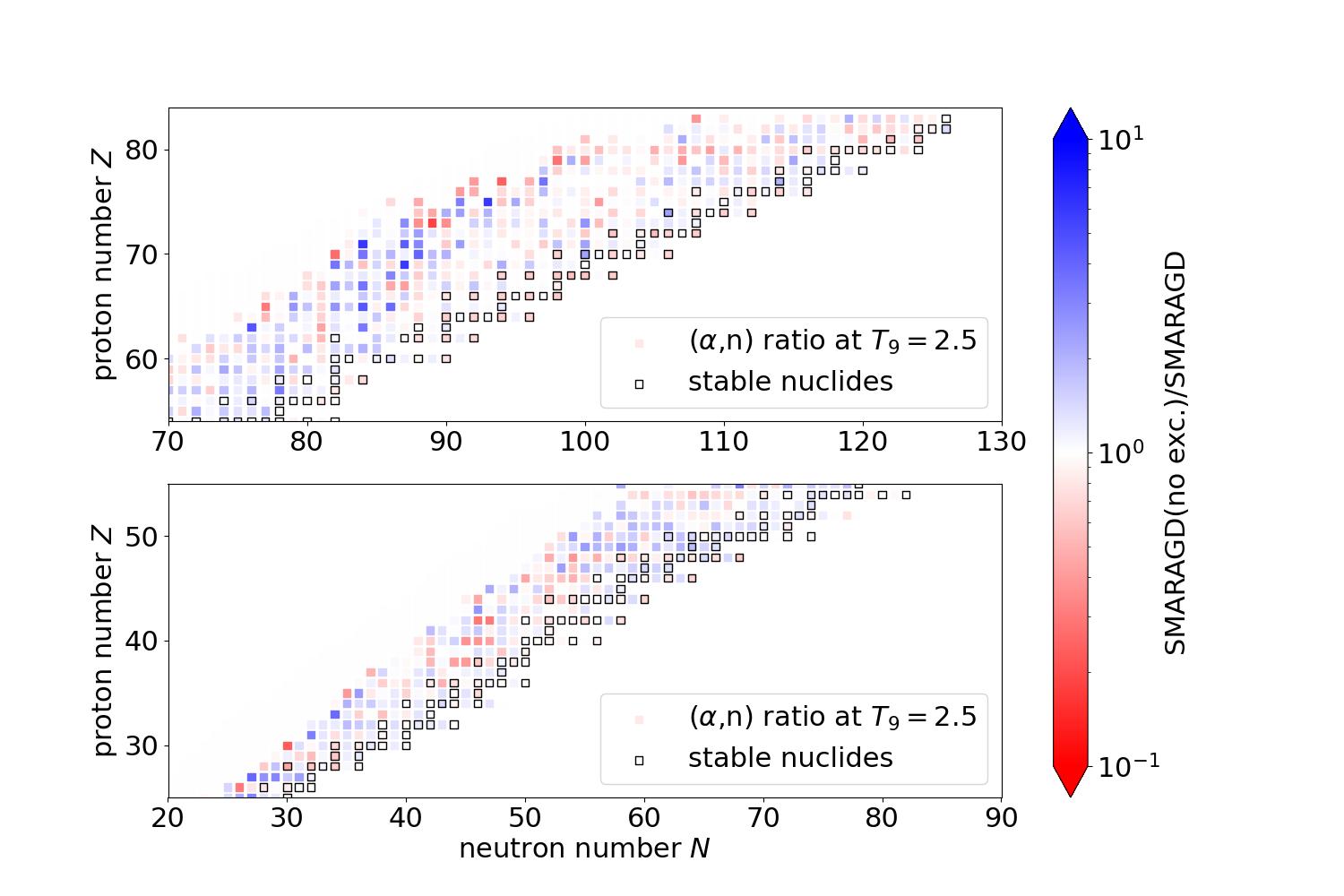}
    \caption{\rev{Comparison of rates obtained with and without the inclusion of experimental excited levels for $\alpha$-induced reactions.}}
    \label{fig:comp_noexc_alpha}
\end{figure*}

\begin{figure*}
    \centering
    \includegraphics[width=\columnwidth]{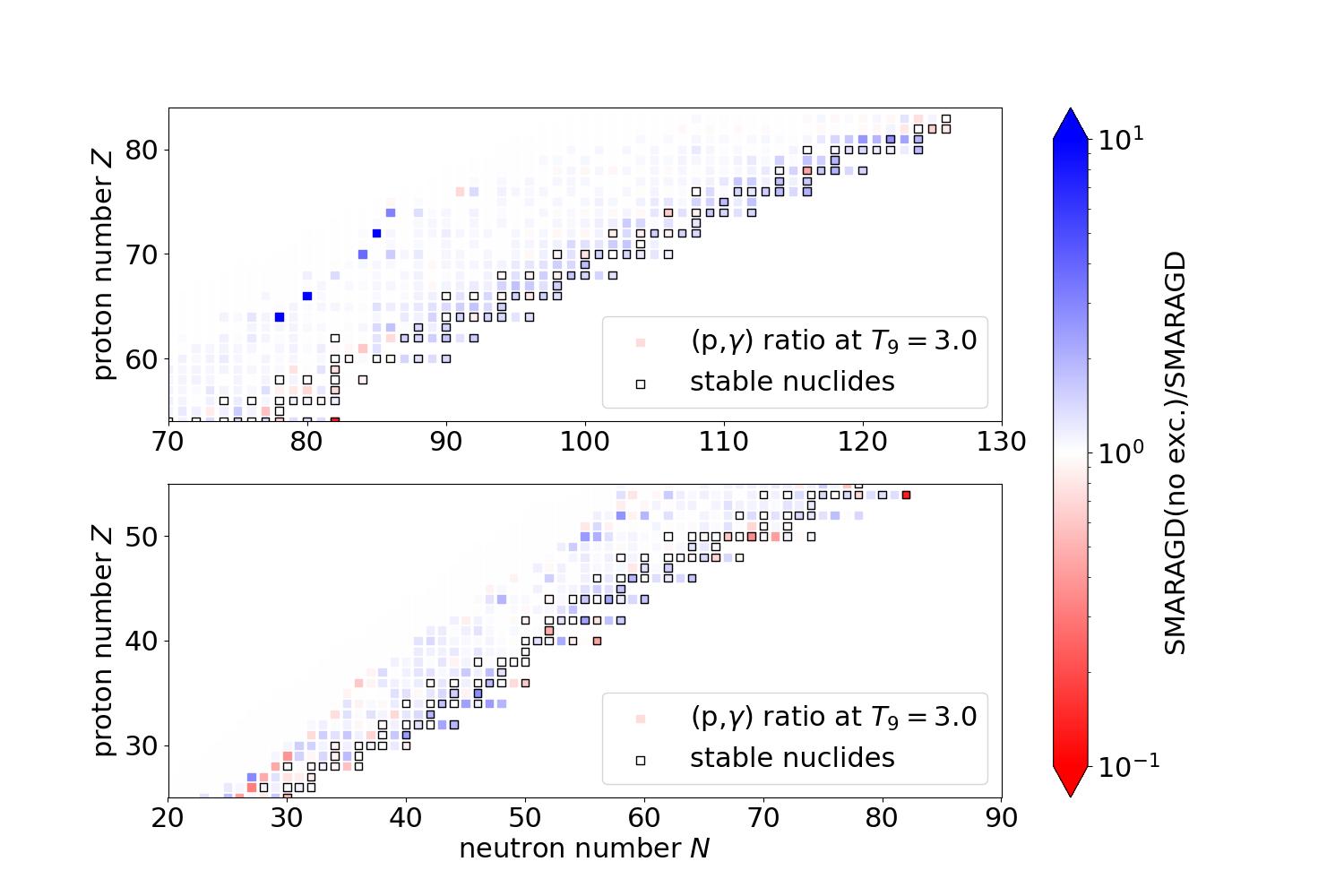}
    \includegraphics[width=\columnwidth]{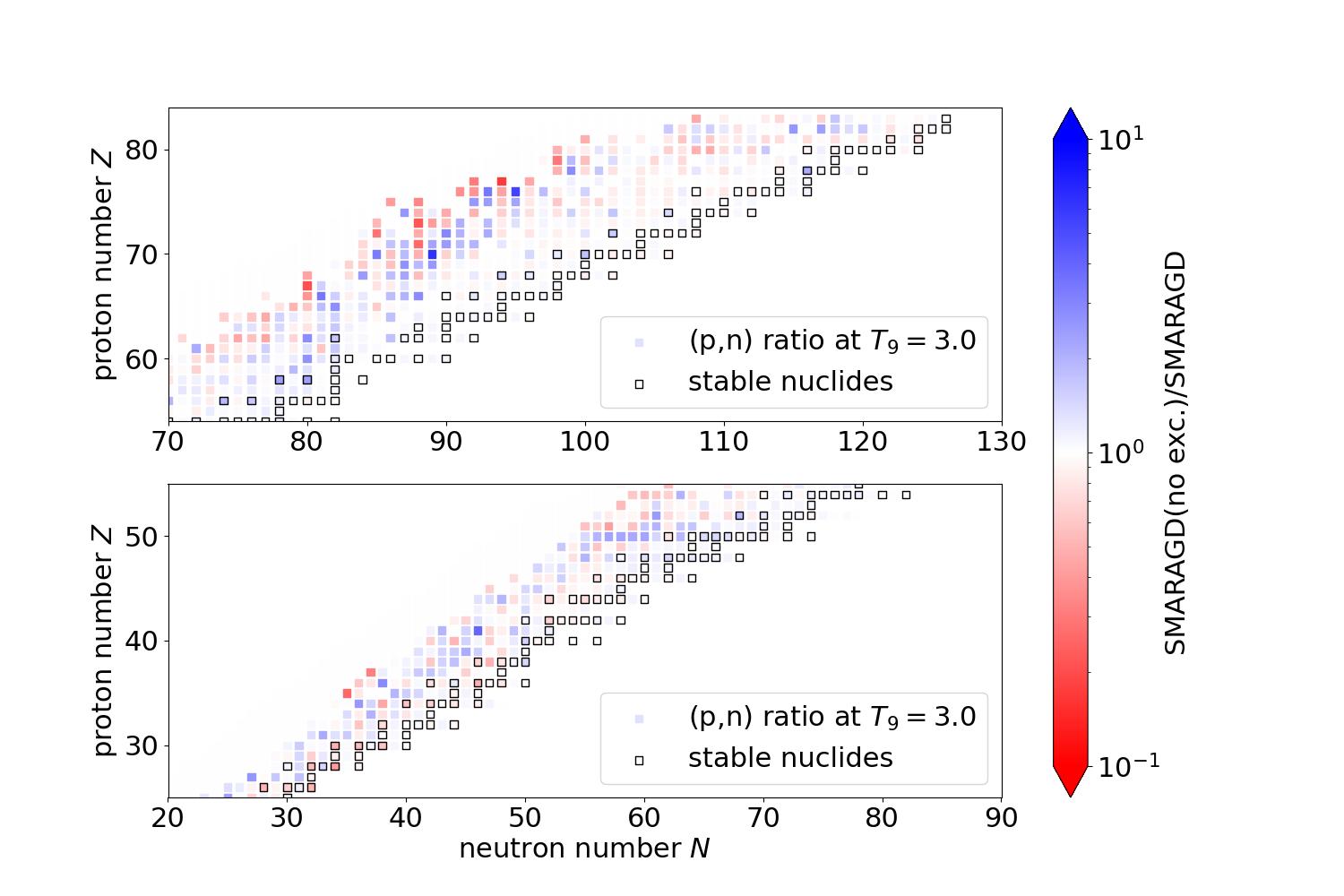}
    \caption{\rev{Comparison of rates obtained with and without the inclusion of experimental excited levels for proton-induced reactions.}}
    \label{fig:comp_noexc_proton}
\end{figure*}

Figures \ref{fig:comp_proton} and \ref{fig:comp_alpha} show a comparison of the present SMARAGD rates with the previous NON-SMOKER results. As the differences seen in these figures originate from a convolution of several changes in code and input, Figures \ref{fig:comp_alphapot}$-$\ref{fig:comp_noexc_proton} selectively show the changes arising from a modification in a single "ingredient". The impact of the changes can be understood better by also consulting the figures and tables showing the sensitivities to different (averaged) resonance widths.

It is obvious that the new choice of the optical $\alpha$+nucleus potential has the largest impact, affecting all reactions with an $\alpha$ particle in either the entrance or the exit channel. This can be clearly seen in Figure \ref{fig:comp_alphapot} -- comparing calculations with the previously used $\alpha$+nucleus optical potential (McF \cite{1966NucPh..84..177M}) and the recent ATOMKI-V2 potential \cite{MOHR2021101453} -- but is also the major reason for the difference between NON-SMOKER and SMARAGD values for such reactions seen in Figures \ref{fig:comp_proton} and \ref{fig:comp_alpha}. In Figures \ref{fig:comp_alpha} and \ref{fig:comp_alphapot} it is also interesting to see that the changes are more pronounced for the heavier nuclides whereas moderate to no changes are seen for the lower part of the investigated mass range. Two effects are contributing to this result. First, the Coulomb barriers are increasing towards the heavier mass range, amplifying the difference between the two optical potentials. Second, the impact of the $\alpha$ width reduces when moving to the proton-rich side in the lower mass range due to the decrease of the proton width, as can be seen in Figure~\ref{fig:domi_alpha}. As a side note it should be remembered that the ATOMKI-V2 potential was derived by analysis of scattering and reaction data in the lower mass range and it is conceivable that it may need an adjustment for the upper mass range.

Figure~\ref{fig:comp_protpot} shows the impact of different choices of the optical proton+nucleus potential. It is evident that the impact is much smaller than the one of the changed $\alpha$+nucleus potential but nevertheless the modification was necessary to obtain better agreement with experimental reaction cross sections \rev{\cite{2008PhRvL.101s1101K,2009PhRvC..80c5801R,2010JPhCS.202a2013R} (see also Section~\ref{sec:implementation}). As shown in Figure~\ref{fig:exp_proton}, the modified potential also fares well in comparison to the data of Refs.~\cite{2017PhRvC..96c5806K,2024PhRvC.110a5803H,2025PhRvL.134n2701D} which were not available when the modification was suggested originally.}

Finally, Figures \ref{fig:comp_noexc_alpha} and \ref{fig:comp_noexc_proton} compare rates obtained with and without the inclusion of discrete experimental levels above the ground state. The theoretical level density was applied above the g.s. or the last included experimental level, respectively. The moderate change underlines the validity of the level density description. Closer to the driplines the difference seems larger but for such nuclides, although only few experimental states with unambiguously assigned spin and parity values were used there, it cannot be ruled out that the included level scheme is incomplete even up to the cutoff energy $E_\mathrm{last}^\mathrm{x}$.

\section{Summary}\label{sec:conclusion}

Astrophysical reaction rates for reactions with proton-rich nuclides from stability to the proton dripline were calculated with an updated version of the SMARAGD statistical model (Hauser-Feshbach) code. Here, the focus was on reactions with protons or $\alpha$ particles as required for nucleosynthesis in proton-rich matter. For completeness, also neutron-induced reactions are given for the same set of targets. The full sets of astrophysical reaction rates and predicted laboratory cross sections are available online at \url{https://doi.org/10.5281/zenodo.17770707}. It is to be noted that these are purely based on theory. In rate libraries to be used in nucleosynthesis models they should be supplemented or replaced by experimental rates where available.

Some comments on dependencies of rates on various nuclear properties and on the appropriate way to compare to experiments have been given above.

The new rate set for charged-particle induced reactions provides a better description of experimental data than the previously widely used NON-SMOKER rates as well as the rates provided in \cite{2010ApJS..189..240C}, especially for reactions involving $\alpha$ particles.

\bibliography{rauscher}


\begin{thebibliography}{53}
\ifx \bisbn   \undefined \def \bisbn  #1{ISBN #1}\fi
\ifx \binits  \undefined \def \binits#1{#1}\fi
\ifx \bauthor  \undefined \def \bauthor#1{#1}\fi
\ifx \batitle  \undefined \def \batitle#1{#1}\fi
\ifx \bjtitle  \undefined \def \bjtitle#1{#1}\fi
\ifx \bvolume  \undefined \def \bvolume#1{\textbf{#1}}\fi
\ifx \byear  \undefined \def \byear#1{#1}\fi
\ifx \bissue  \undefined \def \bissue#1{#1}\fi
\ifx \bfpage  \undefined \def \bfpage#1{#1}\fi
\ifx \blpage  \undefined \def \blpage #1{#1}\fi
\ifx \burl  \undefined \def \burl#1{\textsf{#1}}\fi
\ifx \doiurl  \undefined \def \doiurl#1{\url{https://doi.org/#1}}\fi
\ifx \betal  \undefined \def \betal{\textit{et al.}}\fi
\ifx \binstitute  \undefined \def \binstitute#1{#1}\fi
\ifx \binstitutionaled  \undefined \def \binstitutionaled#1{#1}\fi
\ifx \bctitle  \undefined \def \bctitle#1{#1}\fi
\ifx \beditor  \undefined \def \beditor#1{#1}\fi
\ifx \bpublisher  \undefined \def \bpublisher#1{#1}\fi
\ifx \bbtitle  \undefined \def \bbtitle#1{#1}\fi
\ifx \bedition  \undefined \def \bedition#1{#1}\fi
\ifx \bseriesno  \undefined \def \bseriesno#1{#1}\fi
\ifx \blocation  \undefined \def \blocation#1{#1}\fi
\ifx \bsertitle  \undefined \def \bsertitle#1{#1}\fi
\ifx \bsnm \undefined \def \bsnm#1{#1}\fi
\ifx \bsuffix \undefined \def \bsuffix#1{#1}\fi
\ifx \bparticle \undefined \def \bparticle#1{#1}\fi
\ifx \barticle \undefined \def \barticle#1{#1}\fi
\bibcommenthead
\ifx \bconfdate \undefined \def \bconfdate #1{#1}\fi
\ifx \botherref \undefined \def \botherref #1{#1}\fi
\ifx \url \undefined \def \url#1{\textsf{#1}}\fi
\ifx \bchapter \undefined \def \bchapter#1{#1}\fi
\ifx \bbook \undefined \def \bbook#1{#1}\fi
\ifx \bcomment \undefined \def \bcomment#1{#1}\fi
\ifx \oauthor \undefined \def \oauthor#1{#1}\fi
\ifx \citeauthoryear \undefined \def \citeauthoryear#1{#1}\fi
\ifx \endbibitem  \undefined \def \endbibitem {}\fi
\ifx \bconflocation  \undefined \def \bconflocation#1{#1}\fi
\ifx \arxivurl  \undefined \def \arxivurl#1{\textsf{#1}}\fi
\csname PreBibitemsHook\endcsname

\bibitem[\protect\citeauthoryear{{Bohr}}{1936}]{1936Natur.137..344B}
\begin{barticle}
\bauthor{\bsnm{{Bohr}}, \binits{N.}}:
\batitle{{Neutron Capture and Nuclear Constitution}}.
\bjtitle{\nat}
\bvolume{137}(\bissue{3461}),
\bfpage{344}--\blpage{348}
(\byear{1936})
\doiurl{10.1038/137344a0}
\end{barticle}
\endbibitem

\bibitem[\protect\citeauthoryear{{Weisskopf} and
  {Ewing}}{1940}]{1940PhRv...57..472W}
\begin{barticle}
\bauthor{\bsnm{{Weisskopf}}, \binits{V.F.}},
\bauthor{\bsnm{{Ewing}}, \binits{D.H.}}:
\batitle{{On the Yield of Nuclear Reactions with Heavy Elements}}.
\bjtitle{Physical Review}
\bvolume{57}(\bissue{6}),
\bfpage{472}--\blpage{485}
(\byear{1940})
\doiurl{10.1103/PhysRev.57.472}
\end{barticle}
\endbibitem

\bibitem[\protect\citeauthoryear{{Hauser} and
  {Feshbach}}{1952}]{1952PhRv...87..366H}
\begin{barticle}
\bauthor{\bsnm{{Hauser}}, \binits{W.}},
\bauthor{\bsnm{{Feshbach}}, \binits{H.}}:
\batitle{{The Inelastic Scattering of Neutrons}}.
\bjtitle{Physical Review}
\bvolume{87}(\bissue{2}),
\bfpage{366}--\blpage{373}
(\byear{1952})
\doiurl{10.1103/PhysRev.87.366}
\end{barticle}
\endbibitem

\bibitem[\protect\citeauthoryear{{Rauscher}}{2011}]{2011IJMPE..20.1071R}
\begin{barticle}
\bauthor{\bsnm{{Rauscher}}, \binits{T.}}:
\batitle{{The Path to Improved Reaction Rates for Astrophysics}}.
\bjtitle{International Journal of Modern Physics E}
\bvolume{20}(\bissue{5}),
\bfpage{1071}--\blpage{1169}
(\byear{2011})
\doiurl{10.1142/S021830131101840X}
{\href{https://arxiv.org/abs/1010.4283}{{arXiv:1010.4283}}}
{[nucl-th]}
\end{barticle}
\endbibitem

\bibitem[\protect\citeauthoryear{{Rauscher}}{2020}]{2020entn.book.....R}
\begin{bbook}
\bauthor{\bsnm{{Rauscher}}, \binits{T.}}:
\bbtitle{Essentials of Nucleosynthesis and Theoretical Nuclear Astrophysics}.
\bpublisher{IOP Publishing},
\blocation{Bristol, UK}
(\byear{2020}).
\doiurl{10.1088/2514-3433/ab8737}
\end{bbook}
\endbibitem

\bibitem[\protect\citeauthoryear{{Truran} et~al.}{1966}]{1966CaJPh..44..151T}
\begin{barticle}
\bauthor{\bsnm{{Truran}}, \binits{J.W.}},
\bauthor{\bsnm{{Hansen}}, \binits{C.J.}},
\bauthor{\bsnm{{Cameron}}, \binits{A.G.W.}},
\bauthor{\bsnm{{Gilbert}}, \binits{A.}}:
\batitle{{Thermonuclear Reactions in Medium and Heavy Nuclei}}.
\bjtitle{Canadian Journal of Physics}
\bvolume{44}(\bissue{1}),
\bfpage{151}--\blpage{174}
(\byear{1966})
\doiurl{10.1139/p66-011}
\end{barticle}
\endbibitem

\bibitem[\protect\citeauthoryear{{Arnould}}{1972}]{1972A&A....19...92A}
\begin{barticle}
\bauthor{\bsnm{{Arnould}}, \binits{M.}}:
\batitle{{Influence of the Excited States of Target Nuclei in the Vicinity of
  the Iron Peak on Stellar Reaction Rates}}.
\bjtitle{\aap}
\bvolume{19},
\bfpage{92}
(\byear{1972})
\end{barticle}
\endbibitem

\bibitem[\protect\citeauthoryear{{Holmes} et~al.}{1976}]{1976ADNDT..18..305H}
\begin{barticle}
\bauthor{\bsnm{{Holmes}}, \binits{J.A.}},
\bauthor{\bsnm{{Woosley}}, \binits{S.E.}},
\bauthor{\bsnm{{Fowler}}, \binits{W.A.}},
\bauthor{\bsnm{{Zimmerman}}, \binits{B.A.}}:
\batitle{{Tables of Thermonuclear-Reaction-Rate Data for Neutron-Induced
  Reactions on Heavy Nclei}}.
\bjtitle{Atomic Data and Nuclear Data Tables}
\bvolume{18},
\bfpage{305}
(\byear{1976})
\doiurl{10.1016/0092-640X(76)90011-5}
\end{barticle}
\endbibitem

\bibitem[\protect\citeauthoryear{{Woosley} et~al.}{1978}]{1978ADNDT..22..371W}
\begin{barticle}
\bauthor{\bsnm{{Woosley}}, \binits{S.E.}},
\bauthor{\bsnm{{Fowler}}, \binits{W.A.}},
\bauthor{\bsnm{{Holmes}}, \binits{J.A.}},
\bauthor{\bsnm{{Zimmerman}}, \binits{B.A.}}:
\batitle{{Semiempirical Thermonuclear Reaction-Rate Data for Intermediate-Mass
  Nuclei}}.
\bjtitle{Atomic Data and Nuclear Data Tables}
\bvolume{22},
\bfpage{371}
(\byear{1978})
\doiurl{10.1016/0092-640X(78)90018-9}
\end{barticle}
\endbibitem

\bibitem[\protect\citeauthoryear{{Thielemann}
  et~al.}{1986}]{1986ana..work..525T}
\begin{bchapter}
\bauthor{\bsnm{{Thielemann}}, \binits{F.-K.}},
\bauthor{\bsnm{{Arnould}}, \binits{M.}},
\bauthor{\bsnm{{Truran}}, \binits{J.W.}}:
\bctitle{{Thermonuclear reaction rates from statistical model calculations.}}
In: \beditor{\bsnm{{Vangioni-Flam}}, \binits{E.}},
\beditor{\bsnm{{Audouze}}, \binits{J.}},
\beditor{\bsnm{{Casse}}, \binits{M.}},
\beditor{\bsnm{{Chieze}}, \binits{J.-P.}},
\beditor{\bsnm{{Tran Thanh Van}}, \binits{J.}} (eds.)
\bbtitle{Advances in Nuclear Astrophysics},
pp. \bfpage{525}--\blpage{540}
(\byear{1986})
\end{bchapter}
\endbibitem

\bibitem[\protect\citeauthoryear{{Rauscher} and
  {Thielemann}}{2000}]{2000ADNDT..75....1R}
\begin{barticle}
\bauthor{\bsnm{{Rauscher}}, \binits{T.}},
\bauthor{\bsnm{{Thielemann}}, \binits{F.-K.}}:
\batitle{{Astrophysical Reaction Rates From Statistical Model Calculations}}.
\bjtitle{Atomic Data and Nuclear Data Tables}
\bvolume{75}(\bissue{1-2}),
\bfpage{1}--\blpage{351}
(\byear{2000})
\doiurl{10.1006/adnd.2000.0834}
{\href{https://arxiv.org/abs/astro-ph/0004059}{{arXiv:astro-ph/0004059}}}
{[astro-ph]}
\end{barticle}
\endbibitem

\bibitem[\protect\citeauthoryear{{Rauscher} and
  {Thielemann}}{2001}]{2001ADNDT..79...47R}
\begin{barticle}
\bauthor{\bsnm{{Rauscher}}, \binits{T.}},
\bauthor{\bsnm{{Thielemann}}, \binits{F.-K.}}:
\batitle{{Tables of Nuclear Cross Sections and Reaction Rates: AN Addendum to
  the Paper ``ASTROPHYSICAL Reaction Rates from Statistical Model
  Calculations'' ()}}.
\bjtitle{Atomic Data and Nuclear Data Tables}
\bvolume{79}(\bissue{1}),
\bfpage{47}--\blpage{64}
(\byear{2001})
\doiurl{10.1006/adnd.2001.0863}
{\href{https://arxiv.org/abs/nucl-th/0104003}{{arXiv:nucl-th/0104003}}}
{[nucl-th]}
\end{barticle}
\endbibitem

\bibitem[\protect\citeauthoryear{{Goriely} et~al.}{2008}]{2008A&A...487..767G}
\begin{barticle}
\bauthor{\bsnm{{Goriely}}, \binits{S.}},
\bauthor{\bsnm{{Hilaire}}, \binits{S.}},
\bauthor{\bsnm{{Koning}}, \binits{A.J.}}:
\batitle{{Improved predictions of nuclear reaction rates with the TALYS
  reaction code for astrophysical applications}}.
\bjtitle{\aap}
\bvolume{487}(\bissue{2}),
\bfpage{767}--\blpage{774}
(\byear{2008})
\doiurl{10.1051/0004-6361:20078825}
{\href{https://arxiv.org/abs/0806.2239}{{arXiv:0806.2239}}}
{[astro-ph]}
\end{barticle}
\endbibitem

\bibitem[\protect\citeauthoryear{{Cyburt} et~al.}{2010}]{2010ApJS..189..240C}
\begin{barticle}
\bauthor{\bsnm{{Cyburt}}, \binits{R.H.}},
\bauthor{\bsnm{{Amthor}}, \binits{A.M.}},
\bauthor{\bsnm{{Ferguson}}, \binits{R.}},
\bauthor{\bsnm{{Meisel}}, \binits{Z.}},
\bauthor{\bsnm{{Smith}}, \binits{K.}},
\bauthor{\bsnm{{Warren}}, \binits{S.}},
\bauthor{\bsnm{{Heger}}, \binits{A.}},
\bauthor{\bsnm{{Hoffman}}, \binits{R.D.}},
\bauthor{\bsnm{{Rauscher}}, \binits{T.}},
\bauthor{\bsnm{{Sakharuk}}, \binits{A.}},
\bauthor{\bsnm{{Schatz}}, \binits{H.}},
\bauthor{\bsnm{{Thielemann}}, \binits{F.K.}},
\bauthor{\bsnm{{Wiescher}}, \binits{M.}}:
\batitle{{The JINA REACLIB Database: Its Recent Updates and Impact on Type-I
  X-ray Bursts}}.
\bjtitle{\apjs}
\bvolume{189}(\bissue{1}),
\bfpage{240}--\blpage{252}
(\byear{2010})
\doiurl{10.1088/0067-0049/189/1/240}
\end{barticle}
\endbibitem

\bibitem[\protect\citeauthoryear{{Rauscher}}{2010}]{2010PhRvC..81d5807R}
\begin{barticle}
\bauthor{\bsnm{{Rauscher}}, \binits{T.}}:
\batitle{{Relevant energy ranges for astrophysical reaction rates}}.
\bjtitle{\prc}
\bvolume{81}(\bissue{4}),
\bfpage{045807}
(\byear{2010})
\doiurl{10.1103/PhysRevC.81.045807}
{\href{https://arxiv.org/abs/1003.2802}{{arXiv:1003.2802}}}
{[astro-ph.SR]}
\end{barticle}
\endbibitem

\bibitem[\protect\citeauthoryear{{Fowler}}{1974}]{1974QJRAS..15...82F}
\begin{barticle}
\bauthor{\bsnm{{Fowler}}, \binits{W.A.}}:
\batitle{{High Temperature Nuclear Astrophysics}}.
\bjtitle{\qjras}
\bvolume{15},
\bfpage{82}
(\byear{1974})
\end{barticle}
\endbibitem

\bibitem[\protect\citeauthoryear{{Rauscher} et~al.}{1997}]{1997PhRvC..56.1613R}
\begin{barticle}
\bauthor{\bsnm{{Rauscher}}, \binits{T.}},
\bauthor{\bsnm{{Thielemann}}, \binits{F.-K.}},
\bauthor{\bsnm{{Kratz}}, \binits{K.-L.}}:
\batitle{{Nuclear level density and the determination of thermonuclear rates
  for astrophysics}}.
\bjtitle{\prc}
\bvolume{56}(\bissue{3}),
\bfpage{1613}--\blpage{1625}
(\byear{1997})
\doiurl{10.1103/PhysRevC.56.1613}
{\href{https://arxiv.org/abs/astro-ph/9706294}{{arXiv:astro-ph/9706294}}}
{[astro-ph]}
\end{barticle}
\endbibitem

\bibitem[\protect\citeauthoryear{{Rauscher}}{2003}]{2003ApJS..147..403R}
\begin{barticle}
\bauthor{\bsnm{{Rauscher}}, \binits{T.}}:
\batitle{{Nuclear Partition Functions at Temperatures Exceeding {}10$^{10}$
  K}}.
\bjtitle{\apjs}
\bvolume{147}(\bissue{2}),
\bfpage{403}--\blpage{408}
(\byear{2003})
\doiurl{10.1086/375733}
{\href{https://arxiv.org/abs/astro-ph/0304047}{{arXiv:astro-ph/0304047}}}
{[astro-ph]}
\end{barticle}
\endbibitem

\bibitem[\protect\citeauthoryear{{Rauscher}}{2012}]{2012ApJS..201...26R}
\begin{barticle}
\bauthor{\bsnm{{Rauscher}}, \binits{T.}}:
\batitle{{Sensitivity of Astrophysical Reaction Rates to Nuclear
  Uncertainties}}.
\bjtitle{\apjs}
\bvolume{201}(\bissue{2}),
\bfpage{26}
(\byear{2012})
\doiurl{10.1088/0067-0049/201/2/26}
{\href{https://arxiv.org/abs/1205.0685}{{arXiv:1205.0685}}}
{[astro-ph.SR]}
\end{barticle}
\endbibitem

\bibitem[\protect\citeauthoryear{{Bardin} et~al.}{1972}]{1972CoPhC...3...73B}
\begin{barticle}
\bauthor{\bsnm{{Bardin}}, \binits{C.}},
\bauthor{\bsnm{{Dandeu}}, \binits{Y.}},
\bauthor{\bsnm{{Gauthier}}, \binits{L.}},
\bauthor{\bsnm{{Guillermin}}, \binits{J.}},
\bauthor{\bsnm{{Lena}}, \binits{T.}},
\bauthor{\bsnm{{Pernet}}, \binits{J.-M.}},
\bauthor{\bsnm{{Wolter}}, \binits{H.H.}},
\bauthor{\bsnm{{Tamura}}, \binits{T.}}:
\batitle{{Coulomb functions in entire (eta,rho) - plane}}.
\bjtitle{Computer Physics Communications}
\bvolume{3}(\bissue{2}),
\bfpage{73}--\blpage{87}
(\byear{1972})
\doiurl{10.1016/0010-4655(72)90057-4}
\end{barticle}
\endbibitem

\bibitem[\protect\citeauthoryear{{Schmid} et~al.}{1988}]{1988SchmidSpitzLosch}
\begin{bbook}
\bauthor{\bsnm{{Schmid}}, \binits{E.W.}},
\bauthor{\bsnm{{Spitz}}, \binits{G.}},
\bauthor{\bsnm{{L{\"o}sch}}, \binits{W.}}:
\bbtitle{Theoretical Physics on the Personal Computer}.
\bpublisher{Springer},
\blocation{New York}
(\byear{1988})
\end{bbook}
\endbibitem

\bibitem[\protect\citeauthoryear{{Huang} et~al.}{2021}]{2021ChPhC..45c0002H}
\begin{barticle}
\bauthor{\bsnm{{Huang}}, \binits{W.J.}},
\bauthor{\bsnm{{Wang}}, \binits{M.}},
\bauthor{\bsnm{{Kondev}}, \binits{F.G.}},
\bauthor{\bsnm{{Audi}}, \binits{G.}},
\bauthor{\bsnm{{Naimi}}, \binits{S.}}:
\batitle{{The AME 2020 atomic mass evaluation (I). Evaluation of input data,
  and adjustment procedures}}.
\bjtitle{Chinese Physics C}
\bvolume{45}(\bissue{3}),
\bfpage{030002}
(\byear{2021})
\doiurl{10.1088/1674-1137/abddb0}
\end{barticle}
\endbibitem

\bibitem[\protect\citeauthoryear{{Wang} et~al.}{2021}]{2021ChPhC..45c0003W}
\begin{barticle}
\bauthor{\bsnm{{Wang}}, \binits{M.}},
\bauthor{\bsnm{{Huang}}, \binits{W.J.}},
\bauthor{\bsnm{{Kondev}}, \binits{F.G.}},
\bauthor{\bsnm{{Audi}}, \binits{G.}},
\bauthor{\bsnm{{Naimi}}, \binits{S.}}:
\batitle{{The AME 2020 atomic mass evaluation (II). Tables, graphs and
  references}}.
\bjtitle{Chinese Physics C}
\bvolume{45}(\bissue{3}),
\bfpage{030003}
(\byear{2021})
\doiurl{10.1088/1674-1137/abddaf}
\end{barticle}
\endbibitem

\bibitem[\protect\citeauthoryear{{M{\"o}ller}
  et~al.}{2016}]{2016ADNDT.109....1M}
\begin{barticle}
\bauthor{\bsnm{{M{\"o}ller}}, \binits{P.}},
\bauthor{\bsnm{{Sierk}}, \binits{A.J.}},
\bauthor{\bsnm{{Ichikawa}}, \binits{T.}},
\bauthor{\bsnm{{Sagawa}}, \binits{H.}}:
\batitle{{Nuclear ground-state masses and deformations: FRDM(2012)}}.
\bjtitle{Atomic Data and Nuclear Data Tables}
\bvolume{109},
\bfpage{1}--\blpage{204}
(\byear{2016})
\doiurl{10.1016/j.adt.2015.10.002}
{\href{https://arxiv.org/abs/1508.06294}{{arXiv:1508.06294}}}
{[nucl-th]}
\end{barticle}
\endbibitem

\bibitem[\protect\citeauthoryear{{Mason} et~al.}{2023}]{2023APS..APRH16005M}
\begin{bchapter}
\bauthor{\bsnm{{Mason}}, \binits{D.}},
\bauthor{\bsnm{{McCutchan}}, \binits{E.}},
\bauthor{\bsnm{{Sonzogni}}, \binits{A.}}:
\bctitle{{From ENSDF to NuDat: Search, Filter and Visualize Nuclear Data}}.
In: \bbtitle{APS April Meeting Abstracts}.
\bsertitle{APS Meeting Abstracts},
vol. \bseriesno{2023},
pp. \bfpage{16}--\blpage{005}
(\byear{2023})
\end{bchapter}
\endbibitem

\bibitem[\protect\citeauthoryear{{Mocelj} et~al.}{2007}]{2007PhRvC..75d5805M}
\begin{barticle}
\bauthor{\bsnm{{Mocelj}}, \binits{D.}},
\bauthor{\bsnm{{Rauscher}}, \binits{T.}},
\bauthor{\bsnm{{Mart{\'\i}nez-Pinedo}}, \binits{G.}},
\bauthor{\bsnm{{Langanke}}, \binits{K.}},
\bauthor{\bsnm{{Pacearescu}}, \binits{L.}},
\bauthor{\bsnm{{Faessler}}, \binits{A.}},
\bauthor{\bsnm{{Thielemann}}, \binits{F.-K.}},
\bauthor{\bsnm{{Alhassid}}, \binits{Y.}}:
\batitle{{Large-scale prediction of the parity distribution in the nuclear
  level density and application to astrophysical reaction rates}}.
\bjtitle{\prc}
\bvolume{75}(\bissue{4}),
\bfpage{045805}
(\byear{2007})
\doiurl{10.1103/PhysRevC.75.045805}
{\href{https://arxiv.org/abs/nucl-th/0703033}{{arXiv:nucl-th/0703033}}}
{[nucl-th]}
\end{barticle}
\endbibitem

\bibitem[\protect\citeauthoryear{{Lejeune}}{1980}]{1980PhRvC..21.1107L}
\begin{barticle}
\bauthor{\bsnm{{Lejeune}}, \binits{A.}}:
\batitle{{Low-energy optical model potential in finite nuclei from Reid's hard
  core interaction}}.
\bjtitle{\prc}
\bvolume{21}(\bissue{3}),
\bfpage{1107}--\blpage{1108}
(\byear{1980})
\doiurl{10.1103/PhysRevC.21.1107}
\end{barticle}
\endbibitem

\bibitem[\protect\citeauthoryear{{Jeukenne} et~al.}{1977}]{1977PhRvC..16...80J}
\begin{barticle}
\bauthor{\bsnm{{Jeukenne}}, \binits{J.-P.}},
\bauthor{\bsnm{{Lejeune}}, \binits{A.}},
\bauthor{\bsnm{{Mahaux}}, \binits{C.}}:
\batitle{{Optical-model potential in finite nuclei from Reid's hard core
  interaction}}.
\bjtitle{\prc}
\bvolume{16}(\bissue{1}),
\bfpage{80}--\blpage{96}
(\byear{1977})
\doiurl{10.1103/PhysRevC.16.80}
\end{barticle}
\endbibitem

\bibitem[\protect\citeauthoryear{{Kiss} et~al.}{2008}]{2008PhRvL.101s1101K}
\begin{barticle}
\bauthor{\bsnm{{Kiss}}, \binits{G.G.}},
\bauthor{\bsnm{{Rauscher}}, \binits{T.}},
\bauthor{\bsnm{{Gy{\"u}rky}}, \binits{G.}},
\bauthor{\bsnm{{Simon}}, \binits{A.}},
\bauthor{\bsnm{{F{\"u}l{\"o}p}}, \binits{Z.}},
\bauthor{\bsnm{{Somorjai}}, \binits{E.}}:
\batitle{{Coulomb Suppression of the Stellar Enhancement Factor}}.
\bjtitle{\prl}
\bvolume{101}(\bissue{19}),
\bfpage{191101}
(\byear{2008})
\doiurl{10.1103/PhysRevLett.101.191101}
{\href{https://arxiv.org/abs/0809.2676}{{arXiv:0809.2676}}}
{[astro-ph]}
\end{barticle}
\endbibitem

\bibitem[\protect\citeauthoryear{{Rauscher} et~al.}{2009}]{2009PhRvC..80c5801R}
\begin{barticle}
\bauthor{\bsnm{{Rauscher}}, \binits{T.}},
\bauthor{\bsnm{{Kiss}}, \binits{G.G.}},
\bauthor{\bsnm{{Gy{\"u}rky}}, \binits{G.}},
\bauthor{\bsnm{{Simon}}, \binits{A.}},
\bauthor{\bsnm{{F{\"u}l{\"o}p}}, \binits{Z.}},
\bauthor{\bsnm{{Somorjai}}, \binits{E.}}:
\batitle{{Suppression of the stellar enhancement factor and the reaction
  Rb85(p,n)Sr85}}.
\bjtitle{\prc}
\bvolume{80}(\bissue{3}),
\bfpage{035801}
(\byear{2009})
\doiurl{10.1103/PhysRevC.80.035801}
{\href{https://arxiv.org/abs/0908.3195}{{arXiv:0908.3195}}}
{[astro-ph.SR]}
\end{barticle}
\endbibitem

\bibitem[\protect\citeauthoryear{{Rauscher}}{2010}]{2010JPhCS.202a2013R}
\begin{bchapter}
\bauthor{\bsnm{{Rauscher}}, \binits{T.}}:
\bctitle{{Differences between stellar and laboratory reaction cross sections}}.
In: \bbtitle{Journal of Physics Conference Series}.
\bsertitle{Journal of Physics Conference Series},
vol. \bseriesno{202},
p. \bfpage{012013}
(\byear{2010}).
\doiurl{10.1088/1742-6596/202/1/012013}
\end{bchapter}
\endbibitem

\bibitem[\protect\citeauthoryear{{Mohr} et~al.}{2025}]{2025EPJA...61...89M}
\begin{barticle}
\bauthor{\bsnm{{Mohr}}, \binits{P.}},
\bauthor{\bsnm{{F{\"u}l{\"o}p}}, \binits{Z.}},
\bauthor{\bsnm{{Gy{\"u}rky}}, \binits{G.}},
\bauthor{\bsnm{{Hal{\'a}sz}}, \binits{Z.}},
\bauthor{\bsnm{{Kiss}}, \binits{G.G.}},
\bauthor{\bsnm{{Kov{\'a}cs}}, \binits{S.R.}},
\bauthor{\bsnm{{M{\'a}tyus}}, \binits{Z.}},
\bauthor{\bsnm{{Szegedi}}, \binits{T.N.}},
\bauthor{\bsnm{{Sz{\"u}cs}}, \binits{T.}}:
\batitle{{The {\ensuremath{\alpha}}-nucleus potential: towards a solution of a
  long-standing problem}}.
\bjtitle{European Physical Journal A}
\bvolume{61}(\bissue{4}),
\bfpage{89}
(\byear{2025})
\doiurl{10.1140/epja/s10050-025-01551-3}
\end{barticle}
\endbibitem

\bibitem[\protect\citeauthoryear{{McFadden} and
  {Satchler}}{1966}]{1966NucPh..84..177M}
\begin{barticle}
\bauthor{\bsnm{{McFadden}}, \binits{L.}},
\bauthor{\bsnm{{Satchler}}, \binits{G.R.}}:
\batitle{{Optical-model analysis of the scattering of 24.7 MeV alpha
  particles}}.
\bjtitle{Nuclear Physics}
\bvolume{84}(\bissue{1}),
\bfpage{177}--\blpage{200}
(\byear{1966})
\doiurl{10.1016/0029-5582(66)90441-X}
\end{barticle}
\endbibitem

\bibitem[\protect\citeauthoryear{{Mohr} et~al.}{2020}]{2020PhRvL.124y2701M}
\begin{barticle}
\bauthor{\bsnm{{Mohr}}, \binits{P.}},
\bauthor{\bsnm{{F{\"u}l{\"o}p}}, \binits{Z.}},
\bauthor{\bsnm{{Gy{\"u}rky}}, \binits{G.}},
\bauthor{\bsnm{{Kiss}}, \binits{G.G.}},
\bauthor{\bsnm{{Sz{\"u}cs}}, \binits{T.}}:
\batitle{{Successful Prediction of Total {\ensuremath{\alpha}} -Induced
  Reaction Cross Sections at Astrophysically Relevant Sub-Coulomb Energies
  Using a Novel Approach}}.
\bjtitle{\prl}
\bvolume{124}(\bissue{25}),
\bfpage{252701}
(\byear{2020})
\doiurl{10.1103/PhysRevLett.124.252701}
{\href{https://arxiv.org/abs/2006.03885}{{arXiv:2006.03885}}}
{[nucl-th]}
\end{barticle}
\endbibitem

\bibitem[\protect\citeauthoryear{Mohr et~al.}{2021}]{MOHR2021101453}
\begin{barticle}
\bauthor{\bsnm{Mohr}, \binits{P.}},
\bauthor{\bsnm{Fülöp}, \binits{Z.}},
\bauthor{\bsnm{Gyürky}, \binits{G.}},
\bauthor{\bsnm{Kiss}, \binits{G.G.}},
\bauthor{\bsnm{Szücs}, \binits{T.}},
\bauthor{\bsnm{Arcones}, \binits{A.}},
\bauthor{\bsnm{Jacobi}, \binits{M.}},
\bauthor{\bsnm{Psaltis}, \binits{A.}}:
\batitle{Astrophysical reaction rates of $\alpha$-induced reactions for nuclei
  with $26\leq z\leq 83$ from the new atomki-v2 $\alpha$-nucleus potential}.
\bjtitle{Atomic Data and Nuclear Data Tables}
\bvolume{142},
\bfpage{101453}
(\byear{2021})
\doiurl{10.1016/j.adt.2021.101453}
\end{barticle}
\endbibitem

\bibitem[\protect\citeauthoryear{{Goriely} et~al.}{2007}]{2007PhRvC..75f4312G}
\begin{barticle}
\bauthor{\bsnm{{Goriely}}, \binits{S.}},
\bauthor{\bsnm{{Samyn}}, \binits{M.}},
\bauthor{\bsnm{{Pearson}}, \binits{J.M.}}:
\batitle{{Further explorations of Skyrme-Hartree-Fock-Bogoliubov mass formulas.
  VII. Simultaneous fits to masses and fission barriers}}.
\bjtitle{\prc}
\bvolume{75}(\bissue{6}),
\bfpage{064312}
(\byear{2007})
\doiurl{10.1103/PhysRevC.75.064312}
\end{barticle}
\endbibitem

\bibitem[\protect\citeauthoryear{{Capote} et~al.}{2009}]{2009NDS...110.3107C}
\begin{barticle}
\bauthor{\bsnm{{Capote}}, \binits{R.}},
\bauthor{\bsnm{{Herman}}, \binits{M.}},
\bauthor{\bsnm{{Oblo{\v{z}}insk{\'y}}}, \binits{P.}},
\bauthor{\bsnm{{Young}}, \binits{P.G.}},
\bauthor{\bsnm{{Goriely}}, \binits{S.}},
\bauthor{\bsnm{{Belgya}}, \binits{T.}},
\bauthor{\bsnm{{Ignatyuk}}, \binits{A.V.}},
\bauthor{\bsnm{{Koning}}, \binits{A.J.}},
\bauthor{\bsnm{{Hilaire}}, \binits{S.}},
\bauthor{\bsnm{{Plujko}}, \binits{V.A.}},
\bauthor{\bsnm{{Avrigeanu}}, \binits{M.}},
\bauthor{\bsnm{{Bersillon}}, \binits{O.}},
\bauthor{\bsnm{{Chadwick}}, \binits{M.B.}},
\bauthor{\bsnm{{Fukahori}}, \binits{T.}},
\bauthor{\bsnm{{Ge}}, \binits{Z.}},
\bauthor{\bsnm{{Han}}, \binits{Y.}},
\bauthor{\bsnm{{Kailas}}, \binits{S.}},
\bauthor{\bsnm{{Kopecky}}, \binits{J.}},
\bauthor{\bsnm{{Maslov}}, \binits{V.M.}},
\bauthor{\bsnm{{Reffo}}, \binits{G.}},
\bauthor{\bsnm{{Sin}}, \binits{M.}},
\bauthor{\bsnm{{Soukhovitskii}}, \binits{E.S.}},
\bauthor{\bsnm{{Talou}}, \binits{P.}}:
\batitle{{RIPL - Reference Input Parameter Library for Calculation of Nuclear
  Reactions and Nuclear Data Evaluations}}.
\bjtitle{Nuclear Data Sheets}
\bvolume{110}(\bissue{12}),
\bfpage{3107}--\blpage{3214}
(\byear{2009})
\doiurl{10.1016/j.nds.2009.10.004}
\end{barticle}
\endbibitem

\bibitem[\protect\citeauthoryear{Thielemann and
  Arnould}{1983}]{thielemannarnouldGDR}
\begin{bchapter}
\bauthor{\bsnm{Thielemann}, \binits{F.-K.}},
\bauthor{\bsnm{Arnould}, \binits{M.}}:
\bctitle{Average radiation widths and the giant dipole resonance width}.
In: \beditor{\bsnm{B{\"o}ckhoff}, \binits{K.H.}} (ed.)
\bbtitle{Nuclear Data for Science and Technology},
pp. \bfpage{762}--\blpage{765}.
\bpublisher{Springer},
\blocation{Dordrecht}
(\byear{1983})
\end{bchapter}
\endbibitem

\bibitem[\protect\citeauthoryear{{McCullagh}
  et~al.}{1981}]{1981PhRvC..23.1394M}
\begin{barticle}
\bauthor{\bsnm{{McCullagh}}, \binits{C.M.}},
\bauthor{\bsnm{{Stelts}}, \binits{M.L.}},
\bauthor{\bsnm{{Chrien}}, \binits{R.E.}}:
\batitle{{Dipole radiative strength functions from resonance neutron capture}}.
\bjtitle{\prc}
\bvolume{23}(\bissue{4}),
\bfpage{1394}--\blpage{1403}
(\byear{1981})
\doiurl{10.1103/PhysRevC.23.1394}
\end{barticle}
\endbibitem

\bibitem[\protect\citeauthoryear{{Prestwich}
  et~al.}{1984}]{1984ZPhyA.315..103P}
\begin{barticle}
\bauthor{\bsnm{{Prestwich}}, \binits{W.V.}},
\bauthor{\bsnm{{Islam}}, \binits{M.A.}},
\bauthor{\bsnm{{Kennett}}, \binits{T.J.}}:
\batitle{{Primary E2 transitions observed following neutron capture for the
  mass region 144{\ensuremath{\leqq}}A{\ensuremath{\leqq}}180}}.
\bjtitle{Zeitschrift fur Physik A Hadrons and Nuclei}
\bvolume{315}(\bissue{1}),
\bfpage{103}--\blpage{111}
(\byear{1984})
\doiurl{10.1007/BF01436215}
\end{barticle}
\endbibitem

\bibitem[\protect\citeauthoryear{{Speth} and {van der
  Woude}}{1981}]{1981RPPh...44..719S}
\begin{barticle}
\bauthor{\bsnm{{Speth}}, \binits{J.}},
\bauthor{\bsnm{{van der Woude}}, \binits{A.}}:
\batitle{{Giant resonances in nuclei}}.
\bjtitle{Reports on Progress in Physics}
\bvolume{44}(\bissue{7}),
\bfpage{719}--\blpage{786}
(\byear{1981})
\doiurl{10.1088/0034-4885/44/7/002}
\end{barticle}
\endbibitem

\bibitem[\protect\citeauthoryear{{Rauscher}}{2013}]{2013JPhCS.420a2138R}
\begin{bchapter}
\bauthor{\bsnm{{Rauscher}}, \binits{T.}}:
\bctitle{{General properties of astrophysical reaction rates in explosive
  nucleosynthesis}}.
In: \bbtitle{Journal of Physics Conference Series}.
\bsertitle{Journal of Physics Conference Series},
vol. \bseriesno{420},
p. \bfpage{012138}.
\bpublisher{IOP}, \blocation{???}
(\byear{2013}).
\doiurl{10.1088/1742-6596/420/1/012138}
\end{bchapter}
\endbibitem

\bibitem[\protect\citeauthoryear{{Rauscher}}{2014}]{2014AIPA....4d1012R}
\begin{barticle}
\bauthor{\bsnm{{Rauscher}}, \binits{T.}}:
\batitle{{Challenges in nucleosynthesis of trans-iron elements}}.
\bjtitle{AIP Advances}
\bvolume{4}(\bissue{4}),
\bfpage{041012}
(\byear{2014})
\doiurl{10.1063/1.4868239}
{\href{https://arxiv.org/abs/1403.2015}{{arXiv:1403.2015}}}
{[astro-ph.SR]}
\end{barticle}
\endbibitem

\bibitem[\protect\citeauthoryear{{Rauscher}}{2015}]{2015AIPC.1681e0003R}
\begin{bchapter}
\bauthor{\bsnm{{Rauscher}}, \binits{T.}}:
\bctitle{{Nuclear reactions for nucleosynthesis beyond Fe}}.
In: \bbtitle{Nuclear Structure and Dynamics 2015}.
\bsertitle{American Institute of Physics Conference Series},
vol. \bseriesno{1681},
p. \bfpage{050003}.
\bpublisher{AIP}, \blocation{???}
(\byear{2015}).
\doiurl{10.1063/1.4932278}
\end{bchapter}
\endbibitem

\bibitem[\protect\citeauthoryear{{Rauscher}}{2006}]{2006PhRvC..73a5804R}
\begin{barticle}
\bauthor{\bsnm{{Rauscher}}, \binits{T.}}:
\batitle{{Branchings in the {\ensuremath{\gamma}} process path revisited}}.
\bjtitle{\prc}
\bvolume{73}(\bissue{1}),
\bfpage{015804}
(\byear{2006})
\doiurl{10.1103/PhysRevC.73.015804}
{\href{https://arxiv.org/abs/astro-ph/0510710}{{arXiv:astro-ph/0510710}}}
{[astro-ph]}
\end{barticle}
\endbibitem

\bibitem[\protect\citeauthoryear{{Rauscher} et~al.}{2013}]{2013RPPh...76f6201R}
\begin{barticle}
\bauthor{\bsnm{{Rauscher}}, \binits{T.}},
\bauthor{\bsnm{{Dauphas}}, \binits{N.}},
\bauthor{\bsnm{{Dillmann}}, \binits{I.}},
\bauthor{\bsnm{{Fr{\"o}hlich}}, \binits{C.}},
\bauthor{\bsnm{{F{\"u}l{\"o}p}}, \binits{Z.}},
\bauthor{\bsnm{{Gy{\"u}rky}}, \binits{G.}}:
\batitle{{Constraining the astrophysical origin of the p-nuclei through nuclear
  physics and meteoritic data}}.
\bjtitle{Reports on Progress in Physics}
\bvolume{76}(\bissue{6}),
\bfpage{066201}
(\byear{2013})
\doiurl{10.1088/0034-4885/76/6/066201}
{\href{https://arxiv.org/abs/1303.2666}{{arXiv:1303.2666}}}
{[astro-ph.SR]}
\end{barticle}
\endbibitem

\bibitem[\protect\citeauthoryear{{Krausmann}
  et~al.}{1996}]{1996PhRvC..53..469K}
\begin{barticle}
\bauthor{\bsnm{{Krausmann}}, \binits{E.}},
\bauthor{\bsnm{{Balogh}}, \binits{W.}},
\bauthor{\bsnm{{Oberhummer}}, \binits{H.}},
\bauthor{\bsnm{{Rauscher}}, \binits{T.}},
\bauthor{\bsnm{{Kratz}}, \binits{K.-L.}},
\bauthor{\bsnm{{Ziegert}}, \binits{W.}}:
\batitle{{Direct neutron capture for magic-shell nuclei}}.
\bjtitle{\prc}
\bvolume{53}(\bissue{1}),
\bfpage{469}--\blpage{474}
(\byear{1996})
\doiurl{10.1103/PhysRevC.53.469}
{\href{https://arxiv.org/abs/nucl-th/9511027}{{arXiv:nucl-th/9511027}}}
{[nucl-th]}
\end{barticle}
\endbibitem

\bibitem[\protect\citeauthoryear{{Khaliel} et~al.}{2017}]{2017PhRvC..96c5806K}
\begin{barticle}
\bauthor{\bsnm{{Khaliel}}, \binits{A.}},
\bauthor{\bsnm{{Mertzimekis}}, \binits{T.J.}},
\bauthor{\bsnm{{Asimakopoulou}}, \binits{E.-M.}},
\bauthor{\bsnm{{Kanellakopoulos}}, \binits{A.}},
\bauthor{\bsnm{{Lagaki}}, \binits{V.}},
\bauthor{\bsnm{{Psaltis}}, \binits{A.}},
\bauthor{\bsnm{{Psyrra}}, \binits{I.}},
\bauthor{\bsnm{{Mavrommatis}}, \binits{E.}}:
\batitle{{First cross-section measurements of the reactions
  $^{Ag}$,{}109$_{107}$(p ,{\ensuremath{\gamma}} )$^{Cd}$,{}110$_{108}$ at
  energies relevant to the p process}}.
\bjtitle{\prc}
\bvolume{96}(\bissue{3}),
\bfpage{035806}
(\byear{2017})
\doiurl{10.1103/PhysRevC.96.035806}
\end{barticle}
\endbibitem

\bibitem[\protect\citeauthoryear{{Harissopulos}
  et~al.}{2024}]{2024PhRvC.110a5803H}
\begin{barticle}
\bauthor{\bsnm{{Harissopulos}}, \binits{S.}},
\bauthor{\bsnm{{Vagena}}, \binits{E.}},
\bauthor{\bsnm{{Spyrou}}, \binits{A.}},
\bauthor{\bsnm{{Axiotis}}, \binits{M.}},
\bauthor{\bsnm{{Kotsina}}, \binits{Z.}},
\bauthor{\bsnm{{Tsampa}}, \binits{K.}},
\bauthor{\bsnm{{Lagoyannis}}, \binits{A.}},
\bauthor{\bsnm{{Dimitriou}}, \binits{P.}},
\bauthor{\bsnm{{Becker}}, \binits{H.-W.}},
\bauthor{\bsnm{{Foteinou}}, \binits{V.}}:
\batitle{{(p ,{\ensuremath{\gamma}} ) cross section measurements on Sn isotopes
  relevant to the p process}}.
\bjtitle{\prc}
\bvolume{110}(\bissue{1}),
\bfpage{015803}
(\byear{2024})
\doiurl{10.1103/PhysRevC.110.015803}
\end{barticle}
\endbibitem

\bibitem[\protect\citeauthoryear{{Dellmann} et~al.}{2025}]{2025PhRvL.134n2701D}
\begin{barticle}
\bauthor{\bsnm{{Dellmann}}, \binits{S.F.}},
\bauthor{\bsnm{{Glorius}}, \binits{J.}},
\bauthor{\bsnm{{Litvinov}}, \binits{Y.A.}},
\bauthor{\bsnm{{Reifarth}}, \binits{R.}},
\bauthor{\bsnm{{Varga}}, \binits{L.}},
\bauthor{\bsnm{{Aliotta}}, \binits{M.}},
\bauthor{\bsnm{{Amjad}}, \binits{F.}},
\bauthor{\bsnm{{Blaum}}, \binits{K.}},
\bauthor{\bsnm{{Bott}}, \binits{L.}},
\bauthor{\bsnm{{Brandau}}, \binits{C.}},
\bauthor{\bsnm{{Br{\"u}ckner}}, \binits{B.}},
\bauthor{\bsnm{{Bruno}}, \binits{C.G.}},
\bauthor{\bsnm{{Chen}}, \binits{R.-J.}},
\bauthor{\bsnm{{Davinson}}, \binits{T.}},
\bauthor{\bsnm{{Dickel}}, \binits{T.}},
\bauthor{\bsnm{{Dillmann}}, \binits{I.}},
\bauthor{\bsnm{{Dmytriev}}, \binits{D.}},
\bauthor{\bsnm{{Erbacher}}, \binits{P.}},
\bauthor{\bsnm{{Forstner}}, \binits{O.}},
\bauthor{\bsnm{{Freire-Fern{\'a}ndez}}, \binits{D.}},
\bauthor{\bsnm{{Geissel}}, \binits{H.}},
\bauthor{\bsnm{{G{\"o}bel}}, \binits{K.}},
\bauthor{\bsnm{{Griffin}}, \binits{C.J.}},
\bauthor{\bsnm{{Grisenti}}, \binits{R.E.}},
\bauthor{\bsnm{{Gumberidze}}, \binits{A.}},
\bauthor{\bsnm{{Haettner}}, \binits{E.}},
\bauthor{\bsnm{{Hagmann}}, \binits{S.}},
\bauthor{\bsnm{{Heftrich}}, \binits{T.}},
\bauthor{\bsnm{{Heil}}, \binits{M.}},
\bauthor{\bsnm{{He{\ss}}}, \binits{R.}},
\bauthor{\bsnm{{Hillenbrand}}, \binits{P.-M.}},
\bauthor{\bsnm{{Hornung}}, \binits{C.}},
\bauthor{\bsnm{{Joseph}}, \binits{R.}},
\bauthor{\bsnm{{Jurado}}, \binits{B.}},
\bauthor{\bsnm{{Kazanseva}}, \binits{E.}},
\bauthor{\bsnm{{Khasawneh}}, \binits{K.}},
\bauthor{\bsnm{{Kn{\"o}bel}}, \binits{R.}},
\bauthor{\bsnm{{Kostyleva}}, \binits{D.}},
\bauthor{\bsnm{{Kozhuharov}}, \binits{C.}},
\bauthor{\bsnm{{Kulikov}}, \binits{I.}},
\bauthor{\bsnm{{Kuzminchuk}}, \binits{N.}},
\bauthor{\bsnm{{Kurtulgil}}, \binits{D.}},
\bauthor{\bsnm{{Langer}}, \binits{C.}},
\bauthor{\bsnm{{Leckenby}}, \binits{G.}},
\bauthor{\bsnm{{Lederer-Woods}}, \binits{C.}},
\bauthor{\bsnm{{Lestinsky}}, \binits{M.}},
\bauthor{\bsnm{{Litvinov}}, \binits{S.}},
\bauthor{\bsnm{{L{\"o}her}}, \binits{B.}},
\bauthor{\bsnm{{Lorentz}}, \binits{B.}},
\bauthor{\bsnm{{Lorenz}}, \binits{E.}},
\bauthor{\bsnm{{Marsh}}, \binits{J.}},
\bauthor{\bsnm{{Menz}}, \binits{E.}},
\bauthor{\bsnm{{Morgenroth}}, \binits{T.}},
\bauthor{\bsnm{{Mukha}}, \binits{I.}},
\bauthor{\bsnm{{Petridis}}, \binits{N.}},
\bauthor{\bsnm{{Popp}}, \binits{U.}},
\bauthor{\bsnm{{Psaltis}}, \binits{A.}},
\bauthor{\bsnm{{Purushothaman}}, \binits{S.}},
\bauthor{\bsnm{{Rocco}}, \binits{E.}},
\bauthor{\bsnm{{Roy}}, \binits{P.}},
\bauthor{\bsnm{{Sanjari}}, \binits{M.S.}},
\bauthor{\bsnm{{Scheidenberger}}, \binits{C.}},
\bauthor{\bsnm{{Sguazzin}}, \binits{M.}},
\bauthor{\bsnm{{Sidhu}}, \binits{R.S.}},
\bauthor{\bsnm{{Spillmann}}, \binits{U.}},
\bauthor{\bsnm{{Steck}}, \binits{M.}},
\bauthor{\bsnm{{St{\"o}hlker}}, \binits{T.}},
\bauthor{\bsnm{{Surzhykov}}, \binits{A.}},
\bauthor{\bsnm{{Swartz}}, \binits{J.A.}},
\bauthor{\bsnm{{Tanaka}}, \binits{Y.}},
\bauthor{\bsnm{{T{\"o}rnqvist}}, \binits{H.}},
\bauthor{\bsnm{{Vescovi}}, \binits{D.}},
\bauthor{\bsnm{{Volknandt}}, \binits{M.}},
\bauthor{\bsnm{{Weick}}, \binits{H.}},
\bauthor{\bsnm{{Weigand}}, \binits{M.}},
\bauthor{\bsnm{{Woods}}, \binits{P.J.}},
\bauthor{\bsnm{{Yamaguchi}}, \binits{T.}},
\bauthor{\bsnm{{Zhao}}, \binits{J.}}:
\batitle{{First Proton-Induced Cross Sections on a Stored Rare Ion Beam:
  Measurement of Te118(p,{\ensuremath{\gamma}}) for Explosive
  Nucleosynthesis}}.
\bjtitle{\prl}
\bvolume{134}(\bissue{14}),
\bfpage{142701}
(\byear{2025})
\doiurl{10.1103/PhysRevLett.134.142701}
{\href{https://arxiv.org/abs/2510.08817}{{arXiv:2510.08817}}}
{[nucl-ex]}
\end{barticle}
\endbibitem

\bibitem[\protect\citeauthoryear{{Kiss} et~al.}{2025}]{2025ApJ...988..170K}
\begin{barticle}
\bauthor{\bsnm{{Kiss}}, \binits{G.G.}},
\bauthor{\bsnm{{Kov{\'a}cs}}, \binits{S.R.}},
\bauthor{\bsnm{{Szegedi}}, \binits{T.N.}},
\bauthor{\bsnm{{Mohr}}, \binits{P.}},
\bauthor{\bsnm{{Montes}}, \binits{F.}},
\bauthor{\bsnm{{Arcones}}, \binits{A.}},
\bauthor{\bsnm{{T{\'o}th}}, \binits{{\'A}.}},
\bauthor{\bsnm{{N{\'e}meth}}, \binits{A.}},
\bauthor{\bsnm{{Szil{\'a}gyi}}, \binits{E.}},
\bauthor{\bsnm{{Pal}}, \binits{M.K.}},
\bauthor{\bsnm{{Braun}}, \binits{M.}},
\bauthor{\bsnm{{Hal{\'a}sz}}, \binits{Z.}},
\bauthor{\bsnm{{Elekes}}, \binits{Z.}},
\bauthor{\bsnm{{Gy{\"u}rky}}, \binits{G.}},
\bauthor{\bsnm{{Sz{\"u}cs}}, \binits{T.}}:
\batitle{{Low-energy Measurement of the $^{86}$Kr({\ensuremath{\alpha}},
  n)$^{89}$Sr Reaction Cross Section and Its Impact on Weak R-process
  Nucleosynthesis}}.
\bjtitle{\apj}
\bvolume{988}(\bissue{2}),
\bfpage{170}
(\byear{2025})
\doiurl{10.3847/1538-4357/ade4c9}
\end{barticle}
\endbibitem

\bibitem[\protect\citeauthoryear{{Kiss} et~al.}{2021}]{2021ApJ...908..202K}
\begin{barticle}
\bauthor{\bsnm{{Kiss}}, \binits{G.G.}},
\bauthor{\bsnm{{Szegedi}}, \binits{T.N.}},
\bauthor{\bsnm{{Mohr}}, \binits{P.}},
\bauthor{\bsnm{{Jacobi}}, \binits{M.}},
\bauthor{\bsnm{{Gy{\"u}rky}}, \binits{G.}},
\bauthor{\bsnm{{Husz{\'a}nk}}, \binits{R.}},
\bauthor{\bsnm{{Arcones}}, \binits{A.}}:
\batitle{{Low-energy Measurement of the
  $^{96}$Zr({\ensuremath{\alpha}},n)$^{99}$Mo Reaction Cross Section and Its
  Impact on Weak r-process Nucleosynthesis}}.
\bjtitle{\apj}
\bvolume{908}(\bissue{2}),
\bfpage{202}
(\byear{2021})
\doiurl{10.3847/1538-4357/abd2bc}
{\href{https://arxiv.org/abs/2009.13169}{{arXiv:2009.13169}}}
{[nucl-ex]}
\end{barticle}
\endbibitem

\bibitem[\protect\citeauthoryear{{Szegedi} et~al.}{2021}]{2021PhRvC.104c5804S}
\begin{barticle}
\bauthor{\bsnm{{Szegedi}}, \binits{T.N.}},
\bauthor{\bsnm{{Kiss}}, \binits{G.G.}},
\bauthor{\bsnm{{Mohr}}, \binits{P.}},
\bauthor{\bsnm{{Psaltis}}, \binits{A.}},
\bauthor{\bsnm{{Jacobi}}, \binits{M.}},
\bauthor{\bsnm{{Barnaf{\"o}ldi}}, \binits{G.G.}},
\bauthor{\bsnm{{Sz{\"u}cs}}, \binits{T.}},
\bauthor{\bsnm{{Gy{\"u}rky}}, \binits{G.}},
\bauthor{\bsnm{{Arcones}}, \binits{A.}}:
\batitle{{Activation thick target yield measurement of
  $^{100}$Mo({\ensuremath{\alpha}} ,n )$^{103}$Ru for studying the weak r
  -process nucleosynthesis}}.
\bjtitle{\prc}
\bvolume{104}(\bissue{3}),
\bfpage{035804}
(\byear{2021})
\doiurl{10.1103/PhysRevC.104.035804}
\end{barticle}
\endbibitem

\end{thebibliography}

\end{document}